\documentclass[12pt]{article}
\usepackage{graphicx,amssymb,amsmath,tensor,empheq,color}

\usepackage{physics}

\usepackage{hyperref}
\hypersetup{colorlinks = true, linkcolor=black, citecolor=black, urlcolor=blue}
\usepackage{url}

\makeatletter

\newlength{\fighskip} \fighskip=2pt
\newlength{\figvskip} \figvskip=3pt

\newcommand*{\figbox}[2]{{
  \def\figscale{#1}
  \def\arraystretch{0.8}
  \arraycolsep=0pt
  \begin{array}{c}
    \vbox{\vskip\figscale\figvskip
      \hbox{\hskip\figscale\fighskip
        \includegraphics[scale=\figscale]{#2}}}
  \end{array}}}

\newcommand*{\widebox}[1]{\setlength{\fboxsep}{1ex}%
  \fbox{#1}}
\newcommand*{\wideboxed}[1]{\setlength{\fboxsep}{1ex}%
  \fbox{\m@th$\displaystyle#1$}}

\newcommand*{\useshortskip}[1]{{%
\setlength\abovedisplayskip\abovedisplayshortskip#1}\ignorespaces}

\def\ubrace#1_#2{%
  \underbrace{#1}_{\hb@xt@\z@{\hss$\scriptstyle#2$\hss}}}

\newcommand*{\V}[1]{\boldsymbol{#1}}

\newcommand{\hgf}{%
\,\tensor[_{2\kern-1.2pt}]{F}{_{\kern-0.8pt 1}}\kern-1.2pt}
\newcommand{\hgfs}{\mathbf{F}}

\newcommand{\blangle}{\bigl\langle}
\newcommand{\brangle}{\bigr\rangle}
\newcommand{\dlangle}{\langle\kern-1.5pt\langle}
\newcommand{\drangle}{\rangle\kern-1.5pt\rangle}
\newcommand{\bdlangle}{\blangle\kern-3pt\blangle}
\newcommand{\bdrangle}{\brangle\kern-3pt\brangle}

\newcommand*{\braa}[1]{\langle{#1}|}
\newcommand*{\kett}[1]{|{#1}\rangle}
\newcommand*{\brakett}[2]{\langle{#1}|{#2}\rangle}

\newcommand*{\bbraa}[1]{\blangle{#1}\big|}
\newcommand*{\bkett}[1]{\big|{#1}\brangle}
\newcommand*{\bbrakett}[2]{\blangle{#1}\big|{#2}\brangle}

\newcommand*{\corr}[1]{\langle{#1}\rangle}
\newcommand*{\bcorr}[1]{\blangle{#1}\brangle}
\newcommand*{\ccorr}[1]{\dlangle{#1}\drangle}
\newcommand*{\bccorr}[1]{\bdlangle{#1}\bdrangle}

\renewcommand{\le}{\leqslant}
\renewcommand{\ge}{\geqslant}

\newcommand{\eps}{\varepsilon}
\newcommand{\ph}{\varphi}
\newcommand{\kap}{\varkappa}
\newcommand{\vth}{\vartheta}

\newcommand{\calA}{\mathcal{A}}

\newcommand{\calC}{\mathcal{C}}
\newcommand{\calD}{\mathcal{D}}

\newcommand{\calF}{\mathcal{F}}

\newcommand{\calH}{\mathcal{H}}

\newcommand{\calO}{\mathcal{O}}

\newcommand{\calU}{\mathcal{U}}

\newcommand{\ZZ}{\mathbb{Z}}

\newcommand{\RR}{\mathbb{R}}
\newcommand{\CC}{\mathbb{C}}

\DeclareMathOperator{\Trr}{Tr}
\DeclareMathOperator{\sgn}{sgn}

\DeclareMathOperator{\Pf}{Pf}
\DeclareMathOperator{\Ress}{Res}
\DeclareMathOperator{\Ld}{\mathcal{L}}
\newcommand{\const}{\mathrm{const}}
\newcommand{\Tt}{\mathrm{T}}

\DeclareMathOperator{\SO}{SO}

\DeclareMathOperator{\SU}{SU}

\DeclareMathOperator{\SL}{SL}
\DeclareMathOperator{\PSL}{PSL}
\DeclareMathOperator{\sL}{\mathfrak{sl}}

\DeclareMathOperator{\Diff}{Diff}

\DeclareMathOperator{\Sch}{Sch}
\DeclareMathOperator{\TT}{\mathbf{T}}
\DeclareMathOperator{\AdS}{AdS}

\newcommand{\GG}{\mathfrak{G}}

\newcommand{\bb}{\mathrm{b}}
\newcommand{\cc}{\mathrm{c}}
\newcommand{\eff}{\text{eff}}
\newcommand{\loc}{\text{local}}
\newcommand{\nloc}{\text{non-local}}
\newcommand{\IR}{\text{IR}}
\newcommand{\UV}{\text{UV}}
\newcommand{\Ret}{\text{R}}
\newcommand{\Adv}{\text{A}}
\newcommand{\Wig}{\text{W}}
\newcommand{\IN}{\text{in}}
\newcommand{\OUT}{\text{out}}
\newcommand{\naive}{\text{naive}}
\newcommand{\bath}{\text{bath}}
\newcommand{\other}{\text{other}}

\newcommand{\Grav}{\mathrm{G}}
\newcommand{\Matter}{\mathrm{M}}

\newcommand{\tI}{\tilde{I}}
\newcommand{\tG}{\widetilde{G}}
\newcommand{\tF}{\widetilde{\calF}}
\newcommand{\tSig}{\widetilde{\Sigma}}
\newcommand{\tsig}{\tilde{\sigma}}
\newcommand{\tkap}{\tilde{\kap}}
\newcommand{\tUp}{\widetilde{\Upsilon}}

\newcommand{\OO}{\mathcal{O}}
\newcommand{\unit}{\mathbf{1}}

\newcommand{\al}{\alpha}

\newcommand{\tht}{\theta}

\newcommand{\ga}{\gamma}

\newcommand{\de}{\delta}
\newcommand{\De}{\Delta}

\newcommand{\vt}{\vartheta}
\newcommand{\vep}{\varepsilon}
\newcommand{\tvep}{\tilde{\varepsilon}}
\newcommand{\vp}{\varphi}

\newcommand{\wc}{\overline{w}}

\newcommand{\be}{\begin{equation}}
\newcommand{\ee}{\end{equation}}
\newcommand{\bea}{\begin{eqnarray}}
\newcommand{\eea}{\end{eqnarray}}
\newcommand{\nn}{\nonumber\\}

\def\eg{e.g.\ }
\def\ie{i.e.\ }
\def\cf{cf.\ }

\newcommand{\ov}{\over}
\newcommand{\p}{\partial}

\makeatother

\title {The soft mode in the Sachdev-Ye-Kitaev model\\ and its gravity dual}

\author{Alexei Kitaev\footnote{kitaev@caltech.edu}\\
\normalsize\it California Institute of Technology, Pasadena, CA 91125, U.S.A.
\\[5pt]
S.\ Josephine Suh\footnote{suh@caltech.edu}\\
\normalsize\it University of British Columbia, Vancouver, BC V6T 0C2, Canada,\\
\normalsize\it California Institute of Technology, Pasadena, CA 91125, U.S.A.\vspace{0.5cm}}

\date{May 14, 2018}

\begin{document}

\setcounter{tocdepth}{2}

\maketitle
\begin{abstract}
We give an exposition of the SYK model with several new results. A non-local correction to the Schwarzian effective action is found. The same action is obtained by integrating out the bulk degrees of freedom in a certain variant of dilaton gravity. We also discuss general properties of out-of-time-order correlators.
\end{abstract}

\newpage
\tableofcontents
\newpage

\section{Introduction}

\paragraph{The model:}
The SYK model is a quantum system with many degrees of freedom and random all-to-all interactions. It is analytically solvable and exhibits interesting properties at low temperatures. In particular, it has a collective mode that is similar to the Dray-t'Hooft shock waves at the black hole horizon. The original model of Sachdev and Ye \cite{SaYe93} consists of pairwise coupled $\SU(M)$ spins. Kitaev \cite{Kit.KITP.1,Kit.KITP} proposed a simpler Hamiltonian with $N\gg 1$ Majorana sites and four-body interactions:
\begin{equation}\label{H_SYK4}
H = -\sum_{j<k<l<m}J_{jklm}\, \chi_j\chi_k\chi_l\chi_m,
\qquad \text{where}\quad
\chi_j\chi_k+\chi_k\chi_j=\delta_{jk}.
\end{equation}
The couplings $J_{jklm}$ are independent random variables with zero mean and the following variance:
\begin{equation}
\overline{J_{jklm}^2}=\frac{3!J^2}{N^3}.
\end{equation}
They may be regarded as elements of an antisymmetric tensor such that $\frac{1}{3!}\sum_{k,l,m}J_{jklm}^2\approx J^2$ for each $j$. The number $J$ is the characteristic energy scale. This variant of the model is also more convenient because disorder effects are weaker than in systems with pairwise interactions. A slight generalization involves interactions of order $q$:
\begin{equation}\label{H_SYK}
\wideboxed{
H = \frac{i^{q/2}}{q!}\sum_{j_1,\dots,j_q}J_{j_1\cdots j_q}\,
\chi_{j_1}\dots\chi_{j_q},
\qquad\quad
\overline{J_{j_1\cdots j_q}^2}=\frac{(q-1)!J^2}{N^{q-1}}
}
\end{equation}

In the $N\to\infty$ limit, the model is solved using dynamical mean field theory. Indeed, each variable $\chi_j$ is driven by the effective fermionic field $\xi_j=-i\,\partial H/\partial\chi_j \propto\sum_{j_2,\dots,j_q} J_{j_2\cdots j_q}\chi_{j_2}\dots\chi_{j_q}$. Being a sum of many random terms, $\xi_j(\tau)$ is Gaussian. Furthermore, $\chi_{j_2}(\tau),\dots,\chi_{j_q}(\tau)$ are almost uncorrelated. Thus, one can write the self-consistency (Schwinger-Dyson) equations for the imaginary time correlation functions
\begin{equation}
G(\tau_1,\tau_2)=-\bcorr{\TT\chi_j(\tau_1)\chi_j(\tau_2)},\qquad
\Sigma(\tau_1,\tau_2)=-\bcorr{\TT\xi_j(\tau_1)\xi_j(\tau_2)},
\end{equation} 
in a closed form:
\begin{equation}\label{Schwinger-Dyson}
\hat{G}^{-1}=-\partial_\tau-\hat{\Sigma},\qquad
\Sigma(\tau_1,\tau_2)=J^2G(\tau_1,\tau_2)^{q-1}.
\end{equation}
The Green function $G$ and self-energy $\Sigma$ are antisymmetric functions of $\tau_1,\tau_2\in[0,\beta]$ with antiperiodic boundary conditions, where $\beta$ is the inverse temperature. These equations can also be obtained from the high temperature diagrammatic expansion. It begins with the bare Green function $\hat{G}_{\bb}=(-\partial_\tau)^{-1}$, \ie $G_{\bb}(\tau,\tau')=-\frac{1}{2}\sgn(\tau-\tau')$, denoted by a thin solid line. Neglecting the diagrams that are suppressed by $1/N$, the full Green function is:
\begin{equation}\label{G_hte}
\figbox{1.0}{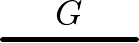}\: =\: \figbox{1.0}{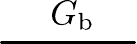} +\figbox{1.0}{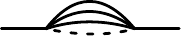}
+\figbox{1.0}{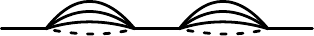} +\figbox{1.0}{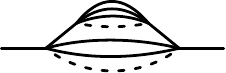} +\cdots
\end{equation}

The Schwinger-Dyson equations can be solved numerically; an analytic solution exists for $\beta,|\tau_1-\tau_2|\gg J^{-1}$. Sachdev and Ye \cite{SaYe93} originally found the Green function at zero temperature. When adapted to Hamiltonian \eqref{H_SYK}, the solution (in the $J|\tau_1-\tau_2|\to \infty$ limit) reads:
\begin{equation}\label{Ginfty}
G_{\beta=\infty}(\tau_1,\tau_2)=
-b^{\Delta}\bigl|J(\tau_1-\tau_2)\bigr|^{-2\Delta} \sgn(\tau_1-\tau_2),
\qquad\quad \Delta=\frac{1}{q},
\end{equation}
where $b$ is some numerical factor (see Table~\ref{tab_coeff} on page~\pageref{tab_coeff}). Parcollet and Georges extended this result to finite values of $\beta$ and argued that the form of the Green function indicates an emergent conformal symmetry \cite{PaGe99}:
\begin{equation}\label{Gbeta}
G(\tau_1,\tau_2)
\approx\biggl(\frac{2\pi}{\beta J}\biggr)^{2\Delta}
\tG_{\cc}\biggl(\frac{2\pi\tau_1}{\beta},\,\frac{2\pi\tau_2}{\beta}\biggr)\qquad
\text{if}\quad |\tau_1-\tau_2|\gg J^{-1},
\end{equation}
where\footnote{The subscript $\cc$ means ``conformal'' and the tilde ``renormalized'' (in this case, using a dimensionless time).}\vspace{-5pt}
\begin{equation}\label{Gc}
\wideboxed{
\tG_{\cc}(\vp_1,\vp_2)
=-b^{\Delta}|\vp_{12}|^{-2\Delta} \sgn\vp_{12},\qquad
\vp_{12}=2\sin\frac{\vp_1-\vp_2}{2}
}
\end{equation}

Note that self-consistency equations similar to \eqref{Schwinger-Dyson} can be written for any model with all-to-all interactions. However, their solution may not be physical if some ordering occurs, such as in spin glasses. For the $q=4$ SYK model, the transition to a glassy phase is expected at extremely low temperature, $T_{\text{glass}}\sim Je^{-\sqrt{N}}$\, \cite{GePaSa00}, so one may assume that $T\gg T_{\text{glass}}$ for almost all purposes. The mean field solution is accurate if $T\gg J/N$; at lower temperatures, quantum fluctuations should be taken into account \cite{BaAlKa16}.

\paragraph{Relation to black holes:}
A connection between this type of models and two-dimensional gravity was first noted in \cite{Sach10}. Indeed, the Green function \eqref{Gbeta} can be interpreted as a propagator of a fermion with certain mass and boundary conditions, between two points on the asymptotic boundary of the hyperbolic plane. Correlation functions in real (rather than imaginary) time are obtained by replacing the hyperbolic plane with a two-dimensional anti-de Sitter space. More recently, a holographic correspondence has been found that involves the dynamics and quantum fluctuations of space-time. To introduce it, let us review some facts about classical gravity.

\begin{figure}
\centerline{\(\displaystyle
\begin{array}{@{}c@{\quad\:}c@{\qquad}c@{}}
\displaystyle d\ell^2=\frac{-4\,du\,dv}{(1+uv)^2}\hspace{-5pt}
\figbox{1.0}{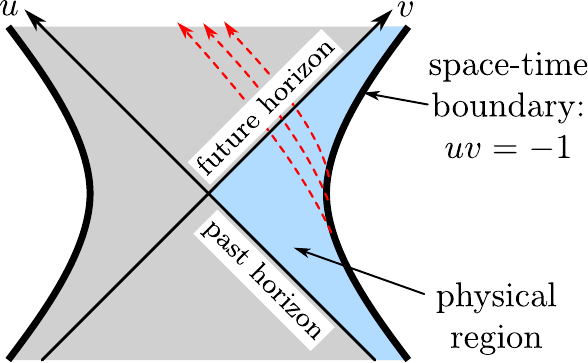} &
\figbox{1.0}{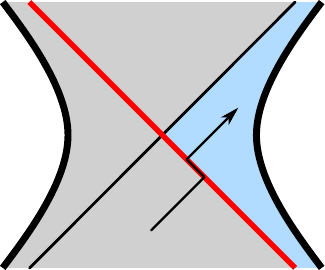} & \figbox{1.0}{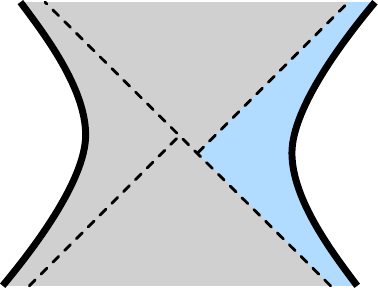}\vspace{5pt}\\
\text{a)} & \text{b)} & \text{c)}
\end{array}
\)}
\caption{($1+1$)-dimensional black hole: a) Structure of space-time and infalling matter (shown as red dotted lines); b) The gravitational perturbations has turned into a ``shock wave''; c) The same geometry in different coordinates.}
\label{fig_black_hole}
\end{figure}

The Einstein theory is well-defined if the number of space-time dimensions $d$ is $3$ or greater. For $d=2$, an interesting theory can be obtained by replacing Newton's constant $G_{N}$ with a dynamical field called a dilaton. For example, the static solution of dilaton gravity with a linear potential is the anti-de Sitter space $\AdS_2$, which may be regarded as an ``eternal'' black hole. Gravitational waves only exist if $d\ge 4$. However, black holes have another kind of gravitational mode \cite{DtH85}. A ``shock wave'' at the past horizon can be caused by any infalling object, as shown in Figure~\ref{fig_black_hole}. Such an object, even a very small one, produces some perturbation of the metric that evolves in time. The passing of time is represented by the transformation $(u,v)\mapsto\bigl(e^{\kap t}u,e^{-\kap t}v\bigr)$ in Kruskal-Szekeres coordinates, where $\kap$ is the surface gravity. Thus, the metric perturbation becomes localized at the past horizon and amplified by the factor $e^{\kap t}$. The effect of such a ``shock wave'' on a passing particle appears as a kink on the particle's worldline, see Figure~\ref{fig_black_hole}b. However, this apparent space-time discontinuity can be removed by a coordinate change. In the $\AdS_2$ case, the new coordinates can be chosen such that the metric remains the same but the boundary is shifted as shown in Figure~\ref{fig_black_hole}c.

T'Hooft considered a pair of gravitational modes localized on the past and the future horizons, wrote an effective action, and quantized it \cite{tH87,tH90, tH96}. In this formalism, the infalling matter directly interacts with the future horizon mode, which mediates its effect on other modes. Likewise, any Hawking radiation particle must cross the past horizon, and thus, interacts with the past horizon mode. This model provides, at least, a partial solution to the black hole information paradox, showing how infalling objects can influence the outgoing radiation. However, for a long time it remained controversial whether a shock wave on the past horizon has any physical effect because it does not change the density matrix of fields in the physical region. It turns out that, indeed, all naturally ordered (Keldysh) correlators of physical observables remain the same, but the gravitational modes have a strong effect on out-of-time-order correlators (OTOCs) of the form $\bcorr{D(t)C(0)B(t)A(0)}$. The latter were first discussed by Larkin and Ovchinnikov for a single particle in the semiclassical approximation \cite{LaOv69}; they characterize the sensitive dependence of the particle's trajectory on initial conditions. In the black hole context, this effect was studied by Shenker and Stanford, first with a classical perturbation source \cite{ShSt13}, and then in the fully quantum setting \cite{ShSt14}.

OTOCs provide a basis for comparison between black holes and conventional many-body systems while not requiring a complete quantum theory of gravity. The black hole OTOCs at early times (but after all two-point correlators have decayed) have some characteristic properties that reflect the physics near the horizon \cite{Kit.BPS}. One salient feature is their time dependence:
\begin{equation}
\bcorr{D(t)C(0)B(t)A(0)}-\corr{DB}\corr{CA}
\propto e^{\kap t},\qquad
\kap=2\pi T,
\end{equation}
where $T$ is the Hawking temperature. At later times, the exponential growth saturates. This general behavior, indicating fast scrambling \cite{SeSu08}, is common in models with all-to-all interactions, but the Lyapunov exponent $\kap$ is usually smaller. In fact, Maldacena, Shenker, and Stanford \cite{MSS15} showed that $\kap\le 2\pi T$ for any quantum system at temperature $T$ (under some natural assumptions). Thus, the condition $\kap=2\pi T$ is nontrivial. It is, actually, rather difficult to satisfy. For example, consider the Heisenberg model with random Gaussian couplings, $\overline{J_{jk}^2}=J^2/N$. If the temperature is high, $T\gg J$, then $\kap\sim J\ll T$; in the opposite limit, the system freezes into a spin glass \cite{BrMo80}. In general, low temperatures are more favorable for saturating the Maldacena-Shenker-Stanford bound. However, when $T$ is small, most systems either develop some ordering or enter the Fermi liquid regime, where energy relaxation and other nontrivial dynamics are slow. For example, the random Hubbard model exhibits the first behavior if the on-site repulsion is strong and the second one if the repulsion is weak.

The SYK model has the maximum Lyapunov exponent,
\begin{equation}
\kap\approx \frac{2\pi}{\beta}\qquad
\text{if }\, \beta J\gg 1.
\end{equation}
This was reported by Kitaev \cite{Kit.KITP.1} along with other results \cite{Kit.KITP}: an approximate reparametrization symmetry, the existence of a soft (pseudo-Goldstone) mode, and its effective action. Polchinski and Rosenhaus \cite{PoRo16} studied the conformal four-point function, which is complementary to the soft mode. Maldacena and Stanford undertook a thorough analysis of the SYK model \cite{MS16}. They found the conformal four-point function in an explicit form, calculated a finite-temperature correction to the Lyapunov exponent, $-(\delta\kap)/\kap \sim(\beta J)^{-1}$, as well as giving detailed derivations of the previous results, computing various numerical factors, and studying the $q\to\infty$ limit. In \cite{Jen16, MSY16, EMV16} the authors made an explicit connection between the SYK model and $d=2$ dilaton gravity, identifying the soft mode with t'Hooft's gravitational modes. However, it remains unknown how to obtain the $(\beta J)^{-1}$ correction to $\kap$ using this picture.

\paragraph{Further properties of the SYK model:}
Reparametrizations of time and the related soft mode will play an important role, so let us describe them in some detail. When $\beta J\gg 1$, the derivative term $-\partial_\tau$ in the Schwinger-Dyson equations \eqref{Schwinger-Dyson} is relatively small. Without this term, the equations are invariant under arbitrary changes of the time coordinate:
\begin{equation}
\begin{aligned}
G(\tau_1,\tau_2) \quad &\longrightarrow\quad
G\bigl(f(\tau_1),f(\tau_2)\bigr)\,
f'(\tau_1)^{\Delta}f'(\tau_2)^{\Delta}
\\[3pt]
\Sigma(\tau_1,\tau_2) \quad &\longrightarrow\quad
\Sigma\bigl(f(\tau_1),f(\tau_2)\bigr)\,
f'(\tau_1)^{1-\Delta}f'(\tau_2)^{1-\Delta}.
\end{aligned}
\end{equation}
For example, $G_{\beta=\infty}$ defined by \eqref{Ginfty} can be transformed into the equilibrium Green function at finite $\beta$ (see \eqref{Gbeta}, \eqref{Gc}) if we use $f(\tau)=\tau_{*}e^{2\pi i\tau/\beta}$, where $\tau_*$ is an arbitrary constant with dimension of time. (If one is uncomfortable with complex functions, $f(\tau)=\tau_{*}\tan\frac{\pi\tau}{\beta}$ also works.)

The soft mode manifold consists of all solutions of the approximate Schwinger-Dyson equations for a given $\beta$. The equilibrium Green function is a solution, as is any function of the form
\begin{equation}
\label{Gphi-1}
\begin{aligned}
G(\tau_1,\tau_2) 
&=G_{\beta=\infty}\bigl(f(\tau_1),f(\tau_2)\bigr)\,
f'(\tau_1)^{\Delta}f'(\tau_2)^{\Delta}\qquad
\text{with}\quad f(\tau)=\const\cdot e^{i\vp(\tau)}
\\[3pt]
&=\tG_{\cc}\bigl(\vp(\tau_1),\vp(\tau_2)\bigr)\,
\bigl(J^{-1}\vp'(\tau_1)\bigr)^{\Delta}
\bigl(J^{-1}\vp'(\tau_2)\bigr)^{\Delta},
\end{aligned}
\end{equation}
where $\vp(\tau)$ takes values in the interval $[0,2\pi]$ with the endpoints glued. The above expression is invariant under certain  transformations $V$ of $\vp(\tau)$ so that the functions $\vp$ and $V\circ\vp$ define the same $G$. These transformations act on the variable $z=e^{i\vp(\tau)}$ by linear fractional maps preserving the unit circle, $z\mapsto\frac{az+b}{cz+d}$. We call such maps ``conformal''; they form a group isomorphic to $\operatorname{PGL}(2,\RR)$. Thus, the manifold of distinct $G$'s is $\operatorname{PGL}(2,\RR)\backslash\Diff(S^1) \cong\PSL(2,\RR)\backslash\Diff^{+}(S^1)$.

When the derivative term is taken into account, only the equilibrium Green function (whose actual form is different at short times, $|\tau_1-\tau_2|\sim J^{-1}$) satisfies the Schwinger-Dyson equations. The general quasi-solution \eqref{Gphi-1} can be characterized by this effective action \cite{Kit.KITP, MS16}
\begin{equation}\label{Sch_intro}
\wideboxed{
I_{\eff}=-N\alpha_S J^{-1}
\int_{0}^{\beta}\Sch\bigl(e^{i\vp(\tau)},\tau\bigr)\,d\tau,\qquad
\text{where}\quad
\Sch\bigl(f(x),x\bigr)
=\frac{f'''}{f'}-\frac{3}{2}\left(\frac{f''}{f'}\right)^2
}
\end{equation}
To see where the Schwarzian derivative $\Sch(\cdots)$ comes from, let $f(\tau)=\const\cdot e^{i\vp(\tau)}$ and let us expand $G(\tau_1,\tau_2) =G_{\beta=\infty}\bigl(f(\tau_1),f(\tau_2)\bigr)\, f'(\tau_1)^{\Delta}f'(\tau_2)^{\Delta}$ in powers of $\tau_1-\tau_2$. Using the fact that $G_{\beta=\infty}(t_1,t_2)\propto (t_1-t_2)^{-2\Delta}$ and Taylor expanding $f$ and $f'$ at $\tau_+=(\tau_1+\tau_2)/2$, we find that
\begin{equation}\label{Grepar_intro}
\wideboxed{
G(\tau_1,\tau_2) =
G_{\beta=\infty}(\tau_1,\tau_2)\biggl(1+\frac{\Delta}{6}\,
\Sch\bigl(f(\tau_+),\tau_+\bigr)\,(\tau_1-\tau_2)^2
+O\bigl((\tau_1-\tau_2)^4\bigr)\biggr)
}
\end{equation}
The physical effect of the second term (proportional to $(\tau_1-\tau_2)^{-2\Delta+2}$ times the Schwarzian) will be explained later. The coefficient $\alpha_S$ in the action \eqref{Sch_intro} can be determined numerically \cite{MS16}, while an analytic expression exists in a special case \cite{MS16,JeSuYo16}:
\begin{equation}
\alpha_S=\frac{1}{24\pi}\qquad \text{for}\quad q\to 2.
\end{equation}
The action achieves its minimum when $\vp(\tau)=\frac{2\pi\tau}{\beta}$ up to a conformal map.

Much of the SYK physics is reflected in the expansion of the free energy in terms of the large parameters $N$ and $\beta J$:
\begin{equation}\label{F_SYK}
\wideboxed{
\beta(F-E_0)= N\Bigl(-s_0-2\pi^2\alpha_S(\beta J)^{-1}
+\frac{\pi^2}{6} \kern1pt\gamma\kern1pt (\beta J)^{-2}+\cdots\Bigr)
+\frac{3}{2}\ln(\beta J)+\const+o(1)
}
\end{equation}
Here $E_0$ is the ground state energy, which is proportional to $N$ but subject to $1/N$ corrections. The number $s_0$, dubbed ``zero-temperature entropy'', is the entropy per Majorana site at very low temperatures (but above $T_{\text{glass}}$). It was first found  \cite{GePaSa00} in the original Sachdev-Ye model. For the SYK model, the ``zero-temperature entropy'' is \cite{Kit.KITP, MS16} 
\begin{equation}
s_0=\pi\int_{\Delta}^{1/2}\bigl(\tfrac{1}{2}-x\bigr)\tan(\pi x)\,dx.
\end{equation}
The next term on the right-hand side of equation \eqref{F_SYK}, $N\bigl(-2\pi^2\alpha_S(\beta J)^{-1}\bigr)$, is simply the value of the Schwarzian action \eqref{Sch_intro} for $\vp(\tau)=2\pi\tau/\beta$. The $(\beta J)^{-2}$ term (with the coefficient $\gamma$ given in Table~\ref{tab_coeff}) was reported in \cite{randmat,JeSu16}. We will derive it from a correction to the Schwarzian action that is non-local in time. Note that the omitted terms in parentheses are expected to contain non-integer powers of $\beta J$, \eg $(\beta J)^{-2.77}$ for $q=4$ \cite{randmat}.

The terms that do not scale with $N$, namely $\frac{3}{2}\ln(\beta J)+\const+o(1)$, are due to quantum fluctuations \cite{MS16}. Remarkably, this expression is valid even for $\beta J\gtrsim N$, when the fluctuations of the soft mode are strong \cite{randmat, BaAlKa17,StWi17}. At such low temperatures, the $(\beta J)^{-2}$ and higher-order terms are negligible, and the thermodynamics is more conveniently described using the density of states \cite{randmat}:
\begin{equation}
e^{-\beta F}=Z=\int \rho(E)\,e^{-\beta E}\,dE,\qquad\quad
\rho(E)\propto e^{s_{0}N}\sinh\Bigl(2\pi\sqrt{2N\alpha_{S}(E-E_0)/J}\Bigr).
\end{equation}

Variants of the SYK model have been studied in \cite{Sach15, FuGaMaSa16, GrRo16, Wi16, KlGr16, Gu16, PeSpVo17}.

\begin{table}[t]
\def\arraystretch{0}
\def\NL#1{\\ & \rule{0pt}{#1}\\}
\def\HL{\hline & \rule{0pt}{5pt}\\}
\centerline{\begin{tabular}{|p{6.5cm}|c|}
\HL
\centering Meaning and where it appears & Analytic expression
\NL{5pt}\HL
Scaling dimension of $\chi_j(\tau)$ & \(\Delta=1/q\)
\NL{10pt}
Overall factor in $G(\tau_1,\tau_2)^q$ &
\(\displaystyle b=\frac{(q-2)\tan(\pi/q)}{2\pi q} \)
\NL{10pt}
Eigenvalue of the conformal kernel and its derivative at $h=2$
(Section~\ref{sec_geneig})& 
\(\begin{gathered}[t]
k_{\cc}(h)=\frac{p(h)}{p(2)},\quad\:
p(h)=\frac
{\Gamma\bigl(\Delta+\frac{h}{2}\bigr)\,
\Gamma\bigl(\Delta+\frac{1-h}{2}\bigr)}
{\Gamma\bigl(1-\Delta+\frac{h}{2}\bigr)\,
\Gamma\bigl(1-\Delta+\frac{1-h}{2}\bigr)}\\[3pt]
-k'_{\cc}(2)=\frac{\pi}{\sin(2\pi/q)}-\frac{q(q^2-6q+6)}{2(q-1)(q-2)}
\end{gathered}\)
\NL{10pt}
\multicolumn{1}{|l|}{\parbox[c]{6.5cm}
{Amplitude of the leading non-conformal perturbation 
(Section~\ref{sec_renorm1})}}&
$a_0$ (fitted to numerical data)
\NL{10pt}
Coefficient in the UV correction to the Green function
(Section~\ref{sec_renorm1})&
\(\displaystyle \alpha_G=\frac{a_0}{(-k'_{\cc}(2))\sqrt{(q-1)b}} \)
\NL{10pt}
Coefficient in the Schwarzian action (Section~\ref{sec: Schw_der})&
\(\displaystyle \alpha_S=a_{0}\,\frac{\sqrt{(q-1)b}}{6q}\)
\NL{8pt}
Coefficient in the non-local effective action (Section~\ref{sec_nonlocal})&
\(\displaystyle \gamma=\frac{2a_0^2}{-k'_{\cc}(2)} \)
\NL{5pt}\hline
\end{tabular}}
\caption{Common numerical coefficients and functions.}
\label{tab_coeff}
\end{table}

\subsection{Outline of the paper}

Let $\beta J\gg 1$ and let us consider all quantities in the $N\to\infty$ limit, or in the leading $1/N$ order. We are still interested in subleading terms with respect to the small parameter $(\beta J)^{-1}$. For the free energy, that means all terms in parentheses in \eqref{F_SYK}. Similarly, one may try to derive a systematic $(\beta J)^{-1}$ expansion for all tree-level correlators. The goal of this paper is more modest. We will construct an effective theory of the soft mode that is one order more accurate than the Schwarzian action and make a similar improvement on the gravity side (compared with \cite{Jen16, MSY16, EMV16}).

We first discuss the RG flow and interaction between UV and IR degrees of freedom for the SYK model. Given the $\beta J\gg 1$ condition, it is useful to separate short times (UV), where all calculations have to be done numerically, from the analytically tractable region, $\tau\gg J^{-1}$. The latter can be further subdivided into long times (IR) and intermediate time scales, $J^{-1}\ll\tau\ll\beta$ (where $\tau$ is understood as $|\tau_1-\tau_2|$). The intermediate asymptotics are particularly simple and insightful. In this regime, the Green function has the form $G=G_{\IR}+G_{\UV}$, where
\begin{align}
G_{\IR}(\tau_1,\tau_2) &
\approx\biggl(1+\frac{\Delta}{12}
\bigl(\tfrac{2\pi}{\beta}(\tau_1-\tau_2)\bigr)^2\biggr)\kern1pt
G_{\beta=\infty}(\tau_1,\tau_2),
\\[3pt]
\label{deltaGUV}
G_{\UV}(\tau_1,\tau_2) &
\approx -\alpha_{G}\,\bigl|J(\tau_1-\tau_2)\bigr|^{-1}\kern1pt
G_{\beta=\infty}(\tau_1,\tau_2).
\end{align}
The UV part results from an irrelevant perturbation, namely, the $-\partial_{\tau}$ term in the Schwinger-Dyson equations. In the UV region, this perturbation is strongly nonlinear. However, we may replace it with a weak perturbation source that is concentrated on the lower end of the intermediate region and has the same response (as defined by $G_{\UV}$) for longer time intervals. Indeed, this response is characterized by the scaling dimension $h_0=2$ (related to the exponent $-1$ in \eqref{deltaGUV}) and the overall factor $\alpha_{G}$. The effective perturbation source and the corresponding response will be described in Section~\ref{sec_renormalization}. In practice, $\alpha_{G}$ is fitted to the numerically computed Green function \cite{MS16}. Note that the nonlinear UV dynamics also sources perturbations with scaling dimensions 
\begin{equation}
h_1\approx 3.77,\quad\: h_2 \approx 5.68,\ldots\qquad (\text{for } q=4).
\end{equation}

The $h=2$ perturbation source couples to the $(\tau_1-\tau_2)^2$ term in $G_{\IR}$, contributing to the free energy. The same coupling is applicable to the more general Green function deformed by the soft mode (see \eqref{Gphi-1}, \eqref{Grepar_intro}), resulting in the Schwarzian effective action. Thus, the coefficient $\alpha_S$ in front of the Schwarzian is proportional to the the UV perturbation amplitude $a_0$, as is $\alpha_G$ (see Table~\ref{tab_coeff}). This leads to a linear relation between $\alpha_G$ and $\alpha_S$, which was originally obtained by a different method \cite{MS16}.

The Schwarzian action can also be written using $\vp=\vp(\tau)$ as a time variable:
\begin{equation} \label{I_loc_conf}
\wideboxed{
\frac{I_{\loc}}{N}=-\alpha_S\int
\underbrace{\left(\frac{\vep^2}{2}-\frac{\vep'^2}{2}+\vep\vep''\right)}
_{J^{-2}\Sch(e^{i\vp},\tau)}\vep^{-1}\,d\vp,\qquad
\text{where}\quad
\vep(\vp)=J^{-1}\frac{d\vp}{d\tau}
}
\end{equation}
(The derivatives are defined with respect to $\vp$.) We will find a non-local correction to it:
\begin{equation} \label{I_nl_conf}
\wideboxed{
\frac{I_{\nloc}}{N}=-\frac{\gamma}{2}
\left[\int\frac{\vep(\vp_1)\vep(\vp_2)}{\vp_{12}^4}
\biggl(\ln\biggl(\frac{\vp_{12}^2}{\vep(\vp_1)\vep(\vp_2)}\biggr)
+c\biggr)\frac{d\vp_1}{2\pi}\frac{d\vp_2}{2\pi}\right]_{\text{fin.}}\!\!,
\quad\, \vp_{12}=2\sin\frac{\vp_1-\vp_2}{2}
}
\end{equation}
where $\gamma$ is given in Table~\ref{tab_coeff}. The number $c$ in this formula depends on the choice of perturbation source and, ultimately, on the definition of the function $\vp$. Indeed, $\vp$ describes the IR part of the Green function. In Section~\ref{sec_gensoft}, we give a certain prescription to extract $\vp$ from $G$ using an integral over time variables. This integral picks up a small contribution from the region $|\tau_1-\tau_2|\sim J^{-1}$, which introduces some ambiguity into the definition of $\vp$. The subscript ``$\text{fin.}$'' in \eqref{I_nl_conf} means excluding the UV divergent part; such a regularization procedure is, actually, unique.

The non-local term in the effective action may be viewed as an intermediate step toward computing physical quantities such as the four-point correlator $\corr{\TT\chi_j(\tau_1)\chi_j(\tau_2)\chi_k(\tau_3)\chi_k(\tau_4)}$. It turns out that for natural time orderings, \eg $\tau_1>\tau_2>\tau_3>\tau_4$, non-local effects mostly cancel: the corrlator depends only on $\tau_1-\tau_2$ and $\tau_3-\tau_4$, with some fast-decaying (as functions of $\tau_2-\tau_3$) terms due to the fields with higher scaling dimensions, $h_1,h_2,\ldots$ This cancellation was first noted by Maldacena and Stanford \cite{MS16}, who found the correlator by a different method (first for $q\to\infty$, and then extrapolating to arbitrary $q$). We give a qualitative explanation of this effect, comparing it with Debye screening, and derive a certain identity for the four-point function. Such deep cancellation does not occur for out-of-time orderings, \eg $\tau_1>\tau_3>\tau_2>\tau_4$.

As a separate problem, we consider a certain variant of Euclidean dilaton theory on the unit disk. Fixing the boundary conditions that depend on an arbitrary function $\vep$ and integrating out the bulk fields, we obtain exactly the same action, $I_{\loc}+I_{\nloc}$. In this case, the parameter $c$ is well-defined, which is due to a more rigorous treatment of the near-boundary region compared with the UV region for the SYK model. However, the definition of the ``conformal time'' $\vp$ is intrinsically non-local; it is related to geodesic distance between boundary points and involves the metric on the whole disc. So it is not surprising that non-locality appears in this context. On the other hand, correlators between boundary observables as functions of the proper time can only contain contact terms (such as $\delta(\tau_1-\tau_2)$) and global (\ie time-independent) terms. This is due to Birkhoff's theorem in dilaton gravity \cite{L-MKu94}, which says that the solution of the classical equations of motion depends on one global parameter and is otherwise completely rigid. Thus, the non-local action may be regarded as an artificial construct, but it is unambigously defined and provides a more detailed correspondence with the SYK model than has previously been known.

\section{Formalism (part 1)}

\subsection{Replica-diagonal action}


\paragraph{Basic form of the action functional:}
The Schwinger-Dyson equations \eqref{Schwinger-Dyson} are exactly the saddle point conditions for this effective action \cite{Kit.KITP, Sach15}:
\begin{equation}\label{rdact0}
I[\Sigma,G]
=N\Biggl(-\ln\Pf(-\partial_\tau-\hat{\Sigma})
+\frac{1}{2}\int d\tau_1\,d\tau_2
\biggl(\Sigma(\tau_1,\tau_2)\,G(\tau_1,\tau_2)
-\frac{J^2}{q}\bigl|G(\tau_1,\tau_2)\bigr|^q\biggr)\!\Biggr).
\end{equation}
As discussed in Appendix~\ref{sec_replicas}, the functional integral of $e^{-I[\Sigma,G]}$ over $\Sigma$ and $G$ gives the disorder-averaged partition function $\overline{Z}$, whereas the physically relevant quantity is $\beta F=-\overline{\ln Z}$. However, the difference between $\overline{\ln Z}$ and $\ln\overline{Z}$ is $O(N^{2-q})$. Furthermore, the diagrammatic expansion around the saddle point correctly reproduces all connected $2n$-point functions in the leading order in $1/N$. (An explicit calculation for $n=3$ was done in \cite{GrRo17}.) If greater accuracy is needed, one may use a similar action with $M$ replicas and do the usual $M\to 0$ trick. The key assumption involved in the derivation of action \eqref{rdact0} is the replica-diagonal ansatz. We also note that the saddle point is a maximum in $G$ and a minimum in $\Sigma$ and that $I[\Sigma,G]$ is well-defined for any real number $q\ge 2$, though the $q=2$ case is degenerate in some respects.

The new formulation of the problem has some subtleties. First, the functional integral should be taken in the complex domain. The exact definition is unclear, but it is not needed for the asymptotic $1/N$-expansion. The integration measure comes with a normalization factor such that
\begin{equation}
\int\calD\Sigma\: \calD G\,\exp\biggl(
-\frac{N}{2}\,\int d\tau_1\,d\tau_2\,
\Sigma(\tau_1,\tau_2)\,G(\tau_1,\tau_2)\biggr) =1.
\end{equation}
Lastly, the Pfaffian should be regularized to eliminate the UV divergence:
\begin{equation}\label{regPf}
\Pf\bigl(-\partial_\tau-\hat{\Sigma}\bigr)\:\to\:
\sqrt{2}\:\frac{\Pf(-\partial_\tau-\hat{\Sigma})}{\Pf(-\partial_\tau)}.
\end{equation}

\paragraph{Tilde notation for IR-normalized quantities:}
Let us transform the action \eqref{rdact0} to a different form so as to have easy access to the IR and intermediate asymptotics. We regard the conformal saddle point $(\Sigma_{\cc},G_{\cc})$ with $G_{\cc}(\tau_1,\tau_2)=\bigl(\frac{2\pi}{\beta J}\bigr)^{2\Delta}
\tG_{\cc}\bigl(\frac{2\pi\tau_1}{\beta},\frac{2\pi\tau_2}{\beta}\bigr)$ as a zeroth approximation. The operator $\hat{\sigma}=\partial_\tau$ is a perturbation. To simplify its treatment, we replace the integral kernel $\sigma(\tau_1,\tau_2)=\delta'(\tau_1-\tau_2)$ with some nonsingular function, which requires a new regularization scheme. So, let us consider the difference $\tI$ between $I[\Sigma,G]$ and $I[\Sigma_{\cc},G_{\cc}]$. It is finite for any nonsingular $\sigma$ and can be written as
\begin{equation}
\tI=I-\beta E_0+N s_0.
\end{equation}
Furthermore, it is convenient to use a dimensionless time coordinate and correspondingly renormalized fermionic fields $\widetilde{\chi}_{j}$. The standard choice is
\begin{equation}\label{theta_frame}
\theta=\frac{2\pi\tau}{\beta},\qquad
\widetilde{\chi}_{j,\theta}(\theta)=\chi_j(\tau)\,\vep_{\theta}^{-\Delta},\qquad
\text{where}\quad \vep_{\theta}=\frac{2\pi}{\beta J}.
\end{equation}
But we can also change the frame $\theta$ to $\vp(\theta)$, where $\vp$ is an arbitrary diffeomorphism of the unit circle (represented by the interval $[0,2\pi]$ with the endpoints glued). In this case,
\begin{equation}
\vp=\vp\bigl(\tfrac{2\pi\tau}{\beta}\bigr),\qquad
\widetilde{\chi}_{j,\vp}(\vp)=\chi_j(\tau)\,\vep_{\vp}(\vp)^{-\Delta},\qquad
\text{where}\quad
\vep_{\vp}(\vp)=\vep_{\theta}\,\frac{d\vp}{d\theta}=J^{-1}\frac{d\vp}{d\tau}.
\end{equation}
(The frame subscript may be omitted when it is clear from the context.) In addition to similar transformations of the Green function and self-energy, we combine the latter with the perturbation source, that is, define $\tSig$ in terms of $\Sigma+\sigma$. We will usually use the inverse transformations:
\begin{equation} \label{tquan}
\begin{aligned}
G(\tau_1,\tau_2) &=\tG_{\vp}(\vp_1,\vp_2)\,
\vep_{\vp}(\vp_1)^{\Delta}\,\vep_{\vp}(\vp_2)^{\Delta},
\\[2pt]
\Sigma(\tau_1,\tau_2)
&=J^2\Bigl(\tSig_{\vp}(\vp_1,\vp_2)-\tsig_{\vp}(\vp_1,\vp_2)\Bigr)\,
\vep_{\vp}(\vp_1)^{1-\Delta}\,\vep_{\vp}(\vp_2)^{1-\Delta},
\\[2pt]
\sigma(\tau_1,\tau_2) &=J^2\tsig_{\vp}(\vp_1,\vp_2)\,
\vep_{\vp}(\vp_1)^{1-\Delta}\,\vep_{\vp}(\vp_2)^{1-\Delta}.
\end{aligned}
\end{equation}
Now, the effective action has exactly the same form in any frame:
\begin{equation}\label{rdact1}
\wideboxed{\begin{aligned}
\frac{\tI[\tSig,\tG]}{N}
=\biggl[&-\ln\Pf(-\hat{\tSig})
+\frac{1}{2}\int d\vp_1\,d\vp_2
\biggl(\tSig(\vp_1,\vp_2)\,\tG(\vp_1,\vp_2)
-\frac{1}{q}\,\bigl|\tG(\vp_1,\vp_2)\bigr|^q\biggr)\biggr]_{\text{reg.}}
\\[3pt]
&-\frac{1}{2}\int d\vp_1\,d\vp_2\,
\tsig(\vp_1,\vp_2)\,\tG(\vp_1,\vp_2)
\end{aligned}}
\end{equation}
where the subscript ``reg.'' indicates the difference between the expression in brackets and its value at the conformal point $(\tSig_{\cc},\tG_{\cc})$.

The change from one frame to another is described by a diffeomorphism $V:\,S^1\to S^1$. A corresponding operator $V_{h}$ acts on functions as follows:
\begin{equation}\label{h-form}
(V_{h}f)(y)=\biggl(\frac{dy}{dx}\biggr)^{-h}f(x)\qquad
\text{for}\quad y=V(x).
\end{equation}
The choice of $h$ depends on context (the physical meaning of $f$). Functions of one variable whose transformation law is characterized by a given $h$ will be called ``$h$-forms''. It is easy to define a similar action $V_{h_1,h_2}$ on functions of two variables. In this notation,
\begin{equation}
\tG_{V\circ\vp}=V_{\Delta,\Delta}\kern1pt\tG_{\vp},\qquad
\tSig_{V\circ\vp}=V_{1-\Delta,1-\Delta}\kern1pt\tSig_{\vp},\qquad
\tsig_{V\circ\vp}=V_{1-\Delta,1-\Delta}\kern1pt\tsig_{\vp}.
\end{equation}

\begin{figure}
\centerline{\(
\begin{array}{@{}l@{\quad}c@{}}
\text{Sheets:}& \displaystyle
-N\figbox{1.0}{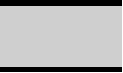}\:,\quad
-N\figbox{1.0}{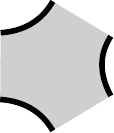}\:,\quad
-N\figbox{1.0}{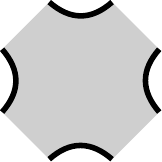}\:,\ldots
\vspace{2pt}\\
& \text{(oriented clockwise or counterclockwise)}
\\[10pt]
\text{Seams:}& \displaystyle
\frac{3}{N}\,\figbox{1.0}{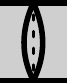}\:\,,\qquad
\frac{3\cdot 2}{N^2}\,\figbox{1.0}{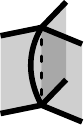}\:\,,\qquad
\frac{3\cdot 2\cdot 1}{N^3}\,\figbox{1.0}{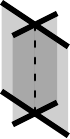}\:\,.
\vspace{2pt}\\
& \text{(oriented up or down)}
\end{array}
\hspace{1cm}
\begin{array}{@{}c@{}}
\\[-3pt]
\raisebox{-5pt}{\hbox{\includegraphics{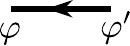}}}\,
=\widetilde{G}_{*}(\vp,\vp')
\vspace{35pt}\\
\text{Template for }\,
\widetilde{\V{G}}(\vp_1,\vp_1';\dots;\vp_n,\vp_n'):\\[5pt]
\figbox{1.0}{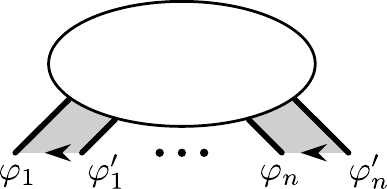}
\end{array}
\)}
\caption{Diagrammatic calculus for the expansion of action \eqref{rdact1} around the saddle point $(\tSig_*,\tG_*)$ for $q=4$. Degenerate $2$-gon sheets representing $\tG_*$ may be attached directly to the template tabs. When the building blocks are composed into a diagram, each orientation conflict between a sheet and an adjacent seam or between a sheet and the template gives a factor of $-1$.}
\label{fig_diag_rd}
\end{figure}

\paragraph{Application to correlation functions:}
We now describe the diagrammatic calculus for quantum fluctuations around the saddle point $(\tSig_*,\tG_*)$ for a fixed $\tsig$. It can be derived by expanding the effective action in $\delta\tSig=\tSig-\tSig_*$ and $\delta \tG=\tG-\tG_*$. The second-order expansion is given in the next subsection. In general, the derivation is similar to that of the high temperature expansion in Appendix~\ref{sec_replicas}. The saddle point expansion can also be obtained by a resummation of high temperature diagrams. As is usual, the sum of closed, connected diagrams represents $\ln\overline{Z}$, whereas the expansion of $\overline{\ln Z}$ includes only those diagrams that are connected along fermionic lines. This difference is not important for our purposes because we work at the tree level. So, let us simply say that we consider the logarithm of the partition function and correlators of the form
\begin{equation}\label{2nfunc}
\widetilde{\V{G}}(\vp_{1},\vp_{1}',\dots,\vp_{n},\vp_{n}')
=(-1)^n\sum_{j_1,\dots,j_n}
\bcorr{\TT\widetilde{\chi}_{j_1}(\vp_{1})\widetilde{\chi}_{j_1}(\vp_{1}')
\dots \widetilde{\chi}_{j_n}(\vp_{n})\widetilde{\chi}_{j_n}(\vp_{n}')}.
\end{equation}
The diagrams for these quantities are built from $2n$-gons, or ``sheets'' (arising from the Taylor expansion of $\ln\Pf(-\tSig)$) that are connected at the sides without a border by ``seams'' (coming from $\tG^q$), see Figure~\ref{fig_diag_rd}. Taking the tubular neighborhood of an embedding of such a diagram into $\mathbb{R}^3$, one obtains a three-dimensional handlebody whose genus $g$ counts the factors of $N$ in $\widetilde{\V{G}}$ as $\widetilde{\V{G}} \sim N^{1-g}$. In fact any closed diagram
that is connected along fermionic lines can be mapped to such a handlebody, see Appendix~\ref{sec_replicas}.

The connected part of correlator \eqref{2nfunc} is $2^{n}$ times the $n$-th variational derivative of $\ln Z$ with respect to $\tsig$. To find it at the tree level, \ie in the leading order in $1/N$, we may approximate $-\ln Z$ by the saddle point value of the action, $\tI_{*}$. Hence,
\begin{equation}\label{Gconn}
\widetilde{\V{G}}_{\text{conn.}}(\vp_{1},\vp_{1}',
\dots,\vp_{n},\vp_{n}')
\approx-2^{n}\,\frac{\delta^{(n)}\tI_{*}}
{\delta\tsig(\vp_{1},\vp_{1}')
\cdots\delta\tsig(\vp_{n},\vp_{n}')}.
\end{equation}
We will need the connected four-point function:
\begin{equation}
\begin{aligned}\label{Gconn1}
\widetilde{\V{G}}_{\text{conn.}}(\vp_1,\vp_2,\vp_3,\vp_4)
&=\widetilde{\V{G}}(\vp_1,\vp_2,\vp_3,\vp_4)
-\widetilde{\V{G}}(\vp_1,\vp_2)\,\widetilde{\V{G}}(\vp_3,\vp_4)
\\[3pt]
&=N\,\tF(\vp_1,\vp_2;\vp_3,\vp_4)
\approx N\,\bigl(\Lambda(\vp_1,\vp_2;\vp_3,\vp_4)
-\Lambda(\vp_1,\vp_2;\vp_4,\vp_3)\bigr),
\end{aligned}
\end{equation}
where the notation $\calF$ has been used in \cite{MS16} (albeit in the standard $\tau$ variables) and $\Lambda$ is a sum of ladder diagrams:
\begin{align}
\Lambda(\vp_1,\vp_2;\vp_3,\vp_4)
&=N^{-1}\biggl(\figbox{1.0}{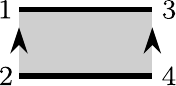}+\figbox{1.0}{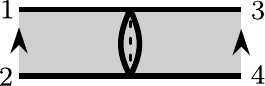}\,+\cdots\biggr)
\qquad
\left[\begin{array}{@{}c@{}}
\text{the coefficients of the}\\ \text{diagrams are not shown}
\end{array}\right]
\nonumber\\[5pt]
&=\figbox{1.0}{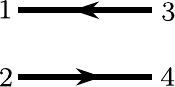}+(q-1)\figbox{1.0}{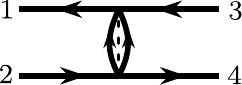}
+(q-1)^2\,\figbox{1.0}{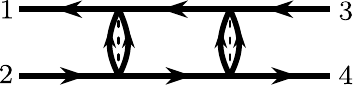}+\cdots
\end{align}
These diagrams have a repeated element, the ladder kernel $K$, which is defined by cutting the ladders along the dotted lines. It is the integral kernel of an operator acting on functions on $S^1\times S^1$. If an extra rung, split in half, is added on both ends of each ladder, the above expression takes the form $L=K+K^2+K^3+\cdots=K(1-K)^{-1}$. Thus,
\begin{equation}
L(\vp_1,\vp_2;\vp_3,\vp_4)
=(q-1)\bigl|\tG_{*}(\vp_1,\vp_2)\bigr|^{\frac{q-2}{2}}
\Lambda(\vp_1,\vp_2,\vp_3,\vp_4)\,
\bigl|\tG_{*}(\vp_3,\vp_4)\bigr|^{\frac{q-2}{2}}
\end{equation}
is the integral kernel of $L=K(1-K)^{-1}$, where
\begin{equation} \label{Kdef}
K(\vp_1,\vp_2;\vp_3,\vp_4)
=(q-1)\bigl|\tG_{*}(\vp_1,\vp_2)\bigr|^{\frac{q-2}{2}}\,
\tG_{*}(\vp_1,\vp_3)\,\tG_{*}(\vp_4,\vp_2)\,
\bigl|\tG_{*}(\vp_3,\vp_4)\bigr|^{\frac{q-2}{2}}.
\end{equation}

\subsection{Conformal kernel and representations of $\PSL(2,\RR)$}

\subsubsection{Conformal kernel}\label{sec_confker}

The action \eqref{rdact1} is suited for the study of the SYK model near the conformal point $(\tSig_{\cc},\tG_{\cc})$, which is a (non-unique) saddle point for $\tsig=0$. We would like to calculate the system response to small perturbations $\tsig(\vp_1,\vp_2)$. In particular, a generic UV perturbation is expected to have a similar effect to that of the kinetic term $\partial_\tau$ in the original model. We can also express various correlation functions by probing the system with additional IR perturbations. So, let us expand the action near the conformal point. If $\tSig=\tSig_{\cc}+\delta\tSig$ and $\tG=\tG_{\cc}+\delta\tG$, then
\begin{align}\label{rdact2a}
\frac{\tI}{N}
\approx{}&\frac{1}{4}\Trr\Bigl(\hat{\tG}_{\cc}\,\delta\hat{\tSig}\Bigr)^2
+\frac{1}{2}\int d\vp_1\,d\vp_2
\biggl(\delta\tSig(\vp_1,\vp_2)\,\delta\tG(\vp_1,\vp_2)
-\frac{q-1}{2}\,\bigl|\tG_{\cc}(\vp_1,\vp_2)\bigr|^{q-2}
\delta\tG(\vp_1,\vp_2)^2\biggr)
\nonumber\\[3pt]
&-\frac{1}{2}\int d\vp_1\,d\vp_2\,
\tsig(\vp_1,\vp_2)\,\bigl(\tG_{\cc}(\vp_1,\vp_2)+\delta\tG(\vp_1,\vp_2)\bigr),
\end{align}
where we have neglected terms of order $3$ and higher. Let
\begin{empheq}[box=\widebox]{gather}
\label{sg_def}
g(\vp_1,\vp_2)=R_{\cc}(\vp_1,\vp_2)\,\tG(\vp_1,\vp_2),\qquad\quad
s(\vp_1,\vp_2)=R_{\cc}(\vp_1,\vp_2)^{-1}\tsig(\vp_1,\vp_2)
\\[8pt]
\label{Rc}
\text{where}\qquad
R_{\cc}(\vp_1,\vp_2)
=\sqrt{q-1}\,\bigl|\tG_{\cc}(\vp_1,\vp_2)\bigr|^{\frac{q-2}{2}}
=-\sqrt{(q-1)b}\kern3pt\vp_{12}^{-1}\,\tG_{\cc}(\vp_1,\vp_2)^{-1}
\end{empheq}
and let us also use the temporary notation
$f(\vp_1,\vp_2)=R_{\cc}(\vp_1,\vp_2)^{-1}\tSig(\vp_1,\vp_2)$. Thus, \eqref{rdact2a} becomes
\begin{equation}
\frac{\tI_2}{N}= -\frac{1}{4}\,\bbraa{\delta f}K_{\cc}\bkett{\delta f}
+\frac{1}{2}\,\brakett{\delta f}{\delta g}
-\frac{1}{4}\,\brakett{\delta g}{\delta g}
-\frac{1}{2}\,\brakett{s}{g_{\cc}+\delta g},
\end{equation}
where the inner product is defined by the integral over $S^1\times S^1$ and
\begin{gather}
g_{\cc}(\vp_1,\vp_2) =-\sqrt{(q-1)b}\kern3pt\vp_{12}^{-1},
\\[8pt]
\label{Kckernel}
\begin{aligned}
K_{\cc}(\vp_1,\vp_2;\vp_3,\vp_4)
&=R_{\cc}(\vp_1,\vp_2)\,
\tG_{\cc}(\vp_1,\vp_3)\,\tG_{\cc}(\vp_4,\vp_2)\,
R_{\cc}(\vp_3,\vp_4)\\[3pt]
&=(q-1)\,b\,|\vp_{12}|^{2\Delta-1}|\vp_{13}|^{-2\Delta}(\sgn\vp_{13})\,
|\vp_{42}|^{-2\Delta}(\sgn\vp_{42})\,|\vp_{34}|^{2\Delta-1}.
\end{aligned}
\end{gather}
Taking the saddle point with respect to $\delta f$, we obtain this very simple result:
\begin{equation} \label{I2}
\wideboxed{
\frac{\tI_2[\delta g]}{N}
=-\frac{1}{2}\,\bbrakett{s}{g_{\cc}+\delta g}
+\frac{1}{4}\,\bbraa{\delta g}K_{\cc}^{-1}-1\bkett{\delta g}
}
\end{equation}
If we also try to take the saddle point with respect to $\delta g$, we should get
\begin{equation}\label{I2star}
\frac{(\tI_2)_{*}}{N}
=-\frac{1}{2}\bbrakett{s}{g_{\cc}}
-\frac{1}{4}\,\bbraa{s}\underbrace{K_{\cc}(1-K_{\cc})^{-1}}_{L_{\cc}}\bkett{s}
=-\frac{1}{2}\bbrakett{\tsig}{\tG_{\cc}}
-\frac{1}{8}\,\bbraa{\tsig}\tF_{\cc}\bkett{\tsig},
\vspace{-15pt}
\end{equation}
where
\begin{equation}\label{Fc0}
\tF_{\cc}(\vp_1,\vp_2;\vp_3,\vp_4)
=R_{\cc}(\vp_1,\vp_2)^{-1}
\bigl(L_{\cc}(\vp_1,\vp_2;\vp_3,\vp_4)-L_{\cc}(\vp_1,\vp_2;\vp_4,\vp_3)\bigr)\,
R_{\cc}(\vp_3,\vp_4)^{-1}.
\end{equation}
The function $\tF_{\cc}$ is a special case of the connected four-point function $\tF$ defined by \eqref{Gconn}, \eqref{Gconn1}. However, the operator $1-K_{\cc}$ has a null subspace that is generated by the soft mode. Therefore, $\tF_{\cc}$ is only defined on the orthogonal complement of the null subspace.

Since the ``conformal kernel'' $K_{\cc}$ is Hermitian with respect to a natural inner product, it can be diagonalized by constructing a basis of normalizable or $\delta$-normalizable eigenfunctions. This has been accomplished in \cite{PoRo16,MS16}. However, the effect of UV perturbations at the intermediate and IR scales can also be studied using non-normalizable eigenfunctions. A helpful analogy is a quantum particle bound to a shallow 1D potential well. Its wave function has long exponential tails on both sides of the well where the potential vanishes. Such a tail, $\psi(x)\propto e^{-\kappa x}$ is a non-normalizable eigenfunction of the kinetic energy operator. (Unlike plane waves, it is not even $\delta$-normalizable.) In our case, the ``potential well'' is the analytically intractable UV region, situated at the time scale $|\vp_1-\vp_2|\sim\vep=2\pi/(\beta J)$. The ``exponential tail'' is the intermediate asymptotics of $\delta G$ or $g$. Such asymptotics are indeed exponential in the variable $\xi=\ln\bigl(|\vp_1-\vp_2|/\vep\bigr)$.

\subsubsection{Normalizable eigenfunctions and decomposition of identity}
\label{sec_normeig}

As mentioned in the introduction, the conformal symmetry is described by the group of linear fractional maps $z\mapsto\frac{az+b}{cz+d}$ preserving the unit circle $z=e^{i\ph}$. For simplicity, we assume that the orientation of the circle is also preserved. The group $\GG$ of such maps is isomorphic to $\PSL(2,\RR)$. The Green function $\tG(\tau_1,\tau_2)$ is transformed under conformal and more general diffeomorphisms as a $\Delta$-form in each variable. (This term is defined below equation \eqref{h-form}.) We simply call $\tG$ a $(\Delta,\Delta)$-form; similarly, the perturbation source $\tsig$ is a $(1-\Delta,1-\Delta)$-form. However, we have replaced $\tsig$ by $s$ and $\tG$ by $g$ using the transformation \eqref{sg_def}. (Note that this transformation commutes with the action of $\GG$ because $\tG_{\cc}$ is invariant under that action.) Both $s$ and $g$ are $(1/2,1/2)$-forms that are antisymmetric and antiperiodic in each variable.

From now on, we use the notation and some results from a companion paper \cite{SL2R} on the representations of $\GG$ and its universal cover. In particular, $\calF^\mu_\lambda$ stands for the space of $\lambda$-forms with the twisted periodicity condition $f(\vp+2\pi)=e^{2\pi i\mu}f(\vp)$. For our purposes, $\mu$ is either $0$ or $1/2$, but $\lambda$ can be any complex number. Note that the space $\calF^\mu_\lambda$ comes with a Hermitian inner product when $\lambda=\frac{1}{2}$ but not in general. Indeed, the expression $\int f_1(\vp)^*f_2(\vp)\,d\vp$ only makes sense (or more precisely, is $\GG$-invariant) if the integrand is a $1$-form, that is, if $\Re\lambda=\frac{1}{2}$. It is interesting that $\calF^{1/2}_{1/2}$ splits into two invariant subspaces that consist of positive and negative Fourier harmonics, respectively:
\begin{equation}
\calF^{1/2}_{1/2}=\calD^{+}_{1/2}\oplus\calD^{-}_{1/2}.
\end{equation}

The conformal kernel $K_{\cc}$ is a Hermitian operator that acts in the space of $(1/2,1/2)$-forms, $\calH=\calF^{1/2}_{1/2}\otimes\calF^{1/2}_{1/2}$, and commutes with the $\GG$-action. Therefore, the study of $K_{\cc}$ can be simplified by reducing it to the intertwiner space $\operatorname{Hom}_{\GG}(\calU,\calH)$ for each unitary irrep $\calU$. In a slightly less abstract language, we should split $\calH$ into isotypic components. Let us first represent this space as follows:
\begin{equation}\label{Hdecomp}
\calH=\calF^{1/2}_{1/2}\otimes\calF^{1/2}_{1/2}
=\bigl(\calD^{+}_{1/2}\otimes\calD^{+}_{1/2}\bigr)
\oplus\bigl(\calD^{+}_{1/2}\otimes\calD^{-}_{1/2}\bigr)
\oplus\bigl(\calD^{-}_{1/2}\otimes\calD^{+}_{1/2}\bigr)
\oplus\bigl(\calD^{-}_{1/2}\otimes\calD^{-}_{1/2}\bigr).
\end{equation}
The tensor products of $\SL(2,\RR)$ irreps have been fully characterized \cite{Rep76,Rep78}. In particular, the first term in the above equation is the sum of discrete series representations $\calD^{+}_{h}$ for $h=1,2,3,\ldots$. The second term splits into the principal series representations $\calC^{0}_{h(1-h)}$ for $h=\frac{1}{2}+is$\, ($s>0$). However, we are actually interested in antisymmetric $(1/2,1/2)$-forms. In the decomposition of ${\calD^{+}_{1/2}\otimes\calD^{+}_{1/2}}$ into $\calD^{+}_{h}$, they correspond to even values of $h$. The full space of antisymmetric forms is represented as follows:
\begin{equation}
\calH^{-}=\Lambda^{2}\bigl(\calF^{1/2}_{1/2}\bigr)\cong
\biggl(\bigoplus_{n=1}^{\infty}\calD^{+}_{2n}\biggr)\oplus
\biggl(\bigoplus_{n=1}^{\infty}\calD^{-}_{2n}\biggr)\oplus
\biggl(\int_{s=0}^{\infty}\calC^{0}_{1/4+s^2}\,ds\biggr).
\end{equation}
Since each irrep occurs with multiplicity $1$, all its elements are eigenfunctions of $K_{\cc}$. To compute the four-point function, we need to find the corresponding eigenvalues as well as the decomposition of the identity operator into projectors onto the irreps.

The identity operator acting in the space of antisymmetric forms has the integral kernel $\frac{1}{2}\bigl(\delta(\vp_1-\vp_3)\,\delta(\vp_2-\vp_4) -\delta(\vp_1-\vp_4)\,\delta(\vp_2-\vp_3)\bigr)$. To represent it as a sum of projectors, one can apply the decomposition of identity for $\calD^{\pm}_{\lambda_1}\otimes\calD^{\pm}_{\lambda_2}$ (see the last section of \cite{SL2R}) to each term in \eqref{Hdecomp}. This gives an expression for $\delta(\vp_1-\vp_3)\,\delta(\vp_2-\vp_4)$ which is then antisymmetrized; it has the form $\sum_{n=1}^{\infty}\Pi^{\text{discr.}}_{2n} +\int_{0}^{\infty}ds\,\Pi^{\text{cont.}}_{1/2+is}$. Maldacena and Stanford in Sections~3.2.2, 3.2.4 of their paper \cite{MS16} did the calculation by a different method. It addition, they expressed the discrete series projectors as residues of the meromorphic function that defines the continuous series projectors:
\begin{equation}\label{decomp_id}
\wideboxed{
\begin{aligned}
&\frac{1}{2}\Bigl(\delta(\vp_1-\vp_3)\,\delta(\vp_2-\vp_4) -\delta(\vp_1-\vp_4)\,\delta(\vp_2-\vp_3)\Bigr)\\[3pt]
&\quad=\frac{1}{2}\,\left[\sum_{n=1}^{\infty}\Ress_{h=2n}
+\int_{1/2-i\infty}^{1/2+i\infty}\frac{dh}{2\pi i}\right]
\biggl(\frac{h-1/2}{\pi\tan(\pi h/2)}\,\Pi_h(\vp_1,\vp_2;\vp_3,\vp_4)\!\biggr)
\end{aligned}
}
\end{equation}
The poles at $h=2n$ come from the normalization factor rather than the unnormalized projector $\Pi_h$, defined below. The projector kernel is expressed in terms of the variables $\vp_{jk}=2\sin\frac{\vp_j-\vp_k}{2}$ and a $\GG$-invariant cross-ratio $\chi$:
\begin{equation}
\Pi_h(\vp_1,\vp_2;\vp_3,\vp_4)=\vp_{12}^{-1}\vp_{34}^{-1}\Psi_{h}(\chi),\qquad
\chi=\frac{\vp_{12}\vp_{34}}{\vp_{13}\vp_{24}}
=\frac{(z_1-z_2)(z_3-z_4)}{(z_1-z_3)(z_2-z_4)},\qquad
z_j=e^{i\vp_j}.
\end{equation}
The function $\Psi_h$ has different expressions depending on the cyclic order of $\vp_1$, $\vp_2$, $\vp_3$, $\vp_4$; they are not related to each other by analytic continuation.\footnote{This is because the decompositions of identity for $\calD^{+}_{1/2}\otimes\calD^{+}_{1/2}$,\, $\calD^{+}_{1/2}\otimes\calD^{-}_{1/2}$, etc.\ have projector kernels with different analytic properties. When they are combined, no single analytic continuation recipe works.} Things are slightly simplified by reducing the six possible cyclic orders down to three cases:
\begin{equation} \label{regions}
\begin{alignedat}{3}
&\text{OPE region I:} \qquad&
&\vp_1>\vp_2>\vp_3>\vp_4 \quad\text{or}\quad \vp_3>\vp_2>\vp_1>\vp_4, \qquad&
&0<\chi<1,\\[3pt]
&\text{OPE region II:} \qquad&
&\vp_2>\vp_1>\vp_3>\vp_4 \quad\text{or}\quad \vp_3>\vp_1>\vp_2>\vp_4, \qquad&
& \chi<0,\\[3pt]
&\text{OTO region:} \qquad&
&\vp_1>\vp_3>\vp_2>\vp_4 \quad\text{or}\quad \vp_2>\vp_3>\vp_1>\vp_4, \qquad&
&\chi>1.
\end{alignedat}
\end{equation}
To describe $\Psi_{h}$, we will use the scaled hypergeometric function $\hgfs(a,b,c;x)=\Gamma(c)^{-1}\,\hgf(a,b,c;x)$ as well as these auxiliary functions:
\begin{equation}
\begin{alignedat}{2}
A_h(u)&=(1-u)^{h}\,\hgfs(h,h,1;u), \qquad&& u<1,\\[3pt] B_h(u)&=(1-u)^h\,\hgfs(h,h,2h;1-u), \qquad&& 0<u<1.
\end{alignedat}
\end{equation}
Note that $A_h(u)=A_{1-h}(u)$ is a linear combination of $B_{h}(u)$ and $B_{1-h}(u)$; any analytic branch of $A_h(u^{-1})=A_{1-h}(u^{-1})$,\, $B_{h}(u^{-1})$, or $B_{1-h}(u^{-1})$ can also be represented as such a combination. In this notation,
\begin{equation}\label{Psi_h}
\Psi_{h}(\chi)=\left\{\begin{array}{@{}l@{}}
\begin{gathered}
\frac{1}{\cos(\pi h)}\Bigl(
\Gamma(h)^2\cos^2\tfrac{\pi h}{2}\,B_{h}(1-\chi)
-\Gamma(1-h)^2\sin^2\tfrac{\pi h}{2}\,B_{1-h}(1-\chi)\Bigr)\\[-2pt]
\text{for }\, 0<\chi<1,
\end{gathered}\vspace{8pt}\\
\begin{gathered}
\frac{1}{\cos(\pi h)}\Bigl(
\Gamma(h)^2\cos^2\tfrac{\pi h}{2}\,
B_{h}\bigl((1-\chi)^{-1}\bigr)
-\Gamma(1-h)^2\sin^2\tfrac{\pi h}{2}\,
B_{1-h}\bigl((1-\chi)^{-1}\bigr)\Bigr)\\[-2pt]
\text{for }\, \chi<0,
\end{gathered}\vspace{8pt}\\
\begin{gathered}
\frac{\pi}{\sin(\pi h)}
\Bigl(A_h(1-\chi)+A_h\bigl((1-\chi)^{-1}\bigr)\Bigr)
=\Gamma\bigl(\tfrac{h}{2}\bigr)\,\Gamma\bigl(\tfrac{1-h}{2}\bigr)\,
\hgfs\Bigl(\tfrac{h}{2},\,\tfrac{1-h}{2},\,\tfrac{1}{2};\,
\bigl(\tfrac{\chi-2}{\chi}\bigr)^2\Bigr)\\[-2pt]
\text{for }\, \chi>1.
\end{gathered}
\end{array}\right.
\end{equation}
The $\vp_1\leftrightarrow\vp_2$ symmetry takes $\chi$ to $\frac{\chi}{\chi-1}$, and thus, $1-\chi$ to $(1-\chi)^{-1}$.

\subsubsection{General eigenfunctions and the corresponding eigenvalues}\label{sec_geneig}

Let us now consider general, not necessarily normalizable, eigenfunctions of $K_{\cc}$. In this setting, $K_{\cc}$ may not be diagonalizable because the inner product is not available. In fact, rank $2$ generalized eivenvectors $\Psi$ (such that $(K_{\cc}-k_{\cc})^2\Psi=0$ but $(K_{\cc}-k_{\cc})\Psi\not=0$) appear in some situations. However, each ordinary or generalized eigenspace is $\GG$-invariant. An effective strategy to search for (generalized) eigenvectors is to consider abstract representations of $\GG$ and their realization by $(1/2,1/2)$ forms. Let us focus on the representation $\calF^0_{1-h}$ for an arbitrary $h$. An intertwiner $W_{h}$ from this representation to $(1/2,1/2)$-forms is given by the following equation:
\begin{gather}
\bigl(W_{h}f\bigr)(\vp_1,\vp_2)
=(2\pi)^{-1}\int W_{h}(\vp_1,\vp_2;\vp_0)\,f(\vp_0)\,d\vp_0,\\[3pt]
\label{W_h}
W_{h}(\vp_1,\vp_2;\vp_0)
=(\sgn\vp_{12})\,|\vp_{12}|^{h-1}|\vp_{10}|^{-h}|\vp_{20}|^{-h},\qquad
\text{where}\quad
\vp_{jk}=2\sin\frac{\vp_j-\vp_k}{2}.
\end{gather}

The integral kernel of $W_{h}$ looks like the conformal three-point function of fields with scaling dimensions $\frac{1}{2}$, $\frac{1}{2}$, and $h$. We may think of $(W_{h}f)(\vp_1,\vp_2)$ as the response to a perturbation of the form $\int f(\vp)\calO(\vp)\,d\vp$, where the field $\calO$ has dimension $h$. This interpretation provides some intuition but should be used with caution because in the present discussion, $h$ is arbitrary, whereas the OPE for the SYK model has a discrete dimension spectrum \cite{MS16}.

Now, the function $W_{h}(\vp_1,\vp_2;\vp_0)$ with a fixed $\vp_0$ is by itself a good candidate for an eigenfunction of $K_{\cc}$. The integral of $K_{\cc}(\vp_1,\vp_2;\vp_3,\vp_4)\, W_{h}(\vp_3,\vp_4;\vp_0)$ over $d\vp_3\,d\vp_4$ is evalueted in two steps:
\begin{equation}
\figbox{1.0}{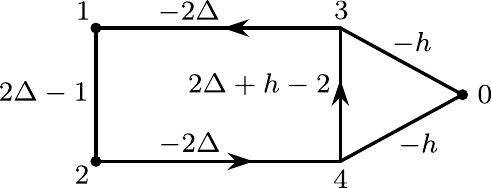} \:\,\propto\:\, \figbox{1.0}{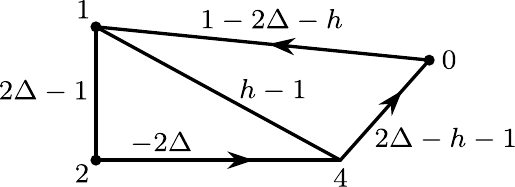}
\:\propto\:\, -\left(\figbox{1.0}{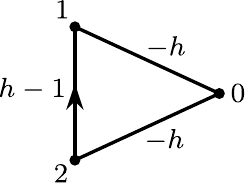}\right).
\end{equation}
In these diagrams, a line with label $a$ stands for $|\ph_{jk}|^a$ and an arrow from $k$ to $j$ for $\sgn\vp_{jk}$. Each step is performed using a star-triangle identity, where the integral is taken over the middle point:
\begin{equation}
\begin{gathered}
\figbox{1.0}{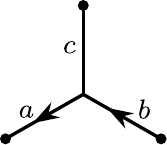}\:=\:
-\frac{4}{\pi}\,\Gamma(1+a)\,\Gamma(1+b)\,\Gamma(1+c)\,
\cos\bigl(\tfrac{\pi}{2}a\bigr)\kern1pt\cos\bigl(\tfrac{\pi}{2}b\bigr)\kern1pt
\sin\bigl(\tfrac{\pi}{2}c\bigr)
\left(\figbox{1.0}{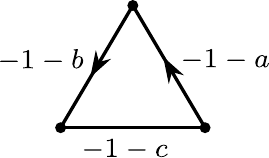}\right)\\
\text{for}\quad a+b+c=-2.
\end{gathered}
\end{equation}
The result is $W_{h}(\vp_1,\vp_2;\vp_0)$ multiplied by the following number (\ie the eigenvalue of the conformal kernel) \cite{Kit.KITP}:
\begin{equation}
k_{\cc}(h)=\frac{u\bigl(\Delta-\frac{1-h}{2}\bigr)\,
u\bigl(\Delta-\frac{h}{2}\bigr)}
{u\bigl(\Delta+\frac{1}{2}\bigr)\,u\bigl(\Delta-1\bigr)},\qquad
\text{where}\quad u(x)=\Gamma(2x)\kern1pt\sin(\pi x).
\end{equation}
Another form of this expression can be found in Table~\ref{tab_coeff} on page~\pageref{tab_coeff}. All solutions of the equation $k_{\cc}(h)=1$, \ie the poles of the function $\frac{k_{\cc}(h)}{1-k_{\cc}(h)}$, are real. We denote the positive solutions by $h_0,h_1,\ldots$ in the increasing order; in particular, $h_0=2$. Since $k_c(1-h)=k_c(h)$, there are also negative solutions.

\section{The SYK model at low temperatures}
\label{sec_renormalization}

We now take a break from the formal style of the previous section and try to describe some interesting physics in the $1\ll\beta J\ll N$ regime using as crude approximations as reasonable. The results concerning renormalization can be generalized and/or derived more rigorously using additional formalism, which will be introduced later.

\subsection{Renormalization scheme}\label{sec_renorm1}

The kinetic term $-\partial_\tau$ in the original effective action $I[\Sigma,G]$ (see \eqref{rdact0}) produces various irrelevant perturbations to the conformal solution. Their intermediate asymptotic form is\footnote{In general, $\delta G$ denotes the difference between the actual and conformal Green functions, whereas $G_{\UV}$ is its part that increases toward the UV region. Thus, $G_{\UV}$ does not include soft mode effects, which are strongest in the IR. The soft mode is absent from the current setting but will be added later.}
\begin{equation}
\delta G(\tau_1,\tau_2)=G_{\UV}(\tau_1,\tau_2)
\approx\const\cdot
|J(\tau_1-\tau_2)|^{1-h}\kern1pt G_{\beta=\infty}(\tau_1,\tau_2)
\qquad\text{if}\quad J^{-1}\ll |\tau_1-\tau_2|\ll \beta,
\end{equation}
where $\Re h>1$. Many such terms contribute to the Green function, but we focus on a single one. Being unable to analytically treat very short times, $|\tau_1-\tau_2|\sim J^{-1}$, we replace $\partial_\tau$ (or equivalently, $\sigma(\tau_1,\tau_2)=\delta'(\tau_1-\tau_2)$) with a suitable source $\tsig$ for the modified action $\tI[\tSig,\tG]$. In doing so, we aim to reproduce the term in $\delta\tG$ that is characterized by a particular exponent $h$.

At first sight, the exponent $h$ seems to be arbitrary. Indeed, $\delta\tG$ is $\tsig$ multiplied by the sum of ladders; if $\tsig$ is proportional to some power of $\tau_1-\tau_2$, then so is $\delta\tG$. To be more precise, let us use the transformation \eqref{sg_def} from $\tG$ and $\tsig$ to $g$ and $s$ so that $\kett{g}=K_{\cc}(1-K_{\cc})^{-1}\kett{s}$. Taking the perturbation source $s(\vp_1,\vp_2)\propto {|\vp_1-\vp_2|^{-h}\sgn(\vp_1-\vp_2)}$, which is the intermediate asymptotics of the unnormalizable eigenfunction $W_{1-h}$ defined by \eqref{W_h}, we obtain the response $\delta g(\vp_1,\vp_2)$ of the same form, multiplied by $\frac{k_{\cc}(h)}{1-k_{\cc}(h)}$. This is equivalent to $\delta G(\tau_1,\tau_2)\propto |\vp_1-\vp_2|^{1-h}G_{\beta=\infty}(\tau_1,\tau_2)$. However, the power law source is not very natural because it directly influences the Green function at intermediate times, whereas the physical effect is due to RG flow. For a clean setting, we should impose the condition that the perturmation source is supported by a slightly extended UV region, where $|\tau_1-\tau_2|$ is bounded from above by $J^{-1}$ times some large constant. With such a cutoff, the response is also concentrated at short time intervals, but only if $\frac{k_{\cc}(h)}{1-k_{\cc}(h)}$ is finite. We will see that in the case of resonance, \ie when $k_{\cc}(h)=1$, the response extends to longer times and, ultimately, contributes to the IR properties of the model.

\begin{figure}
\centerline{\includegraphics{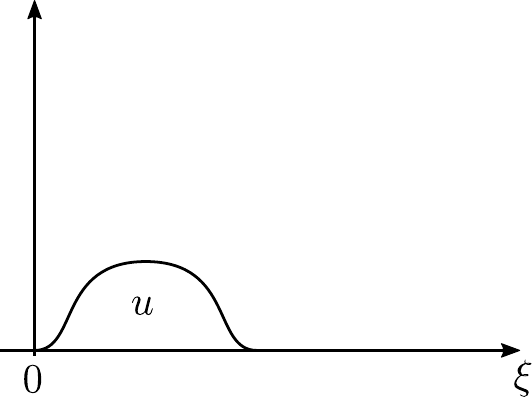} \qquad \qquad \includegraphics{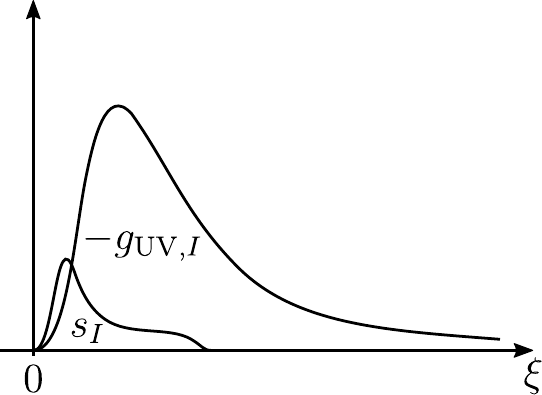}}
\caption{Cartoon of the window function $u$ dressing UV perturbations (left) and the response $\delta g=g_{\UV,I}$ to a UV perturbation $s_{I}$ (right) as a function of $\xi=\ln(|\theta_{12}|/\vep)$.}\label{fig: softsch}
\end{figure}

Let us describe our method in more detail. We work in the physical frame,\footnote{Currently, $\vp(\theta)=\theta$, but we will later use a nontrivial function $\vp$ that represents the soft mode.} $\theta=2\pi\tau/\beta$. The perturbation strength depends on the parameter $\vep=\vep_\theta=2\pi/(\beta J)$, which also sets the UV cutoff for $\theta_1-\theta_2$. We define the renormalization variable to be 
\begin{equation} \label{xidef}
\xi=\ln\frac{|\theta_{12}|}{\vep}.
\end{equation}
Our present analysis is limited to $|\theta_1-\theta_2|\ll 1$, therefore $\theta_{12}=2\sin\frac{\theta_1-\theta_2}{2}$ may be safely replaced with $\theta_1-\theta_2$. To impose gentle UV and IR cutoffs on the perturbation source, we introduce a smooth window function $u$ that spans a sufficiently wide interval of $\xi$ (see Figure~\ref{fig: softsch}) and is normalized as follows:
\begin{equation}
\int_{-\infty}^{\infty}d \xi \, u(\xi)=1.
\end{equation}
The window remains fixed as the maximum value of $\xi$, equal to $\ln(\beta J/\pi)$, goes to infinity.

Let $h=h_I$ be a positive solution of the equation $k_{\cc}(h)=1$ (for example, $h_0=2$). The corresponding perturbation source is
\begin{equation}\label{sIform}
\wideboxed{
s_{I}(\theta_1,\theta_2)
= -a_{I}\kern1pt \vep^{h_I-1}|\theta_1-\theta_2|^{-h_I}
\sgn(\theta_1-\theta_2)\,u(\xi)
}
\end{equation}
where the coefficient $a_I$ can be found by matching the analytically computed response with the numerical solution of the Schwinger-Dyson equations. The factor $\vep^{h_I-1}$ is introduced for the following reason. Translating $s(\theta_1,\theta_2)$ to $\tsig(\theta_1,\theta_2)$ to $\sigma(\tau_1,\tau_2)$ according to equations \eqref{sg_def}, \eqref{tquan} should give an expression that does not involve $\beta$. And indeed, using \eqref{sIform}, we get
\begin{align}
\label{tsigIform}
\tsig_{I}(\theta_1,\theta_2)
&=-a_{I}\kern1pt\sqrt{q-1}\,b^{1/2-\Delta}\,
\vep^{h_I-1}|\theta_1-\theta_2|^{2\Delta-1-h_I}\sgn(\theta_1-\theta_2)\,
u\bigl(\ln|(\theta_1-\theta_2)/\vep|\bigr),\\[6pt]
\sigma_{I}(\tau_1,\tau_2)
&=-a_{I}\kern1pt\sqrt{q-1}\,b^{1/2-\Delta}\kern1pt 
J^2\kern1pt|J(\tau_1-\tau_2)|^{2\Delta-1-h_I}\sgn(\tau_1-\tau_2)\,
u\bigl(\ln|J(\tau_1-\tau_2)|\bigr).
\end{align}
Note that $\int-\sigma_{I}(\tau_1,\tau_2)\,(\tau_1-\tau_2)\,d\tau_1$ is also independent of $J$, and thus, $\sigma_{I}(\tau_1,\tau_2)$ may be regarded as an approximation to the kinetic term $\delta'(\tau_1-\tau_2)$ up to a constant factor. However, this factor need not be $1$ because the perturbation by the kinetic term is nonlinear and because the previously mentioned integral depends on the window function. (The only exception is when $q=2$ and $I=0$, in which case the integral reduces to $2a_0\int_{-\infty}^{\infty}u(\xi)\,d\xi=2a_0$. Hence, $a_0=1/2$ for $q=2$; such linear fitting is implicit in Appendix~C of \cite{JeSuYo16}.)

We will use the fact that for each $h$, the function $f(\theta_1,\theta_2)=|\theta_1-\theta_2|^{-h}\sgn(\theta_1-\theta_2)$ is an approximate eigenfunction of the conformal kernel with the eigenvalue $k_{\cc}(h)$, assuming that ${|\theta_1-\theta_2|}\ll 1$. Indeed, the equation $K_{\cc}f\approx k_{\cc}(h)f$ is made local by interpreting it in terms of the  singularity at $\theta_1-\theta_2\to 0$. As a partial justification, if both $\theta_3$ and $\theta_4$ are far away from $\theta_1,\theta_2$, then the contribution of $f(\theta_3,\theta_4)$ to $(K_{\cc}f)(\theta_1,\theta_2)$ is nonsingular. We take it without proof that if only one of the points is far away, then the contribution may be neglected as well. So, it is sufficient to consider the eigenfunction equation in a small neigborhood of some point $\theta$. In this neighborhood, $f(\theta_1,\theta_2)\approx 4^{1-h}W_{1-h}(\theta_1,\theta_2;\,\theta+\pi)$, where $W_{1-h}$ is the eigenfunction defined by \eqref{W_h}. (As an aside, this argument suggests that the perturbation source may be attributed to a field of conformal dimension $1-h_I$ acting at the point $\theta+\pi$, or even better, at an infinitely distant point.)

Now, we calculate the response $g_{\UV,I}$ to the perturbation $s_I$ in the region $|\theta_1-\theta_2|\ll 1$. In the Fourier representation of the window function,
\begin{equation}
u(\xi)=\int\frac{d\eta}{2\pi}\,\tilde{u}(\eta)\,e^{i\eta\xi},
\end{equation}
$\tilde{u}(\eta)$ is concentrated at small values of $\eta$ because the window is wide. Plugging the Fourier expansion into \eqref{sIform}, we get
\begin{equation}\label{sIasym}
s_{I}(\theta_1,\theta_2)
= -a_I\kern1pt \vep^{h_I-1}
\int \frac{d\eta}{2\pi}\,\vep^{-i \eta}\,
|\theta_1-\theta_2|^{-(h_I-i\eta)}
\sgn(\theta_1-\theta_2)\,\tilde{u}(\eta).
\end{equation}
Thus, the window function serves to smear the power $h_I$ over a small imaginary width. For each given $\eta$, the expression on the right-hand side is an eigenfunction of the conformal kernel with the eigenvalue $k_{\cc}(h)$, where $h=h_I-i\eta$. Therefore, the response is obtained by multiplying the integrand by $\frac{k_{\cc}(h)}{1-k_{\cc}(h)}$. Expanding $k_{\cc}(h)$ to the first order in $h-h_I$, we find that
\begin{equation}
g_{\UV,I}(\theta_1,\theta_2)
\approx -a_I\kern1pt \vep^{h_I-1}
|\theta_1-\theta_2|^{-h_I}\sgn(\theta_1-\theta_2)\,v(\xi),
\end{equation}
where
\begin{equation}
v(\xi)=\frac{1}{-k'_{\cc}(h_I)}\int\frac{d\eta}{2\pi}\,\tilde{u}(\eta)\,
\frac{1}{-i\eta}\,e^{i\eta\xi},
\end{equation}
and there is seen to be an RG equation relating the envelope function $v$ to $u$:
\begin{equation} \label{RGeq}
{d v(\xi) \ov d \xi}= \frac{1}{k'_{\cc}(h_I)}\, u(\xi).
\end{equation}
Integrating with boundary condition $v(-\infty)=0$, we have $v(\xi)=(k_c'(h_I))^{-1}\int_{-\infty}^{\xi}u(\zeta)\,d\zeta$, and in the intermediate region in which all of $u$ has been integrated over in $v$,
\begin{equation} \label{intresp}
\wideboxed{
g_{\UV,I}(\theta_1, \theta_2)
\approx  \frac{a_I}{-k'_{\cc}(h_I)}\,\vep^{h_I-1}  |\theta_1-\theta_2|^{-h_I}\sgn(\theta_1-\theta_2) \qquad
\text{for}\quad \vep \ll |\theta_1-\theta_2| \ll 1
}
\end{equation}
Equivalently,
\begin{equation}
\frac{G_{\UV,I}(\tau_1,\tau_2)}{G_{\beta=\infty}(\tau_1,\tau_2)}
\approx -\frac{a_I}{(-k'_{\cc}(h_I))\sqrt{(q-1)b}}\,
\bigl|J(\tau_1-\tau_2)\bigr|^{1-h}\qquad
\text{for}\quad J^{-1}\ll |\tau_1-\tau_2|\ll \beta.
\end{equation}
In particular, the coefficient for the leading UV correction, denoted by $\alpha_{G}$ in \cite{MS16} and in \eqref{deltaGUV}, is
\begin{equation}\label{alpha_G}
\alpha_{G}=\frac{a_0}{(-k'_{\cc}(2))\sqrt{(q-1)b}}\,.
\end{equation}

\subsection{Derivation of the Schwarzian action} \label{sec: Schw_der}

Let us consider the Green function (or rather, the dynamical variable $G$ in the action $I[\Sigma,G]$) that is deformed by the soft mode. In the introduction, we both expressed it exactly and found two main terms in the $\tau_1-\tau_2$ expansion, see \eqref{Gphi-1} and \eqref{Grepar_intro}. Now we are using slightly different notation, representing the soft mode by a function $\vp$ of the variable $\theta=2\pi\tau/\beta$ and denoting the deformed Green function by $\tG_{\IR}$ (because it does not include the UV corrections):
\begin{equation}
\begin{aligned}
\tG_{\IR}(\theta_1,\theta_2)
&=\tG_{\cc}\bigl(\vp(\theta_1),\vp(\theta_2)\bigr)\,
\vp'(\theta_1)^{\Delta}\vp'(\theta_2)^{\Delta}\\[3pt]
\label{Grepar}
&\approx \tG_{\beta=\infty}(\theta_1,\theta_2)
\biggl(1+\frac{\Delta}{6}\,
\Sch\bigl(e^{i\vp \left( \theta_+ \right)},\theta_+\bigr)\,(\theta_1-\theta_2)^2\biggr)\quad\:
\text{for }\, |\theta_1-\theta_2|\ll 1,
\end{aligned}
\end{equation}
where
\begin{equation}
\tG_{\beta=\infty}(\theta_1,\theta_2)
=-b^{\Delta} |\theta_1-\theta_2|^{-2\Delta}\sgn(\theta_1-\theta_2),\qquad
\theta_+=\frac{\theta_1+\theta_2}{2}.
\end{equation}

The effective action for the soft mode arises due to the coupling between $\tG\approx\tG_{\IR}$ and a UV perturbation source in the full action $\tI[\tSig,\tG]$ as described by the last term in \eqref{rdact1}. The leading source $\tsig=\tsig_0$ is given by \eqref{tsigIform} with $I=0$, \ie $h_I=2$. Integrating it against $\tG_{\beta=\infty}$  gives some number that is independent of $\vp$ and proportional to $\vep^{-1}\propto\beta$. That is just a contribution to the ground state energy, which can be subtracted. So, let us integrate the source with the second term in \eqref{Grepar}:
\begin{equation}\label{Schwarz_2}
\begin{aligned}
I_{\loc}&=-\frac{N}{2}\int d\theta_1\,d\theta_2\,\tsig_{0}(\theta_1,\theta_2)\,
\bigl(\tG_{\IR}(\theta_1,\theta_2)-\tG_{\beta=\infty}(\theta_1,\theta_2)\bigr)
\\[3pt]
&=-N\alpha_{S}\,\vep \int_{0}^{2\pi}
\Sch\bigl(e^{i\vp(\theta)},\theta\bigr)\,d\theta,
\end{aligned}
\end{equation}
where the coefficient $\alpha_S$ is expressed as an integral over $\theta_{12}\approx\theta_1-\theta_2$:
\begin{equation}
\alpha_{S}=a_{0}\kern1pt \sqrt{(q-1)b}\kern3pt \frac{\Delta}{12}
\int_{-\infty}^{\infty}d\theta_{12}\,|\theta_{12}|^{-1}
u\bigl(\ln|\theta_{12}/\vep|\bigr)
=a_{0}\,\frac{\sqrt{(q-1)b}}{6q}.
\end{equation}
Together with \eqref{alpha_G}, this formula gives a relation between two physically significant numbers:
\begin{equation}
\frac{\alpha_G}{\alpha_S}
=\frac{12\pi q^2}{(-k_{\cc}'(2))(q-1)(q-2)\tan(\pi/q)}\,.
\end{equation}

\subsection{Leading four-point function}

The leading contribution to the four-point function is proportional to $\vep^{-1}\propto\beta J$ and comes from the fluctuating soft mode. The general expression is
\begin{equation}\label{F-1_def}
\tF^{(-1)}(\theta_1,\theta_2;\theta_3,\theta_4)
=N\bcorr{\delta\tG_{\IR}(\theta_1,\theta_2)\,
\delta\tG_{\IR}(\theta_3,\theta_4)}_{\loc}\,,
\end{equation}
where the expectation value involves the functional integral over $\vp$ with the Schwarzian action $I_{\loc}$. We assume that $1\ll\beta J\ll N$; the second inequality guarantees that the fluctuations are small so that the Gaussian approximation works. The calculation was first carried out in \cite{MS16}. Our method is very similar, except that we eliminate gauge degrees of freedom at the very beginning. Specifically, we express the Green function $\tG_{\IR}(\theta_1,\theta_2) =\tG_{\cc}\bigl(\vp(\theta_1),\vp(\theta_2)\bigr)\,
\vp'(\theta_1)^{\Delta}\vp'(\theta_2)^{\Delta}$ in terms of the $\SL(2,\RR)$-invariant observable
\begin{equation} \label{Odef}
\calO(\theta)=\Sch\bigl(e^{i\vp(\theta)},\theta\bigr).
\end{equation}

Since the fluctuations are small, it is sufficient to expand $\calO(\theta)$ to the first order and the action to the second order in $\delta\vp(\theta)=\vp(\theta)-\theta$. This formula serves both purposes:
\begin{equation}
\Sch\bigl(e^{i\vp(\theta)},\theta\bigr)
=\frac{1}{2}+\delta\vp'+\delta\vp'''
+\frac{1}{2}(\delta\vp')^2
-(\delta\vp')(\delta\vp''')-\frac{3}{2}(\delta\vp'')^2+O(\delta\vp^3).
\end{equation}
Let $\delta\vp(\theta)=\sum_{m}(\delta\vp)_{m}\kern1pt e^{im\theta}$. The Fourier modes $(\delta\vp)_{m}$ with $m=-1,0,1$ are $\SL(2,\RR)$ generators, and therefore, should cancel from physical observables. In particular,
\begin{equation}\label{deltaO}
\delta\calO(\theta) = \calO(\theta)-\frac{1}{2}
\approx (\partial_{\theta}+\partial_{\theta}^3)\kern1pt\delta\vp(\theta)
=-i\sum_{m}m(m^2-1)\,(\delta\vp)_{m}\kern1pt e^{im\theta}.
\end{equation}
An expression for $\delta\tG_{\IR}=\tG_{\IR}-\tG_{\cc}$ (in the linear approximation) will be derived in Section~\ref{sec_h=2}. It is given by \eqref{gIR_O}, which is equivalent to the first line of the equation below; the second line follows from the fact that $\delta\OO$ does not have $m=-1,0,1$ Fourier harmonics:
\begin{align}\label{GIR_O}
\frac{\delta \tG_{\IR}(\theta_1,\theta_2)}{\tG_{\cc}(\theta_1,\theta_2)}
&\approx \frac{\pi}{q}\int
\left|\frac{\theta_{10}\theta_{20}}{\theta_{12}}\right|
\delta\OO(\theta_0)\,\frac{d\theta_0}{2\pi}
\\[3pt]
\label{GIR_O1}
&=\frac{2\pi}{q}\int_{\theta_2}^{\theta_1}
\frac{\theta_{10}\theta_{02}}{\theta_{12}}\,
\delta\OO(\theta_0)\,\frac{d\theta_0}{2\pi}\qquad
\text{if}\quad 0<\theta_1-\theta_2<2\pi.
\end{align}
The last expression will be helpful for the analysis of the four-point function but we will use \eqref{GIR_O} for now. Combining it with \eqref{F-1_def}, we get 
\begin{equation}\label{F-1_0}
\frac{\tF^{(-1)}(\theta_1,\theta_2;\theta_3,\theta_4)}
{\tG_{\cc}(\theta_1,\theta_2)\,\tG_{\cc}(\theta_3,\theta_4)}
=\frac{\pi^2}{q^2}\,
\int\frac{d\theta_{5}\,d\theta_6}{(2\pi)^2}\,
\left|\frac{\theta_{15}\theta_{25}}{\theta_{12}}\right|\,
P^{(-1)}(\theta_5,\theta_6)\,
\left|\frac{\theta_{36}\theta_{46}}{\theta_{34}}\right|,
\end{equation}
where $P^{(-1)}$ is the correlation function of the observable $\delta\calO$ to the $\vep^{-1}$ order:
\begin{equation} \label{Plead}
P^{(-1)}(\theta_1,\theta_2)
=N\bcorr{\delta\calO(\theta_1)\,\delta\calO(\theta_2)}_{\loc}\,.
\end{equation}
(The subleading $\sim\vep^{0}$ term will be considered in Section~\ref{sec_nonlocal}.) We find $P^{(-1)}(\theta_1,\theta_2)$ using the quadratic expansion of the Schwarzian action in $\delta\vp$:
\begin{equation}
\frac{I_{\loc}}{N}
\approx -2\pi\alpha_{S}\vep \int\biggl(\frac{1}{2}
-\frac{(\delta\vp'')^2-(\delta\vp')^2}{2}\biggr) \frac{d\theta}{2\pi}
= -\pi\alpha_{S}\vep + \pi\alpha_{S}\vep\sum_{m}m^2(m^2-1)\kern1pt
(\delta\vp)_{m}(\delta\vp)_{-m}.
\end{equation}
It implies that
\begin{equation} \label{devpcorr}
N\bcorr{(\delta\vp)_{m}(\delta\vp)_{n}}_{\loc}
=\frac{1}{2\pi\alpha_{S}\vep}\kern3pt \frac{\delta_{m,-n}}{m^2(m^2-1)}\qquad
\text{for}\quad m,n\not=-1,0,1,
\end{equation}
and hence,
\begin{equation} \label{Plead1}
P^{(-1)}(\theta_1,\theta_2)
=N\bcorr{\delta\calO(\theta_1)\,\delta\calO(\theta_2)}_{\loc}
=\frac{1}{2\pi\alpha_{S}\vep}
\ubrace{\sum_{m\not=0}(m^2-1)e^{im(\theta_1-\theta_2)}}
_{-2\pi\left(\rule{0pt}{8pt}
\delta''(\theta_1-\theta_2)+\delta(\theta_1-\theta_2)\right)+1}.
\end{equation}

To complete the calculation, let us introduce a set of auxiliary variables and functions. The final expressions will be different in the OPE and OTO regions,
with those for OPE regions I and II related by the symmetry $\tht_1 \leftrightarrow \tht_2$ of \eqref{F-1_0}. However, for a fixed configuration of fermions, \eqref{F-1_0} is invariant under different choices of coordinates (fixing the period of the circle to be $2\pi$), which account for two possible cyclic orderings -- shown for each region in \eqref{regions} -- or eight possible linear orderings. Thus we are free to consider two representative cases:
\begin{equation} \label{repcases}
\begin{array}{@{}l@{\quad}l@{}}
\text{OPE region:} & 2\pi>\theta_1>\theta_2>\theta_3>\theta_4>0,\vspace{2pt}\\
\text{OTO region:} & 2\pi>\theta_1>\theta_3>\theta_2>\theta_4>0.
\end{array}
\end{equation}
In both cases, the following variables are convenient to use:
\begin{equation}
\tht=\theta_1-\theta_2,\qquad \tht'=\theta_3-\theta_4,\qquad
\De \tht_+=\frac{\theta_1+\theta_2}{2}-\frac{\theta_3+\theta_4}{2}.
\end{equation}
Now, we consider the function
\begin{equation}
\begin{aligned}
&Q(\theta_1,\theta_2,\theta_3,\theta_4)
=\int\frac{d\theta_0}{2\pi}\,|\theta_{10}\theta_{20}\theta_{30}\theta_{40}|
\\[3pt]
&=\frac{2}{\pi}\begin{cases}
\bigl(\pi-\tht-\tht'\bigr)\bigl(\cos \De \tht_++2\cos\frac{\tht}{2}\cos\frac{\tht'}{2}\bigr)
+4\sin\frac{\tht+\tht'}{2}+\bigl(\sin \tht+\sin \tht'\bigr)\cos \De \tht_+
&(\text{OPE})\\[3pt]
(\pi-2\De \tht_+)\bigl(\cos \De \tht_++2\cos\frac{\tht}{2}\cos\frac{\tht'}{2}\bigr)
+\bigl(4+\cos \tht+\cos \tht'\bigr)\sin \De \tht_+
&(\text{OTO})
\end{cases}
\end{aligned}
\end{equation}
and subtract its $m=0$ Fourier harmonic with respect to $\De \tht_+$:
\begin{equation} \label{checkQ_calc}
\begin{aligned} 
&\textstyle\check{Q}(\theta_1;\theta_2,\theta_3,\theta_4)
=Q(\theta_1,\theta_2,\theta_3,\theta_4) -\frac{4}{\pi^2}
\left((\pi-\tht)\cos\frac{\tht}{2}+2\sin\frac{\tht}{2}\right)
\left((\pi-\tht')\cos\frac{\tht'}{2}+2\sin\frac{\tht'}{2}\right)
\\[5pt]
&=\frac{2}{\pi}\begin{cases}
\bigl(\pi+\sin \tht-\tht+\sin \tht'-\tht'\bigr)\cos \De \tht_+
-\frac{2}{\pi}
\bigl(2\sin\frac{\tht}{2}-\tht\cos\frac{\tht}{2}\bigr)
\bigl(2\sin\frac{\tht'}{2}-\tht'\cos\frac{\tht'}{2}\bigr)
&\!(\text{OPE})\!\\[5pt]
\begin{array}{@{}c@{}}
(\pi-2\De \tht_+)\bigl(\cos \De \tht_++2\cos\frac{\tht}{2}\cos\frac{\tht'}{2}\bigr)
+\bigl(4+\cos \tht+\cos \tht'\bigr)\sin \De \tht_+\\
-\frac{2}{\pi}
\bigl(2\sin\frac{\tht}{2}+(\pi-\tht)\cos\frac{\tht}{2}\bigr)
\bigl(2\sin\frac{\tht'}{2}+(\pi-\tht')\cos\frac{\tht'}{2}\bigr)
\end{array}
&\!(\text{OTO})\!
\end{cases}
\end{aligned}
\end{equation}
Using this notation and representing $\hat{P}^{(-1)}$ as $(2\pi\alpha_S\vep)^{-1}(-\partial_{\De \tht_+}^2-1)$ up to the $m=0$ harmonic, we finally perform the integral in \eqref{F-1_0}:
\begin{equation}\label{F-1}
\wideboxed{
\begin{aligned}
&\frac{\tF^{(-1)}(\theta_1,\theta_2;\theta_3,\theta_4)}
{\tG_{\cc}(\theta_1,\theta_2)\,\tG_{\cc}(\theta_3,\theta_4)}
=\frac{\pi}{2q^2\alpha_{S}\vep}\, |\theta_{12}|^{-1}|\theta_{34}|^{-1}
(-\partial_{\De \tht_+}^2-1)\check{Q}(\tht,\tht',\De \tht_+)
\\[6pt]
&=\frac{2}{\pi q^2\alpha_{S}\vep}
\begin{dcases}
\left(1-\frac{\tht}{2\tan\frac{\tht}{2}}\right)
\left(1-\frac{\tht'}{2\tan\frac{\tht'}{2}}\right)
&(\text{OPE})\\[3pt]
-\frac{\pi\sin \De \tht_+}{2\sin\frac{\tht}{2}\sin\frac{\tht'}{2}}
-\frac{\pi(\pi-2\De \tht_+)}{4\tan\frac{\tht}{2}\tan\frac{\tht'}{2}}
+\left(1+\frac{\pi-\tht}{2\tan\frac{\tht}{2}}\right)
\left(1+\frac{\pi-\tht'}{2\tan\frac{\tht'}{2}}\right)
&(\text{OTO})
\end{dcases}
\end{aligned}
}\vspace{5pt}
\end{equation}

Let us briefly discuss some features of the four-point function. First, the OPE correlator is independent of $\Delta\theta_{+}$, which can be explained as follows. If in the derivation of equation \eqref{F-1_0} we use \eqref{GIR_O1} instead of \eqref{GIR_O}, we will get
\begin{equation}\label{F-1_1}
\frac{\tF^{(-1)}(\theta_1,\theta_2;\theta_3,\theta_4)}
{\tG_{\cc}(\theta_1,\theta_2)\,\tG_{\cc}(\theta_3,\theta_4)}
=\frac{4\pi^2}{q^2}\,
\int_{\theta_2}^{\theta_1}\int_{\theta_4}^{\theta_3}
\frac{d\theta_5\,d\theta_6}{(2\pi)^2}\,
\frac{\theta_{15}\theta_{52}}{\theta_{12}}\,
P^{(-1)}(\theta_5,\theta_6)\,
\frac{\theta_{36}\theta_{64}}{\theta_{34}}\,.
\end{equation}
In the OPE case, the intervals $[\theta_2,\theta_1]$ and $[\theta_4,\theta_3]$ do not overlap. Therefore, the $\delta$-function terms in $P^{(-1)}(\theta_5,\theta_6)$ (see \eqref{Plead1}) may be dropped and only the constant term remains. A more physical explanation is this \cite{MS16}: both $\delta\tG_{\IR}(\theta_1,\theta_2)$ and $\delta\tG_{\IR}(\theta_3,\theta_4)$ in \eqref{F-1_def} are determined by the total energy of the system, which is subject to thermal fluctuation.

The OTO correlator is most interesting if we analytically continue it to real time, $t=-i\tau=-i\frac{\beta}{2\pi}\theta$. More exactly, let us consider the function
\begin{equation}\label{ccorr}
\begin{gathered}
-N^{-1}\tF(\theta_1,\theta_2;\theta_3,\theta_4)
=\bcorr{\widetilde{\chi}_j(\theta_1)\,\widetilde{\chi}_k(\theta_3)\,
\widetilde{\chi}_j(\theta_2)\,\widetilde{\chi}_k(\theta_4)}
+\bcorr{\widetilde{\chi}_j(\theta_1)\widetilde{\chi}_j(\theta_2)}
\bcorr{\widetilde{\chi}_k(\theta_3)\widetilde{\chi}_k(\theta_4)},
\\[3pt]
2\pi+\Re\theta_4 \ge\Re\theta_1 \ge\Re\theta_3 \ge \Re\theta_2 \ge\Re\theta_4,
\end{gathered}
\end{equation}
where $\theta_1$, $\theta_2$ are close to $i\frac{2\pi}{\beta}t$ with order of $1$ precision, $\theta_3$, $\theta_4$ are close to $0$, and $t$ is large. In this limit,
\begin{equation}\label{OTOC-1}
{-N^{-1}}\tF^{(-1)}(\theta_1,\theta_2;\theta_3,\theta_4)
\approx \frac{i}{2\alpha_S\vep}\,
e^{-i(\theta_1+\theta_2-\theta_3-\theta_4)/2}\,
\biggl(\frac{2b^{1/q}}{q}\theta_{12}^{-2/q-1}\biggr)
\biggl(\frac{2b^{1/q}}{q}\theta_{34}^{-2/q-1}\biggr).
\end{equation}
This expression is a special case of an ansatz that is discussed in the next section. Thus, the out-of-time-order correlator is proportional to $\frac{\beta J}{N}\, e^{\kap t}$, where $\kap=\frac{2\pi}{\beta}$. The exponential growth saturates when $\frac{\beta J}{N}\,e^{2\pi t/\beta}\sim 1$, at which point the ladder diagrams no longer dominate and one has to include multiple parallel ladders.

\section{Discussion of out-of-time-order correlators}\label{sec_OTOC}

This section is a bit of a digression. We speculate about OTOCs in general systems with all-to-all interaction and a single characteristic time. Some of the ideas were part of the original program~\cite{Kit.BPS} that led to the study of the SYK Hamiltonian; other come from \cite{RSS14,ShSt14,MSS15}. We add some new interpretations and a convenient ansatz.

\paragraph{The intuition.}
Maldacena, Shenker, and Stanford \cite{MSS15} proved the upper bound $\kap\le\frac{2\pi}{\beta}$. The saturation of this bound, found in the SYK model at low temperatures, is a signature of quantum coherence. This intuition has been gained from the study of OTOCs in black holes: the Lyapunov exponent $\kap$ has the maximum value when the collision of gravitational shock waves is described by t'Hooft's effective action \cite{tH87,tH90,tH96}, whereas inelastic scattering results in a negative correction to $\kap$ \cite{ShSt14}. In the former case, t'Hooft defined an $S$-matrix that describes the gravitational interaction of infalling matter and outgoing radiation; it has been further discussed in \cite{KVV95,Pol15}. We will define a similar $S$-matrix for a fairly general quantum system at finite temperature. It characterizes the discrepancy between the full theory (\ie the SYK or a similar many-body Hamiltonian) and its naive version that ignores $1/N$ effects. Such an $S$-matrix is not unitary, but it is almost unitary if the Lyapunov exponent is close to the upper bound.

The naive model includes a small fraction of the actual degrees of freedom, \eg $\chi_1,\dots,\chi_n$ for $n\ll N$, while the rest of the system is replaced by an oscillator bath as originally proposed by Feynman and Vernon \cite{FeVe63}. In the SYK case,
\begin{equation}
H_{\naive}=i\sum_{j=1}^{n}\chi_j\xi_j+H_{\bath},
\end{equation}
where $\xi_j$ are some linear combinations of elementary fermionic operators that constitute the bath. The bath Hamiltonian is quadratic in the elementary fermions but its exact form is not important; it is sufficient to assume that $\corr{\TT\xi_j(\tau_1)\xi_k(\tau_2)}_{\bath} =-\Sigma(\tau_1,\tau_2)\,\delta_{jk}$ and that the higher-order correlators are given by Wick's theorem. On the other hand, the bath may be described as $n$ species of Majorana fermion in $\AdS_2$ with Dirichlet-like boundary conditions. Such a model is expected to reproduce equal-time correlation functions of simple observables, or even the observables that are evolved by the Heisenberg equation over a short period. It should also work for correlators of the form $\corr{X_1(t_1)\cdots X_{s}(t_s)}$ with $t_1<\dots<t_r>\dots>t_s$. These are exactly the correlators that can be measured without reversing the arrow of time or calculated using the Keldysh formalism.

\paragraph{Embedding the naive Hilbert space into the full Hilbert space.}
Since the naive model is reasonably accurate for many purposes, one may try to map its Hilbert space $\calH_{\naive}$ to the Hilbert space $\calH$ of the actual system. Let us examine this problem and see how it is related to out-of-time-order correlators. In fact, there is no genuine embedding $\calH_{\naive}\to\calH$ because the bath has continuous spectral density, and therefore, $\calH_{\naive}$ is infinite-dimensional. But we may restrict $\calH_{\naive}$ to those quantum states that are easy to produce within some time and energy constraints.

As is usual, one begins by defining a set of observable and then constructs the Hilbert space. For a sufficiently short time interval $\Delta t$ and an energy bound $E_{\max}$, we consider the operators $X=\int f(t')\,\chi_j(t')\,dt'$, where the function $f$ is concentrated in the interval $[t-\Delta t,\,t+\Delta t]$ and its Fourier transform at energies below $E_{\max}$, with exponentially decaying tails.
Let us include the products of $0,1,2,\ldots$ such operators, up to a given number. Fixing the details of this definition, we obtain a finite-dimensional subspace $\calA_{t}$ of the operator algebra $\calA_\naive$.

Now, the Hilbert space $\calH_\naive$ is defined by the operator algebra and the thermal state via the Gelfand-Naimark-Segal construction. Specifically, we interpret each operator $X$ as a state vector $\kett{X}_\naive$ and define the inner product using the thermal expectation value in the naive model:
\begin{equation}
\brakett{Y}{X}_\naive=\corr{Y^{\dag}X}_\naive.
\end{equation}
A more constructive description is this: $\kett{X}_\naive=(X\otimes I)\kett{I}_\naive$, where $\kett{I}_\naive$ is the thermofield double state. But mathematically, the thermofield double state is simply the vector associated with the identity operator. (How else can we define it if we begin with local observables but no vectors or Hamiltonian?) The same construction is applicable to operators of the full model. It gives the Hilbert space $\calH=\calH_{\text{physical}}\otimes\calH_{\text{physical}}^{*}$. In this case, the thermofield double state has an independent definition, $\kett{I}=Z^{-1/2}\sum_{m}e^{-\beta E_m/2}\kett{m,m}$.

Restricting the space of operators to $\calA_{t}$, we obtain the subspace $\calH_{t}\subseteq\calH_\naive$. Each element $X\in\calA_{t}$ can also be interpreted as an operator acting on the physical system. Thus, $\calH_{t}$ is mapped to the full Hilbert space $\calH$. This map is not unitary because the inner product $\brakett{Y}{X}=\corr{Y^{\dag}X}$ is silghtly different from the previous one. But since the naive model works well for simple observables, the difference should be negligible. More exactly, we assume that\footnote{The bound \eqref{inprod_diff} imposes a constraint even on those states that are produced with difficulty, \ie such that $\bigl\|\kett{X}_\naive\bigr\|$ is much smaller than the operator norm of $X$. For example, the annihilation operator of a particle with energy much higher than the temperature creates an excitation on the other side of the thermofield double, though with a tiny amplitude.}
\begin{equation}\label{inprod_diff}
\brakett{Y}{X}-\brakett{Y}{X}_\naive
= O(N^{-1})\,\bigl\|\kett{Y}_\naive\bigr\|\,\bigl\|\kett{X}_\naive\bigr\|.
\end{equation}

\paragraph{There are actually two embeddings!}
The implicit constant in the big-$O$ notation in \eqref{inprod_diff} depends on the dimension of the Hilbert space $\calH_{t}$. Surely, as we include more observables, the naive model becomes less accurate. However, there is a more serious problem: the accuracy deteriorates dramatically if we allow products of operators $\chi_j(t')$ in a large time window. Fortunately, there is a way to refine the definition of the embedding so as to mitigate this effect. In general, we divide the window into small overlapping intervals centered at $t_1>t_2>\dots>t_s$ and write the naive operator we want to embed as $X=X_1X_2\cdots X_s$, where $X_j\in\calA_{t_j}$. Alternatively, we can use the reverse time order, $\tilde{X}=\tilde{X}_s\cdots\tilde{X}_2\tilde{X}_1$. Any operator of the naive theory can be expressed as a linear combination of operators of the first or the second type using the commutation relations of free fermions. It turns out that the embedding based on either ordering is reasonably good but the errors grow exponentially with time if the two orderings are mixed.

To keep things simple, we will not discuss large continuous windows, but rather, consider the space of operators $\calA_{\{0,t\}}$ that act in two disjoint intervals, $[-\Delta t,\Delta t]$ and $[t-\Delta t,\,t+\Delta t]$. The elements of $\calA_{\{0,t\}}$ are operators of the form $X=X_tX_0$, where $X_t\in\calA_t$ and $X_0\in\calA_0$. We assume that $t$ is sufficiently large so that any connected two-point function between times $0$ and $t$ is very small. In the naive theory, this implies that $X_t$ and $X_0$ almost commute (or anticommute). Furthermore, if $X_t,Y_t\in\calA_t$ and $X_0,Y_0\in\calA_0$, then
\begin{equation}
\bbrakett{Y_tY_0}{X_tX_0}_\naive
=\bcorr{Y_{0}^{\dag}Y_{t}^{\dag}X_{t}X_{0}}_\naive
\approx \corr{Y_{0}^{\dag}X_{0}}_\naive\corr{Y_{t}^{\dag}X_{t}}_\naive.
\end{equation}
Thus, the Hilbert space $\calH_{\{0,t\}}$ is simply the ($\ZZ_2$-graded) tensor product of $\calH_0$ and $\calH_t$. The two embeddings are as follows:
\begin{equation}
\wideboxed{
S_{\OUT},S_{\IN}:\,\calH_{\{0,t\}}\to\calH,\qquad
S_{\OUT}\kett{X_tX_0}_\naive=\kett{X_tX_0},\quad\:
S_{\IN}\kett{X_tX_0}_\naive=\pm\kett{X_0X_t}
}
\end{equation}
(In the last equation, the minus sign is chosen if both $X_0$ and $X_t$ have odd fermionic parity.) To see that $S_{\OUT}$ is indeed an embedding of Hilbert spaces (with reasonable accuracy), we check that it does not distort the inner product too much:
\begin{equation}
\braa{Y_tY_0}_\naive\, S_{\OUT}^{\dag}S_{\OUT} \kett{X_tX_0}_\naive
=\corr{Y_0^{\dag}Y_t^{\dag}X_tX_0}
\approx\corr{Y_0^{\dag}X_0}_\naive\corr{Y_t^{\dag}X_t}_\naive.
\end{equation}
Here, we have used the assumptions that Keldysh correlators are faithfully described by the naive model and that the connected two-point functions like $\ccorr{\chi_j(t)\chi_j(0)}$ are negligible. (Recall that the naive model obeys Wick's theorem.) The same argument is applicable to $S_{\IN}$. However,
\begin{equation}
\braa{Y_tY_0}_\naive\, S_{\OUT}^{\dag}S_{\IN} \kett{X_tX_0}_\naive
=\pm\corr{Y_0^{\dag}Y_t^{\dag}X_0X_t}
\not\approx\corr{Y_0^{\dag}X_0}\corr{Y_t^{\dag}X_t}.
\end{equation}
because OTOCs are not generally reproduced by the naive model. Therefore, $S_{\OUT}\not=S_{\IN}$; the difference is given by
\begin{equation}
\bigl\|(S_{\OUT}-S_{\IN})\kett{X_tX_0}_\naive\bigr\|^2
=\bcorr{[X_t,X_0]^{\dag}[X_t,X_0]}.
\end{equation}

\paragraph{$S$-matrix and quantum coherence.}
The state $S_{\OUT}\kett{X_tX_0}_\naive=(X_tX_0\otimes I)\kett{I}$ (where $\kett{I}$ is the thermofield double of the physical system) may be regarded as an outgoing scattering state. In the bulk picture (if one exists), such a state is described as a pair of wave packets on a future time slice \cite{ShSt13.1,ShSt14}. Similarly, $S_{\IN}\kett{X_tX_0}_\naive$ is an in-state. Thus, the scattering operator is $S=S_{\OUT}^{\dag}S_{\IN}$. Because both $S_{\OUT}$ and $S_{\IN}$ are almost-isometric embeddings, we have
\begin{equation}
\|S\|\le 1+O(N^{-1}).
\end{equation}
The relations between the operator $S$ and OTOCs can be summarized as follows:
\begin{empheq}[box=\widebox]{align}
\label{OTOC_0}
\braa{Y_tY_0}_\naive\,S\kett{X_tX_0}_\naive
&\approx\pm\corr{Y_0^{\dag}Y_t^{\dag}X_0X_t}
\\[5pt]
\label{OTOC_1}
\braa{Y_tY_0}_\naive\,(I-S)\kett{X_tX_0}_\naive
&\approx\corr{Y_0^{\dag}X_0}\corr{Y_t^{\dag}X_t}
\mp\corr{Y_0^{\dag}Y_t^{\dag}X_0X_t}\qquad
\\[5pt]
\label{OTOC_2}
\bbraa{Y_tY_0}_\naive(2I-S-S^\dag)\bkett{X_tX_0}_\naive
&\approx\bcorr{[Y_t,Y_0]^{\dag}[X_t,X_0]}
\end{empheq}
Replacing $S$ with $S^{\dag}$ in \eqref{OTOC_0}, \eqref{OTOC_1} changes the order of times from $0\kern1pt t\kern1pt 0\kern1pt t$ to $t\kern1pt 0\kern1pt t\kern1pt 0$. The square brackets denote the supercommutator, $[A,B]=AB\mp BA$. 

Let us focus on the early times when connected OTOCs of the form \eqref{OTOC_1} are small. For the SYK model, they are of the order of 
$\lambda=\frac{\beta J}{N}e^{(2\pi/\beta)t}$; more generally, $\lambda\propto N^{-1}e^{\kap t}$. We assume this to be an upper bound for all suitably normalized operators $X_tX_0\in\calA_{\{0,t\}}$ so that $\|I-S\|\sim\lambda\ll 1$. If $S$ is unitary, then $2I-S-S^\dag=(I-S)^{\dag}(I-S)=O(\lambda^2)$, implying this property (provided $X_0$ and $X_t$ are suitably normalized):
\begin{equation}\label{coh_S}
\text{Coherent early-time regime:}\qquad
\corr{[X_t,X_0]^{\dag}[X_t,X_0]}=O(\lambda^2).
\end{equation}
On the contrary, the typical behavior in most systems is $\corr{[X_t,X_0]^{\dag}[X_t,X_0]}\sim\lambda$. We cannot exclude the possibility of restoring unitarity by adding new states that are generated by more complex operators of the naive model. However, their connected OTOCs would have to scale as $\sqrt{\lambda}$, which is unlikely. Property \eqref{coh_S} has previously been noticed for black holes \cite{Kit.BPS} and it holds for the SYK model at low temperatures. Douglas Stanford and Yingfei Gu independently showed (in private discussions) that it follows from the condition $\kap=\frac{2\pi}{\beta}$. We will employ the same idea (based on \cite{MSS15}) together with some simplifying assumptions that are natural for systems with all-to-all interactions.

\paragraph{Single-mode ansatz for early-time OTOCs.}
Let us use the variable $\theta=\frac{2\pi}{\beta}\tau =i\frac{2\pi}{\beta}t$, define the dimensionless Lyapunov exponent
\begin{equation}
\tkap=\frac{\beta}{2\pi}\,\kap,\qquad 0<\tkap\le 1,
\end{equation}
and consider four complex times $\theta_1,\theta_2,\theta_3,\theta_4$ such that
\begin{equation}
\theta_1,\theta_2=i\kern1pt\frac{2\pi}{\beta}\kern1pt t+O(1),\quad
\theta_3,\theta_4=O(1),\qquad
2\pi+\Re\theta_4 \ge\Re\theta_1 \ge\Re\theta_3 \ge \Re\theta_2 \ge\Re\theta_4.
\end{equation}
Generalizing \eqref{OTOC-1} and similar equations for gravitational shock waves \cite{ShSt14}, we expect the connected OTOC to have the following form:
\begin{equation}\label{ansatz}
\figbox{1.0}{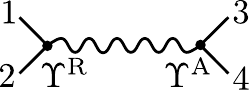}\qquad
\wideboxed{
\begin{gathered}
\bcorr{X_1(\theta_1)\,X_2(\theta_2)}\,\bcorr{X_3(\theta_3)\,X_4(\theta_4)}
\mp\bcorr{X_1(\theta_1)\,X_3(\theta_3)\,X_2(\theta_2)\,X_4(\theta_4)}
\\[5pt]
\approx C^{-1} e^{i\tkap(\pi-\theta_1-\theta_2+\theta_3+\theta_4)/2}\,
\Upsilon_{X_{1},X_{2}}^{\Ret}(\theta_1-\theta_2)\,
\Upsilon_{X_{3},X_{4}}^{\Adv}(\theta_3-\theta_4)
\end{gathered}
}
\end{equation}
with $O(\lambda^2)$ accuracy, where $\lambda=C^{-1}e^{2\pi\tkap t/\beta}$. The diagram on the left conveys the intuition: the process is mediated by some mode (``scramblon'') with the propagator $C^{-1} e^{i\tkap(\pi-\theta_1-\theta_2+\theta_3+\theta_4)/2}$, whereas $\Upsilon^{\Ret}$ and $\Upsilon^{\Adv}$ are the vertex functions. Alternatively, one may regard $C^{-1}e^{i\tkap\pi/2}$ as the propagator and define
\begin{equation}
\!\!\tUp_{X_{1},X_{2}}^{\Ret}(\theta_1,\theta_2)
=e^{-i\tkap(\theta_1+\theta_2)/2}
\Upsilon_{X_{1},X_{2}}^{\Ret}(\theta_1-\theta_2),
\qquad
\tUp_{X_{3},X_{4}}^{\Adv}(\theta_3,\theta_4)
=e^{i\tkap(\theta_3+\theta_4)/2}
\Upsilon_{X_{3},X_{4}}^{\Adv}(\theta_3-\theta_4).
\end{equation}
For the SYK model, one of the vertex functions satisfies the equation $(K^{\Ret}-1)\tUp_{\chi,\chi}^{\Ret}=0$, where $K^{\Ret}(\tht_1, \tht_2, \tht_3, \tht_4)
=(q-1)\tG^{\Ret}(\tht_1,\tht_3)\,\tG^{\Adv}(\tht_4,\tht_2)\,
\bigl|\tG^{\Wig}(\tht_3,\tht_4)\bigr|^{q-2}$
is a retarded kernel (\cf \eqref{Kdef}) where retarded, advanced, and Wightman Green functions are defined with respect to the imaginary part of Euclidean time arguments $\tht_i$. Likewise,  $\tUp_{\chi,\chi}^{\Adv}$ is an eigenvalue $1$ eigenfunction of an analogously defined advanced kernel. This was the first derivation of the Lyapunov exponent in the zero-temperature limit \cite{Kit.KITP.1} and one of the methods used in \cite{MS16} to compute the $(\beta J)^{-1}$ correction.

Applying the arguments of \cite{MSS15} to correlators of the specific form \eqref{ansatz}, one can show that
\begin{gather}
\Upsilon_{X_{1},X_{2}}^{\Ret}(\theta)^*
=\Upsilon_{X_{2}^{\dag},X_{1}^{\dag}}^{\Ret}(\theta^*),\qquad
\Upsilon_{X_{2},X_{1}}^{\Ret}(\theta)
=\Upsilon_{X_{1},X_{2}}^{\Ret}(2\pi-\theta)\qquad
(\text{and similarly, for } \Upsilon^{\Adv});
\\[5pt]
\Upsilon_{X^\dag,X}^{\Ret}(\theta),\,
\Upsilon_{X^\dag,X}^{\Adv}(\theta)\ge 0\quad
\text{for } 0\le\theta\le 2\pi,\qquad\quad C>0.
\end{gather}
It follows that up to $O(\lambda^2)$ terms, 
\begin{equation}
\Bigl\langle
\bigl[X_3(\theta),X_1\bigl(\theta+i\tfrac{2\pi}{\beta}t\bigr)\bigr]\,
\bigl[X_2\bigl(i\tfrac{2\pi}{\beta}t\bigr),X_4(0)\bigr]
\Bigr\rangle
\approx\frac{2\cos(\tkap\pi/2)}{C}\,e^{\tkap(2\pi/\beta)t}\,
\Upsilon_{X_{1},X_{2}}^{\Ret}(\theta)\,
\Upsilon_{X_{3},X_{4}}^{\Adv}(\theta).
\end{equation}
For the SYK model, both the decoherence factor $\cos(\tkap\pi/2)$ and the number $C$ are proportional to $(\beta J)^{-1}$; hence, the OTOC of commutators has a finite limit at zero temperature.

\paragraph{The coherent regime.} In the SYK case (see \eqref{OTOC-1}), the functions $\tUp^{\Ret}$ and $\tUp^{\Adv}$ are obtained by applying the $\sL_2$ generators $L_{-1}$ and $L_{1}$ to one of the variables of the conformal Green function. In Lorentzian time, they generate, respectively, an exponentially growing and an exponentially decreasing perturbation. By definition, the operator $L_{m}$ is minus the Lie derivative along the vector field $v(\theta)=e^{im\theta}$. Its action on $1/q$-forms of the variable $\theta_j$ is given by $L_{m}^{(j)}=-e^{im\theta_j}(\partial_{\theta_j}+im/q)$. Thus,
\begin{equation}
\tUp^{\Ret}_{\tilde{\chi},\tilde{\chi}}(\theta_1,\theta_2)
=\bigl(L_{-1}^{(1)}\kern1pt\tG_{\cc}\bigr)(\theta_1,\theta_2)
=-\frac{2b^{1/q}}{q}\,e^{-i(\theta_1+\theta_2)/2}\,\theta_{12}^{-2/q-1}.
\end{equation}

This result can be generalized. We conjecture that if the Lyapunov exponent is close to the upper bound, then
\begin{empheq}[box=\widebox]{gather}
\label{ansatz_c1}
\tUp^{\Ret}_{X_1,X_2}(\theta_1,\theta_2)
=i\,\bcorr{[p^{\OUT},X_1(\theta_1)]\,X_2(\theta_2)},\qquad
\tUp^{\Adv}_{X_3,X_4}(\theta_1,\theta_2)
=i\,\bcorr{[p^{\IN},X_3(\theta_3)]\,X_4(\theta_4)}
\\[5pt]
\label{ansatz_c2}
C=i\,\bcorr{[p^{\IN},p^{\OUT}]}
\end{empheq}
where $p^\OUT=p^\OUT_\naive+p^\OUT_{\other}$ and $p^\IN=p^\IN_\naive+p^\IN_\other$ are defined in a suitable effective theory. For the SYK model, it is the naive theory augmented with the soft mode, such that $p^\OUT_\naive=iL_{-1}$,\, $p^\IN_\naive=iL_{1}$, and the operators $p^\OUT_\other$, $p^\IN_\other$ act on $\vp\in\Diff^{+}(S^1)$ on the left (\ie by changing the value of $\vp(\theta)$ rather than the time variable $\theta$). The complete picture, including the connection to a 2D black hole, has been worked out in \cite{Jen16, MSY16, EMV16}. In the black hole case, the naive description is the low-energy field theory, whereas the other degrees of freedom are a pair of gravitational shock waves \cite{tH87,tH90,tH96}. Note that in any case, the main contribution to $C$ in \eqref{ansatz_c2} comes from $[p^\IN_\other,p^\OUT_\other]$. Further details will be given elsewhere.

\section{Formalism (part 2)}

The purpose of this section is to develop tools that have (among others) the following applications. By using the conformal three-point function $W_h(\theta_1,\theta_2;\theta_0)$, we will derive the expression for $\delta\tG_{\IR}$ in terms of $\delta\calO$ and extend the formula for $g_{\UV,I}(\theta_1,\theta_2)$ from $|\theta_1-\theta_2|\ll 1$ to the general case. A four-point conformal function will be employed in Section~\ref{sec_next_order} to find a non-local order $\vep^2$ correction to the Schwarzian action, which is itself proportional to $\vep=2\pi/(\beta J)$. We will later study the full four-point function. Its leading $\sim\beta J$ term is not conformal and comes from the soft mode, whereas the subleading $\sim 1$ correction has both conformal and non-conformal pieces. Their calculation is quite technical, and various identities obtained here will come in useful.

\subsection{Properties and some applications of conformal functions} \label{sec: conf_func}

\subsubsection{The 2- and 3-point functions and an elementary 4-point function}

We consider the 2-point function $U_h$ of fields with scaling dimension $h$, the function $W_h$ (which defines an intertwiner from $(1-h)$-forms to antisymmetric $(1/2,1/2)$-forms), and the unnormalized projector kernel $\Pi_h$ discussed in Section~\ref{sec_normeig}. Let us put all definitions in one place:
\begin{empheq}[box=\widebox]{gather}
\label{UWdef}
U_{h}(\vp_1,\vp_2)=|\vp_{12}|^{-2h},\qquad
W_{h}(\vp_1,\vp_2;\vp_0)
=\vp_{12}^{-1}\biggl|\frac{\vp_{10}\vp_{20}}{\vp_{12}}\biggr|^{-h},\qquad
\vp_{jk}=2\sin\frac{\vp_j-\vp_k}{2}
\\[5pt]
\Pi_{h}(\vp_1,\vp_2;\vp_3,\vp_4)
=\vp_{12}^{-1}\vp_{34}^{-1}\Psi_{h}(\chi),\qquad
\text{where}\quad
\chi=\frac{\vp_{12}\vp_{34}}{\vp_{13}\vp_{24}}
\end{empheq}
The function $\Psi_h$, defined by \eqref{Psi_h}, has a succinct integral representation:
\begin{equation}
\label{Psi_h_int}
\Psi_{h}(\chi)
=\pi\int\,
\biggl|\frac{\vp_{10}\vp_{20}}{\vp_{12}}\biggr|^{-h}
\biggl|\frac{\vp_{30}\vp_{40}}{\vp_{34}}\biggr|^{h-1}\,
\frac{d\vp_0}{2\pi}.
\end{equation}
(The integral converges if $0<\Re h<1$, but the resulting expression extends to a meromorphic function of $h$.) Let us write this identity in an operator form, along with a similar relation between the 2- and 3-point functions:
\begin{equation}\label{PiWUrel}
\wideboxed{
\Pi_h=\pi\, W_{h}W_{1-h}^{\Tt},\qquad\quad
W_{h}=\pi\,\frac{\cos(\pi h)\,\Gamma(2h)}{\cos^2\frac{\pi h}{2}\,\Gamma(h)^2}\,
W_{1-h}U_{h}
}
\end{equation}
Here, $W_{1-h}^{\Tt}$ is the operator with the integral kernel $W_{1-h}^{\Tt}(\vp_0;\vp_1,\vp_2)=W_{1-h}(\vp_1,\vp_2;\vp_0)$; it maps antisymmetric $(1/2,1/2)$-forms to ($1-h$)-forms. The operator $U_{h}$ is an intertwiner from $\calF^0_{1-h}$ to $\calF^0_{h}$. The product of operators in the above equations is defined as an integral over $d\vp_{0}/(2\pi)$.\smallskip

\subsubsection{Fourier representation}

Let $f_{h,m}$ be the $m$-th Fourier harmonic, $f_{h,m}(\vp_0)=e^{im\vp_0}$ regarded as an element of $\calF^0_{h}$. Then we can write abstractly:
\begin{equation} \label{UWfourier}
U_{h}f_{1-h,m}=u_{h,m}f_{h,m},\qquad
W_{h}f_{1-h,m}=W_{h,m}.
\end{equation}
Because $W_{h}$ is an intertwiner from $\calF^{0}_{1-h}$, the functions $W_{h,m}$ transform as the basis elements of that space. In particular, the generators $L_{-1}$, $L_{0}$, $L_{1}$ of $\sL_2$ (\ie the complexified Lie algebra of $\GG$) act on the functions $W_{h,m}$ as follows (\cf equation (15) in \cite{SL2R}):
\begin{equation}
L_{0}W_{h,m}=-mW_{h,m},\qquad
L_{\pm1}W_{h,m}=-\bigl(m\pm(1-h)\bigr)W_{h,m\pm1}.
\end{equation}

The calculation of the coefficients $u_{h,m}$ is quite straightforward:
\begin{equation}
\begin{aligned} \label{Ufourier}
u_{h,m}&=\int_{0}^{2\pi}\bigl(2\sin\tfrac{\vp}{2}\bigr)^{-2h}
e^{im\vp}\,\frac{d\vp}{2\pi}
=e^{i\pi m}\,\frac{\Gamma(1-2h)}{\Gamma(1-h+m)\,\Gamma(1-h-m)}\\[3pt]
&=\frac{1}{2\cos(\pi h)\,\Gamma(2h)}\,\frac{\Gamma(h\pm m)}{\Gamma(1-h\pm m)}
\quad \text{since $m$ is an integer}.
\end{aligned}
\end{equation}
Similarly, for the 3-point function,
\begin{equation} \label{W_hm}
W_{h,m}(\vp_1,\vp_2)
=\vp_{12}^{-1}\int\,\biggl|\frac{\vp_{10}\vp_{20}}{\vp_{12}}\biggr|^{-h}
e^{im\vp_0}\,\frac{d\vp_0}{2\pi}
=\vp_{12}^{-1}\,e^{im(\vp_1+\vp_2)/2}\,w_{h,m}(\vp_1-\vp_2),
\end{equation}
where\footnote{Since $W_{h,m}(\vp_1,\vp_2)$ is antiperiodic in each variable, the function $w_{h,m}$ satisfies the condition $w_{h,m}(\vp+2\pi)=(-1)^{m}w_{h,m}(\vp)$. Therefore, it is sufficient to determine it in the fundamental domain $0<\vp<2\pi$.}
\useshortskip{\begin{gather}
\label{w_hm}
w_{h,m}(\vp)
=\frac{1}{2\cos\frac{\pi h}{2}}\,\frac{\Gamma(h\pm m)}{\Gamma(h)}\,
C_{h,\pm m}(e^{i\vp}),
\qquad\quad 0<\vp<2\pi,
\\[5pt]
C_{h,m}(u)=A^{+}_{h,m}(u)+A^{-}_{h,m}(u^{-1}),\qquad
A_{h,m}(u)=u^{m/2}(1-u)^{h}\,\hgfs\bigl(h,\,h+m,\,1+m;\,u\bigr).
\end{gather}}
The superscript ``$\pm$'' in $A_{h,\pm m}^{\pm}(z)$ indicates the function is analytically continued to the domain $\CC-[0,+\infty)$ from the interval $0<z<1$, where it is unambiguously defined, through the upper or lower half-plane for $+$ and $-$, respectively. The sign in front of $m$ does not matter because $\Gamma(h+m)\,A_{h,m}(u)=\Gamma(h-m)\,A_{h,-m}(u)$. Another identity, $A_{h,m}(u)=A_{1-h,m}(u)$ implies that
\begin{equation}\label{w_h_1-h}
w_{h,m}(\vp)=\frac{\pi}{2\cos^2\frac{\pi h}{2}\,\Gamma(h)^2}\,
\frac{\Gamma(h\pm m)}{\Gamma(1-h\pm m)}\,w_{1-h,m}(\vp).
\end{equation}
This relation is just the second equation in \eqref{PiWUrel} written in terms of Fourier coefficients.

Let us give another expression for $w_{h,m}$ under the same restriction on the variable, $0<\vp<2\pi$:
\begin{equation}\label{w_hm_B}
w_{h,m}(\vp)
=\frac{1}{2\cos(\pi h)}\biggl(
\frac{\Gamma(h+m)}{\Gamma(1-h+m)}\,
e^{i\frac{\pi}{2}h}\kern1pt B_{h,m}^{+}(e^{i\vp})
-\frac{\Gamma(1-h)\kern1pt\tan\frac{\pi h}{2}}{\Gamma(h)}\,
e^{i\frac{\pi}{2}(1-h)} B_{1-h,m}^{+}(e^{i\vp})\biggr).
\end{equation}
Here, $B_{h,m}^{+}(u)$ is the analytic continuation of $B_{h,m}(u)=u^{m/2}(1-u)^{h}\,\hgfs(h,\,h+m,\,2h;\,1-u)$ from the interval $0<u<1$ to the domain $\CC-[0,+\infty)$ through the upper half-plane.

The following special cases will be used frequently. (In \eqref{w_01}, we assume that $0<\vp<2\pi$.)
\begin{empheq}[box=\widebox]{gather}
\label{w_2m}
w_{2,m}(\vp)=\frac{\sin\bigl(|m|\vp/2\bigr)}{\tan(\vp/2)}-|m|\cos(m\vp/2)
\\[6pt]
\label{w_-1m}
w_{-1,m}(\vp)
=\begin{dcases}
\,\,\frac{2}{\pi}\,w_{[0]}(\vp) & \text{if } m=0,
\vspace{3pt}\\
-\frac{1}{\pi}\,w_{[1]}(\vp) & \text{if } m=\pm 1,
\vspace{4pt}\\
\,\,\frac{2}{\pi}\,\frac{w_{2,m}(\vp)}{|m|(m^2-1)} & \text{if } |m|\ge 2
\end{dcases}\\[6pt]
\label{w_01}
w_{[0]}(\vp)=\bigl[\partial_{h}w_{h,0}(\vp)\bigr]_{h=2}
=\frac{\pi-\vp}{2\tan\frac{\vp}{2}}+1,\qquad
w_{[1]}(\vp)=\bigl[\partial_{h}w_{h,1}(\vp)\bigr]_{h=2}
=\frac{\pi-\vp}{2\sin\frac{\vp}{2}}+\cos\frac{\vp}{2}
\end{empheq}

\subsubsection{Linearized IR perturbations}\label{sec_h=2}

The soft mode generates perturbations $\delta\tG_{\IR}(\theta_1,\theta_2) =R_{\cc}(\theta_1,\theta_2)^{-1} \de g_{\IR}(\theta_1,\theta_2)$ to the conformal Green function. Note that $\de g_{\IR}\in\calD^{+}_{2}\oplus\calD^{-}_{2}$. Indeed, suppose that $\tG=\tG_{\IR}$ is equal to the conformal Green function $\tG_{\cc}$ in some frame $\ph$ that is very close to the physical frame $\theta=2\pi\tau/\beta$. In the linear order, the difference between the frames,
\begin{equation}
\delta\vp(\theta)=\vp(\theta)-\theta
\end{equation}
is a vector field. Vector fields are $-1$-forms, \ie elements of the representation $\calF^0_{1-h}$ with $h=2$. However, the Fourier harmonics of $\delta\vp$ with $m=-1,0,1$ are symmetry generators; they do not produce any change in $\tG_{\IR}$. The quotient of $\calF^0_{-1}$ by those null modes splits into the representations $\calD^{+}_2$ and $\calD^{-}_2$, which correspond to $m\ge 2$ and $m\le-2$, respectively.

Let us actually calculate $\delta\tG_{\IR}$ for a given $\delta\vp$. The function $\tG_{\IR}$ transforms as a $\Delta$-form (with $\Delta=q^{-1}$) in each variable. Infinitesimal transformations of $\Delta$-forms involve a particular type of Lie derivative:
\begin{equation} \label{LieDdef}
\delta\tG_{\IR}(\theta_1,\theta_2)
=\bigl(\Ld_{\delta\vp}^{(1)}
+\Ld_{\delta\vp}^{(2)}\bigr)\,\tG_{c}(\theta_1,\theta_2),\qquad
\Ld_{v}^{(j)}=v(\theta_j)\,\partial_{\theta_j}+\Delta\, v'(\theta_j)\quad
\text{for } j=1,2.
\end{equation}
Evaluating the Lie derivatives at the conformal point, $\tG_{\cc}(\theta_1,\theta_2) \propto\theta_{12}^{-2\Delta}$, we get
\begin{align}
\frac{\delta\tG_{\IR}(\theta_1,\theta_2)}{\tG_{\cc}(\theta_1,\theta_2)}
&=-q^{-1}
\biggl(\frac{\delta\vp(\theta_1)-\delta\vp(\theta_2)}
{\tan((\theta_1-\theta_2)/2)}
-\delta\vp'(\theta_1)-\delta\vp'(\theta_2)\biggr)
\\[5pt]
\label{deltaG_w2}
&=-2iq^{-1}(\sgn m)\,e^{im(\theta_1+\theta_2)/2}\,w_{2,m}(\theta_1-\theta_2)
\qquad \text{if }\, \delta\vp(\theta)=e^{im\theta},
\end{align}
where the function $w_{2,m}$ is given by \eqref{w_2m}; note that $w_{2,m}=0$ for $m=-1,0,1$. Translating $\delta\tG_{\IR}(\theta_1,\theta_2)$ to
\begin{equation}\label{dg_W2}
\de g_{\IR}(\theta_1,\theta_2)
=-\sqrt{(q-1)b}\,\, \theta_{12}^{-1}\,
\frac{\delta\tG_{\IR}(\theta_1,\theta_2)}{\tG_{\cc}(\theta_1,\theta_2)}
=2i\,\frac{\sqrt{(q-1)b}}{q}\,(\sgn m)\,W_{2,m}(\theta_1,\theta_2),
\end{equation}
we see that the vector field $\delta\vp(\theta)=e^{im\theta}$ produces the perturbation $\de g_{\IR}(\theta_1,\theta_2)$ that is proportional to $(\sgn m)\,W_{2,m}(\theta_1,\theta_2)$. The map $\delta\vp\mapsto g_{\IR}$ is an intertwiner from the space $\calF^0_{-1}$ to antisymmetric $(1/2,1/2)$-forms; it differs from the intertwiner $W_{2}$ by the $(\sgn m)$ factor. For completeness, we also give the normalized antisymmetric $(1/2,1/2)$-forms that transform as the basis vectors $\kett{m}$ of $\calD^{+}_{2}$ (if $m\ge 2$) or $\calD^{-}_{2}$ (if $m\le -2$):
\begin{equation}
W_{2,m}^{\text{normalized}}
=\gamma_m\,\pi^{-1}\sqrt{\frac{3}{|m|(m^2-1)}}\,W_{2,m},\qquad\quad
\gamma_m=\begin{cases}
1 &\text{if }m\ge 2,\\
(-1)^m &\text{if }m\le -2.
\end{cases}
\end{equation}

Finally, let us express $\delta\tG_{\IR}(\theta_1,\theta_2)$ in terms of $\delta\calO(\theta)=\Sch(e^{i\vp},\theta)-\frac{1}{2}$. Recall that in the linear order, $\delta\calO(\theta) =(\partial_{\theta}+\partial_{\theta}^3)\kern1pt\delta\vp(\theta)$. If $\delta\vp(\theta)=e^{im\theta}$, then $\delta\calO(\theta)=-im(m^2-1)e^{im\theta}$. Using \eqref{dg_W2} and the last case of \eqref{w_-1m}, we get:
\begin{equation}\label{gIR_O}
\wideboxed{
\delta g_{\IR}=-\frac{\pi\sqrt{b(q-1)}}{q}\, W_{-1}\kern1pt\delta\OO
}
\end{equation}
Since the result is independent of $m$, it holds for all functions $\delta\vp$.

\subsubsection{General form of UV perturbations}

As discussed in Section~\ref{sec_geneig}, any linear perturbation of the form $g_{\UV}(\theta_1,\theta_2)=(W_{h}f)(\theta_1,\theta_2)$ is allowed by conformal symmetry. It may be thought of as coming from the term $\int f(\theta)\calO(\theta)\,d\theta$ in a suitable effective theory, where $\calO$ is some field of dimension $h$. On the other hand, the physical perturbations have a discrete dimension spectrum, and their asymptotic form at $|\theta_1-\theta_2|\ll 1$ is given by \eqref{intresp}. We now combine these results and derive a more general expression for physical perturbations. The obvious thing to do is to find $W_{h}f$ for a constant function $f$, \ie $f\propto f_{1-h,0}$, and match the asymptotics. The case $h=2$ is special and involves $[\partial_{h}W_{h}]_{h=2}$ instead of $W_2$.

Let $\Re h>1/2$ and let us also assume that $h$ is not an integer. The relevant asymptotic expression is obtained from the second term in \eqref{w_hm_B}:
\begin{equation}
w_{h,m}(\vp)
\approx \frac{\Gamma(2h-1)\kern1pt\tan(\pi h/2)}{\Gamma(h)^2}\,\vp^{1-h}\qquad
\text{for}\quad \vp\to 0.
\end{equation}
(We are using the convention $0<\vp<2\pi$.) Passing to $(W_{h}f_{1-h,0})(\theta_1,\theta_2) =W_{h,0}(\theta_1,\theta_2) =\theta_{12}^{-1}w_{h,0}(\theta_1-\theta_2)$ and matching the corresponding asymptotics with \eqref{intresp}, we get:
\begin{equation}
g_{\UV,I}(\theta_1,\theta_2)
\approx \frac{a_I}{-k'_{\cc}(h_I)}\,\vep^{h_{I}-1}
\left[\frac{\Gamma(h)^2\,W_{h,0}(\theta_1,\theta_2)}
{\Gamma(2h-1)\kern1pt \tan(\pi h/2)}\,\right]_{h=h_I}\qquad
\text{for}\quad |\theta_{12}|\gg\vep.
\end{equation}
In the special case of $I=0$, both $W_{h,0}$ and $\tan(\pi h/2)$ vanish at $h\to h_I=2$, but the whole expression has a finite limit. The result below is equivalent to equation~(3.121) in \cite{MS16}:
\begin{equation} \label{gUV0_const_vep}
\wideboxed{
g_{\UV,0}(\theta_1,\theta_2)
\approx \frac{a_0}{(-k'_{\cc}(2))\kern1pt\pi}\,\vep\,
\theta_{12}^{-1}w_{[0]}(\theta_1-\theta_2)\qquad
\text{for}\quad |\theta_{12}|\gg\vep
}
\end{equation}
where the function $w_{[0]}$ is defined in~(\ref{w_01}).

Formally, one may also consider perturbations by time-dependent sources. They have the form $g_{\UV,I}=W_{h_I}f$ for an arbitrary function $f$, which plays the role of $a_{I}\vep^{h_I-1}$. In the $I=0$ case, one may use the explicit source
\begin{equation}\label{s0form-td}
s(\theta_1,\theta_2)
= -a_{0}\kern1pt \vep(\theta_+)\,\theta_{12}^{-2}
\sgn(\theta_1-\theta_2)\,
u\biggl(\ln\frac{\theta_{12}}{\vep(\theta_+)}\biggr),\qquad
\theta_+=\frac{\theta_1+\theta_2}{2}.
\end{equation}
Assuming a local relation between $\vep(\theta)$ and the singular part of the response, the perturbation to the Green function is given by
\begin{equation}
g=W_{2}f^{(0)}+[\partial_hW_h]_{h=2}f^{(1)},\qquad
f^{(1)}(\theta)=\frac{a_0}{(-k'_{\cc}(2))\kern1pt\pi}\,\vep(\theta).
\end{equation}
The first term is some linear combination of the nonsingular functions $W_{2,m}$ and may be understood as an IR perturbation. In the conformal setting, it is completely arbitrary; its calculation requires minimizing the action of the soft mode. This problem is solved by passing to a suitable frame such that $f^{(1)}$ is constant and then appying equation~(\ref{gUV0_const_vep}) (see Section~\ref{sec_calc_scheme} for more detail). It is interesting to note that $[\partial_hW_{h,m}]_{h=2}$ is a generalized eigenfunction of the conformal kernel because
\begin{equation}
K_{\cc}\,[\partial_hW_{h,m}]_{h=2}
=k_{\cc}(2)\,[\partial_hW_{h,m}]_{h=2}+k_{\cc}'(2)\,W_{2,m}.
\end{equation}
(For $m=-1,0,1$, the second terms vanishes, so $[\partial_hW_{h,m}]_{h=2}$ is an ordinary eigenfunction.)

\subsubsection{The functions $f^{\perp}$ and $f^{\parallel}$}

The conformal four-point function $\tF_{\cc}$ of the SYK model is formally defined by equation \eqref{Fc0}. It involves an antisymmetrized variant of the operator $L_{\cc}=K_{\cc}(1-K_{\cc})^{-1}$, which can be expressed by multiplying each term in the decomposition of identity \eqref{decomp_id} by $2\kern1pt\frac{k_{\cc}(h)}{1-k_{\cc}(h)}$. As already mentioned, the result is divergent because $k_{\cc}(2)=1$. Excluding the $h=2$ term, we obtain the function
\begin{equation} \label{tF_perp_c}
\tF^{\perp}_{\cc}(\vp_1,\vp_2;\vp_3,\vp_4)
=R_{\cc}(\vp_1,\vp_2)^{-1}
f^{\perp}(\vp_1,\vp_2;\vp_3,\vp_4)\,
R_{\cc}(\vp_3,\vp_4)^{-1},
\end{equation}
where
\begin{equation}\label{f_perp}
f^{\perp}=\left[\sum_{n=2}^{\infty}\Ress_{h=2n}
+\int_{1/2-i\infty}^{1/2+i\infty}\frac{dh}{2\pi i}\right]
\biggl(\frac{h-1/2}{\pi\tan(\pi h/2)}\,
\frac{k_{\cc}(h)}{1-k_{\cc}(h)}\,\Pi_h\biggr).
\end{equation}
(For convenience, we put the $\frac{k_{\cc}(h)}{1-k_{\cc}(h)}$ factor under the residue sign.) Maldacena and Stanford \cite{MS16} found alternative expressions for $f^{\perp}$ that are useful for extracting physically relevant asymptotics. In particular, if $0<\chi<1$, then
\begin{equation}\label{f_perp1}
f^{\perp}(\vp_1,\vp_2;\vp_3,\vp_4)
=-\vp_{12}^{-1}\vp_{34}^{-1}\sum_{I=0}^{\infty}\Ress_{h=h_I}
\biggl(\frac{h-1/2}{\pi\tan(\pi h/2)}\,\frac{k_{\cc}(h)}{1-k_{\cc}(h)}\,
\Gamma(h)^2\chi^h\hgfs(h,h,2h;\chi)\!\biggr),
\end{equation}
where $h_0,h_1,\ldots$ are solutions of the equation $k_{\cc}(h)=1$. This formula is, essentially, an operator product expansion. At small $\chi$, the leading contribution comes from the residue at the double pole, $h_0=2$:
\begin{gather}
\Ress_{h=2}
\biggl(\frac{h-1/2}{\pi\tan(\pi h/2)}\,\frac{k_{\cc}(h)}{1-k_{\cc}(h)}\,
\Gamma(h)^2\chi^h\hgfs(h,h,2h;\chi)\!\biggr)
\approx\frac{\chi^2(\ln\chi-c_1)}{2\pi^2(-k_{\cc}'(2))},
\\[5pt]
\text{where}\qquad
c_1=1-k_{\cc}'(2)+\frac{k_{\cc}''(2)}{2k_{\cc}'(2)}.
\end{gather}
Thus,
\begin{equation} \label{f_perp_asym}
f^{\perp}(\vp_1,\vp_2;\vp_3,\vp_4)
\approx\frac{\vp_{12}^{-1}\vp_{34}^{-1}}{2\pi^2(-k_{\cc}'(2))}\,
\chi^2\biggl(\ln\frac{1}{|\chi|}+c_1\biggr)\qquad
\text{for }\chi\to 0.
\end{equation}

Let us now consider the missing $h=2$ term in equation \eqref{f_perp}:
\begin{equation}\label{f_par}
f^{\parallel}=\Ress_{h=2}
\biggl(\frac{h-1/2}{\pi\tan(\pi h/2)}\,
\frac{k_{\cc}(h)}{1-k_{\cc}(h)}\,\Pi_h\biggr)
\end{equation}
It does not have any obvious physical meaning; in any case, the soft mode should be treated separately. However, it so happens that $f^{\parallel}$ appears in the soft mode contribution to the four-point function (along with terms that lack conformal symmetry, see Section~\ref{sec_sub4}). In the OPE region $0<\chi<1$, the addition of $f^{\parallel}$ almost cancels the double pole term from \eqref{f_perp1}. (This cancellation was noted by Maldacena and Stanford \cite{MS16} in the $q\to\infty$ limit, and they argued that it occurs in general.) Actually, the sum of $f^{\parallel}$ and the double pole term is equal to $\vp_{12}^{-1}\vp_{34}^{-1}$ multiplied by
\begin{equation}\label{f_par_plus}
\Ress_{h=2}
\biggl(\frac{h-1/2}{\pi\tan(\pi h/2)}\,
\frac{k_{\cc}(h)}{1-k_{\cc}(h)}\,
\Bigl(\Psi_h(\chi)-\Gamma(h)^2\chi^h\hgfs(h,h,2h;\chi)\Bigr)\biggr)
=\frac{3}{-k_{\cc}'(2)}\,\biggl(\chi^{-1}-\frac{1}{2}\biggr).
\end{equation}
The strong singularity at $\chi=0$ cancels with some other terms.

In the soft mode calculation, $f^{\parallel}$ comes in three pieces. At this point, we can write them formally by representing $\Pi_h$ in \eqref{f_par} using equations \eqref{PiWUrel}:
\begin{equation} \label{f_par_long}
\wideboxed{
f^{\parallel}=\frac{3}{\pi(-k'_{\cc}(2))}
\Bigl(\bigl[\partial_{h}W_{h}\bigr]_{h=2}W_{-1}^{\Tt}
+W_{-1}\bigl[\partial_{h}W_{h}^{\Tt}\bigr]_{h=2}
-6\pi\,W_{-1}\bigl[(\partial_{h}+c_{1})U_{h}\bigr]_{h=2}W_{-1}^{\Tt}\Bigr)
}
\end{equation}

\subsection{General treatment of the soft mode}\label{sec_gensoft}

\paragraph{Separation of the soft mode and other degrees of freedom:}
The dynamical degrees of freedom of the replica-diagonal action \eqref{rdact1} split into the $h=2$ part and its orthogonal complement, $\tG=\tG^{\parallel}+\tG^{\perp}$ (and similarly for $\tSig$). This decomposition depends on the choice of frame. Eventually, we want to write all results in the physical frame, $\theta=2\pi\tau/\beta$. However, there is another special frame $\vp$, called the ``conformal frame'', such that $\tG_{\vp}^{\parallel}=0$. Some calculations are simpler in that frame because we can use the conformal four-point function $\calF^{\perp}_{\cc}$. So, let us represent the set of variables $\tG_{\theta}$ by the the diffeomorphism $\vp$ of the unit circle and the function $\tG_{\vp}^{\perp}$, and also replace $\tSig_{\theta}$ with $\tSig_{\vp}$. (This approach was proposed but not pursued in \cite{JeSuYo16}.) Thus, the action depends on the dynamical variables $\vp$, $\tG_{\vp}^{\perp}$, $\tSig_{\vp}$, as well as the physical perturbation source $\tsig_{\theta}$. The action can be expressed in any frame; for example, we can use $\tG_{\vp}=\tG_{\vp}^{\perp}$ and $\tSig_{\vp}$ directly, and transform $\tsig_{\theta}$ to the conformal frame. However, in any case, the partition function $Z$ depends on $\tsig_{\theta}$:
\begin{equation} \label{fullZ}
Z[\tsig_{\theta}]=\int \calD\vp\,\calD\tG_{\vp}^{\perp}\,\calD\tSig_{\vp}\, \exp\Bigl(-\tI[\tSig_{\vp},\tG_{\vp}^{\perp},\vp;\,\tsig_{\theta}]\Bigr).
\end{equation}

It is interesting to note that the Jacobian
\begin{equation}\label{Jacobian}
\mathbf{J}=\frac{\calD\tG_{\theta}\,\calD\tSig_{\theta}}
{\calD\vp\,\calD\tG_{\vp}^{\perp}\,\calD\tSig_{\vp}}
\end{equation}
is constant if $\calD\vp$ is understood as a right-invariant measure on $\Diff^{+}(S^1)$. To see this, let $\vp$, $\tG_{\vp}^{\perp}$, $\tSig_{\vp}$ be independent variables and $V$ some fixed diffeomorphism. Both the numerator and denominator in the above equation remain the same if we change $\vp$ to $\vp\circ V^{-1}$ and $\theta$ to $V(\theta)$ (which means transforming $\tG_{\theta}$, $\tSig_{\theta}$ with $V$). Thus, we may assume without loss of generality that $\vp$ is infinitesimally close to the identity, $\vp(\theta)=\theta+\delta\vp(\theta)$. In this case, $\delta\tG_{\theta}^{\parallel}$ depends on $\delta\vp$ but not on $\delta\tG_{\vp}^{\perp}$ or $\delta\tSig_{\vp}$. Therefore, the Jacobian is the product of two factors, the first being a constant and the second equal to $1$:
\begin{equation}
\mathbf{J}=\biggl[\frac{\calD\tG_{\theta}^{\parallel}}
{\calD\vp}\biggr]_{\vp=\theta}
\biggl[\frac{\calD\tG_{\theta}^{\perp}\,\calD\tSig_{\theta}}
{\calD\tG_{\vp}^{\perp}\,\calD\tSig_{\vp}}\biggr]_{\vp=\theta}
=\const.
\end{equation}
However, the Jacobian is not important for subsequent calculations, which only include the leading terms in $1/N$. Indeed, the Jacobian makes an $O(1)$ contribution to the logarithm of the integrand in the functional integral, whereas $\tI$ is proportional to $N$.

\paragraph{Effective action to quadratic order in $\tsig_\theta$:}
To simplify the action, we eliminate $\tSig_{\vp}$ and $\tG_{\vp}^{\perp}$ as described in Section~\ref{sec_confker}. In particular, we may use \eqref{I2star}, adapting it to our present notation and replacing $\calF_{\cc}$ with $\calF^{\perp}_{\cc}$:
\begin{equation} \label{effIt}
\wideboxed{
\frac{\tI_{\eff}[\vp;\,\tsig_{\theta}]}{N}
\approx-\frac{1}{2}\bbrakett{\tG_{\cc}}{\tsig_{\vp}}
-\frac{1}{8}\,\bbraa{\tsig_{\vp}}\tF^{\perp}_{\cc}\bkett{\tsig_{\vp}}
}
\end{equation}
where
\begin{equation}
\tsig_{\vp}(\vp_1,\vp_2)
=\vp'(\theta_1)^{\Delta-1}\vp'(\theta_2)^{\Delta-1}
\tsig_{\theta}(\theta_1,\theta_2).
\end{equation}
The first term in~\eqref{effIt} has already been considered in the physical frame, where it has the form $-\frac{1}{2}\brakett{\tG_{\IR}}{\tsig_{\theta}}$, and found to generate the Schwarzian. In the next section, we will show that the second term gives rise to the non-local correction~\eqref{I_nl_conf} to the effective action.

\paragraph{Covariance properties of $\calO$ and $\vep$:}
The observable $\OO_{\theta}(\theta)=\Sch(e^{i\vp},\vth)$ can be represented in any frame. Its transformation law is similar to that of the holomorphic energy-momentum tensor in two-dimensional CFT:
\begin{equation}
\OO_{y}(y)=\biggl(\frac{dy}{dx}\biggr)^{-2}\Bigl(\OO_{x}(x)-\Sch(y,x)\Bigr).
\end{equation}
The pair of fields $(1,\OO)$ forms a representation of $\Diff(S^1)$. In the conformal frame, $\OO$ is constant and equal to $\frac{1}{2}$. The field $\vep_x=\frac{2\pi}{\beta J}(\partial_x\theta)^{-1}$ is the source coupling to $\OO_x$. Indeed, the Schwarzian effective action can be written in a covariant form in any frame:
\begin{equation}\label{Iepsrho}
\frac{\tI_{\loc}}{N}
=-\alpha_S \int(\vep_x\OO_{x}-\rho_{x})\,dx,\qquad\quad
\rho_x=\frac{(\partial_x\vep_x)^2}{2\vep_x}-\partial_x^2\vep_x
\end{equation}
We will see that $\vep_x$ and $\rho_x$ represent, respectively, an approximate perturbation source in frame $x$ and the correction to the soft mode action resulting from that approximation. This pair of fields transforms as follows:
\begin{equation}
\vep_y(y)=\frac{dy}{dx}\,\vep_x(x),\qquad
\rho_y(y)=\biggl(\frac{dy}{dx}\biggr)^{-1}
\Bigl(\rho_{x}(x)-\Sch(y,x)\,\vep_x(x)\Bigr).
\end{equation}

\section{Next order corrections}\label{sec_next_order}

In this section, we derive the order $\vep^2$ non-local correction to the Schwarzian action as well as calculating the order $\vep^0$ term in the four-point function. (As is usual, $\vep=\frac{2\pi}{\beta J}$.)

\subsection{The calculation scheme and physical considerations}\label{sec_calc_scheme}

\paragraph{Perturbation source:}
The required accuracy can still be achieved using the leading perturbation source $\tsig_0$ that corresponds to the $h_0=2$ pole of the conformal kernel. In the physical frame, the source is given by $\tsig_{0,\theta}(\theta_1,\theta_2) =R_{\cc}(\theta_1,\theta_2)\,s_{0,\theta}(\theta_1,\theta_2)$ with
\begin{equation}
s_{0,\theta}(\theta_1,\theta_2)
= -a_{0}\kern1pt\vep\kern1pt\theta_{12}^{-2}
\sgn(\theta_1-\theta_2)\,u(\xi),\qquad
\text{where}\quad
\xi=\ln\frac{|\theta_{12}|}{\vep},\quad\:
\int_{-\infty}^{\infty}u(\xi)\,d\xi=1.
\end{equation}
The function $s_0$ transforms to the conformal frame as a $(1/2,1/2)$-form:
\begin{align}
s_{0,\vp}(\vp_1,\vp_2)
&=\vp'(\theta_1)^{-1/2}\vp'(\theta_2)^{-1/2}s_{0,\theta}(\theta_1,\theta_2)
\nonumber\\[2pt]
\label{s0_approx1}
&\approx -a_{0}\sqrt{\vep_{\vp}(\vp_{1})\vep_{\vp}(\vp_{2})}\,
\vp_{12}^{-2}\sgn(\vp_1-\vp_2)\, u\biggl(\ln\frac{|\vp_{12}|}
{\sqrt{\vep_{\vp}(\vp_{1})\vep_{\vp}(\vp_{2})}}\biggr)
\\[2pt]
\label{s0_approx2}
&\approx -a_{0}\kern1pt\vep_{\vp}(\vp_{+})\kern1pt\vp_{12}^{-2}
\sgn(\vp_1-\vp_2)\,u\biggl(\ln\frac{|\vp_{12}|}{\vep_{\vp}(\vp_{+})}\biggr),
\qquad \vp_{+}=\frac{\vp_1+\vp_2}{2}.
\end{align}
Both approximations are equivalent for our purposes, though the first one has the advantage of being $\SL(2,\RR)$-invariant. Note that these expressions lead to an incorrect result if one plugs them in the first term of \eqref{effIt} and follows the derivation of the Schwarzian action in Section~\ref{sec: Schw_der}. The error comes from the inaccuracy of the source function as well as from neglecting the integral of $\tsig_{0}$ with $\tG_{\beta=\infty}$. The problem and its solution are more evident if we replace the conformal frame $\vp$ with an arbitrary frame $x$. Then the approximation results in the local action $-N\alpha_S\int\vep_x\calO_x\,dx$; the error is accounted for by the field $\rho_x$ in \eqref{Iepsrho}.

\paragraph{Correction to the soft mode action:}
The non-local correction arises from the second term in \eqref{effIt}:
\begin{equation}
I_{\nloc}=-\frac{1}{8}\int d\vp_1\,d\vp_2\,d\vp_3\,d\vp_4\,
f_{\cc}^{\perp}(\vp_1,\vp_2;\vp_3,\vp_4)\,
s_{0,\vp}(\vp_1,\vp_2)\,s_{0,\vp}(\vp_3,\vp_4).
\end{equation}
In this case, both approximate expressions for $s_{0,\vp}$ are good enough. (The actual calculation will be done in a third way, directly in the physical frame.) The relevant contribution to the integral comes from the region where the pairs $(\vp_1,\vp_2)$ and $(\vp_3,\vp_4)$ are much farther apart than the points within each pair.\footnote{The other part of the integral is strongly dependent on the window function $u$, and therefore, cannot be treated consistently within our renormalization scheme. However, it can be excluded to produce an unambiguously defined regularized integral.} Therefore, $f_{\cc}^{\perp}(\vp_1,\vp_2;\vp_3,\vp_4) \propto\vp_{12}^{-1}\vp_{34}^{-1}\chi^2\ln\frac{1}{|\chi|}$; see \eqref{f_perp_asym} for a more accurate expression. We will first integrate over $\vp_3$ and $\vp_4$; this intermediate result may be interpreted as a UV correction to the Green function:
\begin{equation}
g_{\UV,\vp}=\frac{1}{2}\int d\vp_3\,d\vp_4\,
f_{\cc}^{\perp}(\vp_1,\vp_2;\vp_3,\vp_4)\,
s_{0,\vp}(\vp_3,\vp_4).
\end{equation}

\paragraph{Correction to the four-point function:}
The four-point function is obtained by taking a variational derivative with respect to the perturbation source:
\begin{equation}\label{fp_PI}
\tF_{\theta}(\theta_1,\theta_2,\theta_3,\theta_4)=\frac{4}{N}\kern1pt
\left.\frac{\delta^2\ln Z[\tsig_{\theta}]}
{\delta\tsig_{\theta}(\theta_1,\theta_2)\,
\delta\tsig_{\theta}(\theta_3,\theta_4)}\right|_{\tsig_\theta=\tsig_{0,\theta}},
\qquad
Z[\tsig_{\theta}]=\int \calD\vp\,
\exp\bigl(-\tI_{\eff}[\vp;\tsig_{\theta}]\bigr).
\end{equation}
We proceed by introducing the fluctuating quantity $\tG_{\theta}=\tG_{\IR,\theta}+\tG_{\UV,\theta}$. More explicitly,
\begin{equation}
\tG_{\theta}[\vp;\tsig_{\theta}](\theta_1,\theta_2)
=-2\kern1pt\frac{\delta\tI[\vp;\tsig_{\theta}]}
{\delta\tsig_{\theta}(\theta_1,\theta_2)}
=\tG_{\IR,\theta}[\vp](\theta_1,\theta_2)
+\tG_{\UV,\theta}[\vp;\tsig_{\theta}](\theta_1,\theta_2),
\end{equation}
where (omitting the parameters $\vp$, $\tsig_{\theta}$ in square brackets)
\begin{align}
\tG_{\IR,\theta}(\theta_1,\theta_2)
&=\tG_{\cc}\bigl(\vp(\theta_1),\vp(\theta_2)\bigr)\,
\vp'(\theta_1)^{\Delta}\vp'(\theta_2)^{\Delta},
\displaybreak[0]\\[3pt]
\label{tGUV}
\tG_{\UV,\theta}(\theta_1,\theta_2)
&=\frac{1}{2}\int
\tF^{\perp}_{\theta}(\theta_1,\theta_2,\theta_3,\theta_4)\,
\tsig_{\theta}(\theta_3,\theta_4)\,d\theta_3\,d\theta_4,
\displaybreak[0]\\[3pt]
\tF^{\perp}_{\theta}(\theta_1,\theta_2,\theta_3,\theta_4)
&=\tF^{\perp}_{\cc}(\vp_1,\vp_2,\vp_3,\vp_4)\,
\vp'(\theta_1)^{\Delta}\vp'(\theta_2)^{\Delta}
\vp'(\theta_3)^{\Delta}\vp'(\theta_4)^{\Delta}.
\end{align}
Now, the four-point function is expressed in terms of average values over the fluctuating field $\vp$:
\begin{equation} \label{fp}
\wideboxed{
\tF_{\theta}(\theta_1,\theta_2,\theta_3,\theta_4)
=\bcorr{\tF^{\perp}_{\theta}(\theta_1,\theta_2,\theta_3,\theta_4)}
+N\bccorr{\tG_{\theta}(\theta_1,\theta_2)\,\tG_{\theta}(\theta_3,\theta_4)}
}
\end{equation}
The double brackets in the second term denote the connected correlator. The perturbation source $\tsig_{\theta}$ is implicit and may be set to $\tsig_{0,\theta}$. In the first term of \eqref{fp}, the average value of $\tF^{\perp}$ may be replaced with its value at $\vp$ equal to the identity function because the fluctuation corrections are small:
\begin{equation}
\tF^{\perp}_{\theta}(\theta_1,\theta_2,\theta_3,\theta_4)
=\tF^{\perp}_{\cc}(\theta_1,\theta_2,\theta_3,\theta_4)+O(N^{-1}).
\end{equation}

\paragraph{Asymptotics of the four-point function in the OPE region:}
Following the outlined scheme, we will obtain a quite complicated expression, which hides some interesting physics. One salient feature, first noted by Maldacena and Stanford \cite{MS16}, is the cancellation of the $\chi^2\ln(1/|\chi|)$ term in the OPE region. We argue that this phenomenon is similar to the Debye screening of electric charges in plasma. In our case, the $\chi^2\ln(1/|\chi|)$ term in the conformal four-point function plays the role of bare Coulomb interaction. It is screened by the soft mode, which is analogous to the charge density.

To be more concrete, let us examine the asymptotics of
\begin{equation}\label{func_FGcGc}
\frac{\tF_{\theta}(\theta_1,\theta_2;\theta_3,\theta_4)}
{\tG_{\cc}(\theta_1,\theta_2)\,\tG_{\cc}(\theta_3,\theta_4)}
=\frac{1}{(q-1)b}\,\theta_{12}\theta_{34}\kern1pt
f(\theta_1,\theta_2;\theta_3,\theta_4)
\end{equation}
at $\theta,\theta'\to 0$, where $\theta=\theta_1-\theta_2$ and $\theta'=\theta_3-\theta_4$. We are looking for a term that is equal to $\theta^2\theta'^2$ times an arbitrary function of $\Delta\theta_{+} =(\theta_1+\theta_2-\theta_3-\theta_4)/2$. As it turns out, the leading (at $\theta,\theta'\to 0$) part of the function \eqref{func_FGcGc} is proportional to $|\theta|\theta'^2+|\theta'|\theta^2$, but the $\theta^2\theta'^2$ term is more pertinent to the discussion because $\chi^2\ln(1/|\chi|)\approx \theta^2\theta'^2 \ln(\Delta\theta_{+}^2/\vep^2)$ for $\theta,\theta'\sim\vep$. To find this term, we may probe the system with an additional perturbation source $\delta\tsig_{\theta}$ that has the same dependence on $\theta_1-\theta_2$ as $\tsig_{0,\theta}$ but also depends on $\theta_{+}=(\theta_1+\theta_2)/2$. Thus, the full source $\tsig_{\theta}=\tsig_{0,\theta}+\delta\tsig_{\theta}$ has the form \eqref{s0_approx2}, but in the physical rather than conformal frame. The local perturbation strength is given by some function $\vep_{\theta}(\theta)$, and we have
\begin{equation}
\frac{\tF_{\theta}(\theta_1,\theta_2;\theta_3,\theta_4)}
{\tG_{\cc}(\theta_1,\theta_2)\,\tG_{\cc}(\theta_3,\theta_4)}
=\frac{1}{a_0^2(q-1)b}\kern1pt
\left.\frac{\delta^2(N^{-1}\ln Z)}
{\delta\vep_\theta(\theta_{+})\,\delta\vep_\theta(\theta'_{+})}\right|
_{\vep_\theta=\vep} \theta^2\theta'^2
+\text{other terms},
\end{equation}
where $\theta_{+}=(\theta_1+\theta_2)/2$ and $\theta'_{+}=(\theta_3+\theta_4)/2$. Up to $1/N$ corrections and trivial terms (which correspond to the ground state energy and zero-temperature entropy), $-\ln Z$ is equal to $\tI_{*}[\tsig_{\theta}]=\min_{\vp}\tI_{\eff}[\vp,\tsig_{\theta}]$. It is clear that the minimum (or at least an extremum) is achieved when $\vep_{\vp}$ is a constant function. Thus, the equilibrium with the modified source is equivalent to the thermal equilibrium at a slightly different temperature,
\begin{equation}
\frac{\widetilde{\beta}J}{2\pi}=\vep_{\vp}^{-1}
=\int\frac{d\theta}{2\pi}\,\vep_{\theta}^{-1}.
\end{equation}
Furthermore, $-\ln Z[\tsig_{\theta}]=\beta F$, where $F$ is the free energy at the indicated temperature. It follows that for small values of $\theta_1-\theta_2$ and $\theta_3-\theta_4$,
\begin{equation}\label{4p-thermo}
\wideboxed{
\frac{\tF_{\theta}(\theta_1,\theta_2;\theta_3,\theta_4)}
{\tG_{\cc}(\theta_1,\theta_2)\,\tG_{\cc}(\theta_3,\theta_4)}
=\frac{1}{36q^2\alpha_S^2}\biggl(\frac{\beta J}{2\pi}\biggr)^4\,
\frac{\partial^2(-N^{-1}\beta F)}{\partial(\beta J)^2}\,
(\theta_1-\theta_2)^2(\theta_3-\theta_4)^2
+\text{other terms}
}
\end{equation}

\subsection{Non-local action and the soft correlator}\label{sec_nonlocal}

To extract the leading term at small $\vep$ in the second term of the effective action \eqref{effIt}, we can use the leading UV perturbation $\tsig=\tsig_0$. As an intermediate step, let us first obtain the Green function response to the perturbation in the conformal frame, for general $\vp(\tht)$ (see \eqref{tGUV}):
\bea
g_{\UV, \vp}(\vp_1, \vp_2)&=&{1 \ov 2}\int d\vp_3 d\vp_4\, f^{\perp}(\vp_1, \vp_2; \vp_3, \vp_4)s_{\vp}(\vp_3, \vp_4)\nn
&=&{1 \ov 2}\int d\tht_3 d\tht_4\, \vp'(\tht_3)^{1/2}\vp'(\tht_4)^{1/2}f^{\perp}(\vp_1, \vp_2; \vp_3(\tht_3), \vp_4(\tht_4))s_{\tht}(\tht_3, \tht_4).
\eea
In the second line we switched to the physical frame to use $s_{\tht}$ as given in \eqref{sIform} with constant $\vep$. We are only interested in the portion of the integral where the source $s_{\tht}$ is not too close to $\tht_1, \tht_2$, farther than the cutoff $\vep$, and also in the leading, order $\vep$ contribution to $\vp_{12}g_{\UV, \vp} \sim \tG_{\UV, \vp}/\tG_c$. Thus we can use the $\chi \to 0$ asymptotics of $f^{\perp}$ given in \eqref{f_perp_asym}. It is convenient to denote the transform of $\vp_{12}$ to the physical frame $d_{12}(\tht_1, \tht_2)=\vp'(\tht_1)^{-1/2}\vp'(\tht_2)^{-1/2}2\sin\left( (\vp(\tht_1)-\vp(\tht_2)/2\right)$ with leading UV behavior $d_{12} \approx \tht_1-\tht_2$, using which 
\bea \label{gUVdet}
&&\vp'(\tht_3)^{1/2}\vp'(\tht_4)^{1/2}\left(\vp_{12}^{-1}\vp_{34}^{-1}\chi^2 \left( \ln \abs{\chi}^{-1}+c_1\right)\right)\abs{\tht_3-\tht_4}^{-2}\sgn(\tht_3-\tht_4)u\left(\ln  \abs{\tht_{34}}/ \vep\right)\nn
&\approx &\vp_{12}^{-1}\left( {d_{12} \ov d_{13}d_{24}}\right)^2 \left( \ln \abs{{d_{13}d_{24} \ov d_{12}}}+c_1-\ln \abs{\tht_3-\tht_4}\right)\abs{\tht_3-\tht_4}^{-1}u\left( \ln {\abs{\tht_{34}} \ov \vep} \right),
\eea
and so
\be \label{gUV_gen}
g_{\UV, \vp}(\vp_1, \vp_2) \approx -{a_0 \vep \ov \pi(-k_c'(2))}\vp_{12}^{-1}\int {d\tht_0 \ov 2 \pi}\, \left( {d_{12} \ov d_{10}d_{20}}\right)^2 \left( \ln \abs{{d_{10}d_{20} \ov d_{12} \vep}} + c_1-c_2\right)
\ee
where we have introduced the constant
\be
c_2=\int_{-\infty}^{\infty} d\xi\, \xi u(\xi).
\ee
We can also write the integral \eqref{gUV_gen} in the conformal frame in which $\vep(\vp)$ is varying,
\begin{equation}\label{gUV_gen_conf}
\wideboxed{
g_{\UV, \vp}(\vp_1, \vp_2)
\approx-{a_0 \ov \pi (-k_c'(2))}\,\vp_{12}^{-1}
\int\frac{d\vp_0}{2\pi}\,\biggl|\frac{\vp_{10}\vp_{20}}{\vp_{12}}\biggr|^{-2}
\left(\ln\biggl|\frac{\vp_{10}\vp_{20}}{\vep(\vp_0)}\biggr|+c_1-c_2\right)
\vep(\vp_0)
}
\end{equation}
This form will be used in our calculation of the four-point function.

Now we can use the response \eqref{gUV_gen} in the second term of \eqref{effIt}, and from a similar calculation as in \eqref{gUVdet} easily obtain
\bea \label{pre_nonlocal_I}
-{1 \ov 8}\mel**{\tsig_{\tht} }{ \tF_\tht^{\perp}[\vp]}{\tsig_{\tht}}&=&-{1 \ov 4}\int d\tht_1 d\tht_2\, s_{\tht}(\tht_1, \tht_2)\vp'(\tht_1)^{1/2}\vp'(\tht_2)^{1/2}g_{\UV, \vp}(\vp_3, \vp_4)\nn
&\approx &-{\ga \ov 2}\int {d\vp_1 \ov 2 \pi}{d\vp_2 \ov 2\pi}{\vep(\vp_1)\vep(\vp_2) \ov \vp_{12}^4}\left( \ln \left({\vp_{12}^2 \ov \vep(\vp_1)\vep(\vp_2)}\right) + c \right)
\eea
where
\be \label{coeffrel}
c=c_1-2c_2, \qquad  \gamma={2 a_0^2 \ov -k_c'(2)}.
 \ee
The integral diverges near $\vp_{12} \approx 0$ and needs to be regulated. Using some $\PSL(2, \RR)$-invariant cutoff - for example $\vp_{12}^2 > \vep(\vp_1)\vep(\vp_2)$ -- one will obtain the Schwarzian along with other local terms which have cutoff-dependent coefficients. Thus it seems natural to view the Schwarzian action as a UV completion of the non-local action, which we may identify as the order $\vep^2$ cutoff-independent portion of the integral,\footnote{We can see this finite part of the integral is well-defined, as follows. Two cutoff schemes for the integral will result in finite parts that differ at most by a local integral that is order $\vep^2$. The integrand of such an integral must be a $1$-form, and can have at most a singularity going as $\vep^{-3}$ as $\vep \to 0$ -- \ie take the form $p(\vep, \vep', \vep'',\dots)/\vep^3$ where $p$ is some polynomial -- given that in the original integral the $\vp_{12}^{-4}$ singularity in the integrand is integrated with a cutoff going as $\vep \sim (\beta J)^{-1}$. Thus the integrand of the local integral must be $\vep^{(3)}\vep$ or $\vep'' \vep'$, but both integrands give integrals odd under reflection $ \vp \to -\vp$ whereas the original integral is even. \label{nolocarg}}
\be \label{Inlfin}
\wideboxed{{I_{\nloc} \ov N}=-{\ga \ov 2}\left[\int {d\vp_1 \ov 2 \pi}{d\vp_2 \ov 2\pi}{\vep(\vp_1)\vep(\vp_2) \ov \vp_{12}^4}\left( \ln \left({\vp_{12}^2 \ov \vep(\vp_1)\vep(\vp_2)}\right) + c \right)\right]_{\text{fin.}}}
\ee 
However, not having a method to choose one particular cutoff, we have implicitly discarded all cutoff-dependent local terms arising from \eqref{pre_nonlocal_I} in fixing the relation between coefficients of the Schwarzian and non-local action as in \eqref{coeffrel}.

Finally, let us expand $I_{\nloc}$ to quadratic order in the soft fluctuation $\de \vep(\vp)=\vep(\vp)-\vep$:
\be \label{Inlfinexp}
{I_{\nloc} \ov N}\approx \ga\Bigg({\vep^2 \ov 24} + {\vep \ov 12}\int{d\vp \ov 2 \pi}\de \vep(\vp)-{1 \ov 2}\left[\int{d\vp_1 \ov 2 \pi}{d\vp_2 \ov 2 \pi}{\de \vep(\vp_1)\de \vep(\vp_2) \ov \vp_{12}^4} \left( \ln{\vp_{12}^2 \ov \vep^2}-2+c\right)\right]_{\text{fin.}}\Bigg).
\ee
Here we have used the integrals $\left[\int{d\vp \ov 2 \pi}\left(2 \sin{\vp \ov 2}\right)^{-4}\right]_{\text{fin.}}=u_{2,0}=0$ and $\left[\int{d\vp \ov 2 \pi}\left(2 \sin{\vp \ov 2}\right)^{-4}\abs{2 \sin{\vp \ov 2}}\right]_{\text{fin.}}=-{1 \ov 2}\left[\p_h u_{h,0}\right]_{h=2}=-{1 \ov 24}$ (see \eqref{UWdef}, \eqref{UWfourier}, and \eqref{Ufourier}). The first term gives a contribution proportional to $\ga$ to the free energy at finite temperature,
\be
\beta(F -E_0)=N\left(-s_0 - 2 \pi^2 \al_S (\beta J)^{-1}+{\pi^2 \ov 6}\ga(\beta J)^{-2}+\dots\right)+\dots
\ee
The second term is in fact equal to a quadratic contact term up to higher order terms,
\be
\int {d\vp \ov 2\pi} \de \vep(\vp) \approx \int {d\vp \ov 2\pi}\vep^{-1}(\de \vep(\vp))^2
\ee
as $\de\left(\int{d \vp \ov 2 \pi}\vep(\vp)^{-1} \right)=\de\left({\beta J \ov 2 \pi}\right)=0$ and expanding $\vep(\vp)=\vep+ \de \vep(\vp)$,
\bea
\de\left(\int{d \vp \ov 2 \pi}\vep(\vp)^{-1} \right)=\int{d\vp \ov 2 \pi}\vep^{-2}\left( -\de \vep + \vep^{-1}\de \vep^2+O\left(\de \vep^3 \right)\right).
\eea
Then we can write the kernel of the quadratic action for $\de \vep(\vp)$ in terms of the conformal two-point function $U_h$ introduced in \eqref{UWdef},
and the contact two-point function $\sum_m e^{im(\vp_1-\vp_2)}=2\pi\de(\vp_1-\vp_2)$ (for $\vp_1-\vp_2$ considered modulo $2\pi$) which we will denote as the identity $\unit$ in operator form. The quadratic action remains the same after transforming to the physical frame as $\de \vep(\tht) \approx \de \vep(\vp)$ to lowest approximation, and the resulting correction to the correlator $P(\tht_1, \tht_2)$, of order $1$, is given by
\begin{align}
P^{(0)}(\tht_1, \tht_2) \hspace{-30pt}&\hspace{30pt}
=N\expval{\de \OO(\tht_1)\de \OO(\tht_2)}_{\nloc}=N \bigl( \expval{\de \OO(\tht_1)\de \OO(\tht_2)}-\expval{\de \OO(\tht_1)\de \OO(\tht_2)}_{\loc}\bigr)\nonumber\\
\label{Pcorr}
&=-\int d\tht_3 d\tht_4\, \expval{\de \OO(\tht_1)\de \vep(\tht_2)}_{\loc} {\de^2 I_{\nloc} \ov \de \vep(\tht_3)\de \vep(\tht_4)}\expval{\de \vep(\tht_4)\de \OO(\tht_2)}_{\loc}
\\
&=-{18 q^2 \ov \pi^2(-k_c'(2)) (q-1)b}\left(\left[ \p_h U_h(\tht_1, \tht_2)\right]_{h=2}+\left( 2\ln \vep + 2-c\right)U_2(\tht_1, \tht_2) +{1 \ov 6}\unit(\tht_1, \tht_2) -{1 \ov 4} \right).\nonumber
\end{align}
where we have used
\be \label{altlead}
N\expval{\de \OO(\tht_1)\de \vep(\tht_2)}_{\loc}=-{1 \ov 2 \pi \al_S}\sum_{m \neq -1,0,1}e^{im(\tht_1-\tht_2)}
\ee
derived from \eqref{devpcorr}. Note $m=-1,0,1$ Fourier harmonics cancel between terms in \eqref{Pcorr} as they are not present in $\de\OO(\tht)$.

\subsection{Subleading four-point function}\label{sec_sub4}

We now calculate the four-point function subleading in $\beta J$, using \eqref{fp}. Recall that expectation values in \eqref{fp} are taken with respect to the path integral over the soft mode $\vp$, with effective action $\tI_{\eff}[\vp, \tsig_{\tht}]$ obtained from integrating out $\tG^{\perp}_{\vp}$ and $\tSig_{\vp}$ in \eqref{fullZ}. In the following all quantities are to be understood as written in the physical frame unless denoted otherwise. 

Since we work in the large $N$ limit, the first term in \eqref{fp} is just the conformal four-point function $\tF^{\perp}_c(\tht_1, \tht_2, \tht_3, \tht_4)$ given in \eqref{tF_perp_c}, of order $\vep^0$. Meanwhile, the second term is reduced to
\be \label{deGcorr}
N \expval{\de \tG(\tht_1, \tht_2)\de \tG(\tht_3, \tht_4)}
\ee
where $\de \tG=\tG- \tG_*$ is the linearized change in the Green function in small $\de \vp=\vp(\tht)-\tht$. Here we are using $\tG_*$ to denote the Green function at the saddle-point $\vp(\tht)=\tht$ with respect to the physical UV perturbation $\tsig=\tsig_{0}$.

Now, to subleading accuracy in $\vep$,\, $\de \tG=\de \tG_{\IR}+\de \tG_{\UV}$ with
\be
\de \tG_{\IR}=\tG_{\IR}-\tG_c, \qquad \de\tG_{\UV}=\tG_{\UV}-\big(\tG_{*}\big)_{\UV}
\ee
of order $\vep^0$ and $\vep^1$, respectively. As the leading soft two-point function \eqref{Plead} is order $\vep^{-1}$, to evaluate \eqref{deGcorr} to order $\vep^0$, we should i) include the fluctuation of the UV response $\de \tG_{\UV}$ in $\de \tG$ and ii) include in the effective action $\tI_{\eff}$ the non-local action derived in the previous section, or in other words use the two-point function of the soft mode with the correction calculated in \eqref{Pcorr}. The subleading, order $\vep^0$ four-point function can be organized as
\begin{equation}\label{subleadF}
\begin{aligned}
\tF^{(0)}(\tht_1, \tht_2; \tht_3, \tht_4)
={}&\tF_c^{\perp}(\tht_1, \tht_2; \tht_3, \tht_4)+N\expval{\de \tG_{\IR}(\tht_1, \tht_2) \de\tG_{\IR}(\tht_3, \tht_4)}_{\nloc}
\\
&\hspace{-20pt}+N\expval{\de \tG_{\IR}(\tht_1, \tht_2)\de\tG_{\UV}(\tht_3, \tht_4)}_{\loc}+N\expval{\de \tG_{\UV}(\tht_1, \tht_2)\de\tG_{\IR}(\tht_3, \tht_4)}_{\loc}.
\end{aligned}
\end{equation}

Let us work with $\de g$ normalized relative to $\de \tG$ as in \eqref{sg_def}. To calculate the expectation values in \eqref{subleadF}, we may express various terms in $\de g$ as three-points functions $W_{h=-1,2}$ or $\left[\p_h W_h\right]_{h=2}$ (defined in \eqref{UWdef}) acting on $\de \OO$ or $\de \vep$, then use local and non-local parts of two-point functions $\expval{\de \OO\de \OO}$ and $\expval{\de \OO \de \vep}$ as appropriate. The variation $\de g_{\IR}$ was already expressed in the desired form in \eqref{gIR_O},
\be
\delta g_{\IR}=-\frac{\pi\sqrt{b(q-1)}}{q}\, W_{-1}\cdot \delta\OO.
\ee
 Meanwhile, $\de g_{\UV}$ can be divided into two pieces. The first is the variation of the response in the conformal frame
\be
\de g_{\UV, \vp}=g_{\UV, \vp}-(g_{*})_{\UV, \vp}
\ee
due to the dependence of the perturbation $\tsig_{\vp}$ on $\vep(\vp)$, see \eqref{gUV_gen_conf}. Expanding $\vep(\vp)=\vep + \de \vep(\vp)$, we find
\be
\de g_{\UV, \vp}={a_0 \ov \pi (-k_c'(2))}\left( \left[ \p_h W_h\right]_{h=2}+ \left( \ln \vep + 1- {c_1 + c \ov 2}\right)W_2\right)\cdot \de \vep.
\ee
The second is the variation due to the transformation of the response from the conformal to the physical frame (below we are using Lie derivatives acting on $\De$-forms defined in \eqref{LieDdef}),
\be
\de \tG^{(\text{diff)}}_{\UV}=\left(\Ld^{(1)}_{\de \vp} + \Ld^{(2)}_{\de \vp} \right) \big(\tG_*\big)_{\UV}.
\ee
Factoring $\big(\tG_*\big)_{\UV}=\left(\big(\tG_*\big)_{\UV}/\tG_c\right)\tG_c$,
\bea
\de \tG^{(\text{diff)}}_{\UV}(\tht_1, \tht_2)&=&\ubrace{{\big(\tG_*\big)_{\UV}(\tht_1, \tht_2) \ov \tG_c(\tht_1, \tht_2)}\de \tG_{\IR}(\tht_1, \tht_2)}_{\de \tG_{\UV}^{(\text{diff}, c)} (\tht_1, \tht_2)} +\ubrace{\tG_c(\tht_1, \tht_2)\left( \de \vp(\tht_1)\p_{\tht_1} + \de \vp(\tht_2)\p_{\tht_2}\right){\big(\tG_*\big)_{\UV}(\tht_1, \tht_2) \ov \tG_c (\tht_1, \tht_2)}}_{\de \tG_{\UV}^{(\text{diff}, \bar{c})}(\tht_1, \tht_2)}\nn
\eea
where we have denoted the Lie derivative of the $\tG_c$ factor $\de \tG_{\UV}^{(\text{diff}, c)}$ and the variation of the complementary factor $\de \tG_{\UV}^{(\text{diff}, \bar{c})}$. We find using \eqref{gIR_O}
\be
\de g_{\UV}^{(\text{diff}, c)}(\tht_1, \tht_2)={a_0 \vep \ov (-k_c'(2))q}w_{[0]}(\tht_1-\tht_2)\left( W_{-1}\cdot \de \OO\right)(\tht_1, \tht_2),
\ee
and using $(g_*)_{\UV}$ which was given in \eqref{gUV0_const_vep},
\be
\de g_{\UV}^{(\text{diff}, \bar{c})}(\tht_1, \tht_2)=-{a_0 \ov (-k_c'(2))}\tht_{12}^{-1}w_{[1]}(\tht_1-\tht_2)\left( \p_{\tht_1}-\p_{\tht_2}+{2 \ov \tan{\tht_1-\tht_2 \ov 2}}\right)\left( W_{-1}\cdot \de \vep\right)(\tht_1, \tht_2).
\ee
Note the total expression for $\de g_{\UV}$ is $\SL(2,\RR)$ invariant, as $\de \OO$ does not have $m=-1,0, 1$ modes and $\de \vep$ no $m=0$ modes, and $m=\pm 1$ harmonics with respect to $(\tht_1+\tht_2)/2$ cancel between $\de g_{\UV, \vp}$ and $\de g_{\UV}^{(\text{diff}, \bar{c})}$.

Now using expressions for $\de g_{\IR}$, $\de g_{\UV, \vp}$, $\de g_{\UV}^{(\text{diff, c})}$, and $\de g_{\UV}^{(\text{diff}, \bar{c})}$ we have obtained so far together with correlators $\expval{\de \OO \de \OO}_{\nloc}$ and $\expval{\de \OO \de \vep}_{\loc}$ given in \eqref{Pcorr} and \eqref{altlead}, the last three terms in \eqref{subleadF} are expressed as forms bilinear in $W_{-1}$ and $\left[\p_h W_h\right]_{h=2}$ (recall $W_2$ can be related to $W_{-1}$ as in \eqref{PiWUrel}). In particular, we find that in their sum the $h=2$ residue $f^{\parallel}$ \eqref{f_perp_asym} that was missing in $f^{\perp}$ \eqref{f_perp} appears in the form given in \eqref{f_par_long}. The total subleading four-point function including the first term in \eqref{subleadF} is given by
\begin{equation}
\wideboxed{\begin{aligned}
&\left[{\tF(\tht_1, \tht_2; \tht_3, \tht_4) \ov \tG_*(\tht_1, \tht_2)\tG_*(\tht_3, \tht_4)}\right]^{(0)}
\\
&={1 \ov (q-1)b}\Bigg[\tht_{12}\tht_{34}\left( f^{\perp}(\tht_1, \tht_2; \tht_3, \tht_4)+f^{\parallel}(\tht_1, \tht_2; \tht_3, \tht_4)\right)-{6 \ov \pi^2(-k_c(2))}w_{[0]}(\tht_1-\tht_2)w_{[0]}(\tht_3-\tht_4)
\\
&-{3 \abs{\tht_{12}}^{-1}\abs{\tht_{34}}^{-1} \ov (-k'_c(2))} \Bigl( 1+w_{[1]}(\tht_1-\tht_2)\,\tht_{12}^{-1}(\p_{\tht_1}-\p_{\tht_2})
+w_{[1]}(\tht_3-\tht_4)\,\tht_{34}^{-1}(\p_{\tht_3}-\p_{\tht_4})\Bigr)
\check{Q}(\tht_1, \tht_2; \tht_3, \tht_4)\Bigg]
\end{aligned}}
\end{equation}
where 
\be
\check{Q}(\tht_1, \tht_2; \tht_3, \tht_4)=\tht_{12}\abs{\tht_{12}}\tht_{34}\abs{\tht_{34}}\sum_{m \neq 0}W_{-1,m}(\tht_1, \tht_2)W_{-1, -m}(\tht_3, \tht_4).
\ee
was calculated previously in \eqref{checkQ_calc} in the section on the leading four-point function.

We again give explicit expressions in the cases \eqref{repcases} representative of OPE and OTO regions. In the OPE region with $2\pi>\theta_1>\theta_2>\theta_3>\theta_4>0$,
\begin{align}
\hspace{20pt}&\hspace{-20pt}\left[{\tF(\tht_1, \tht_2; \tht_3, \tht_4) \ov \tG_*(\tht_1, \tht_2)\tG_*(\tht_3, \tht_4)}\right]^{(0)}={1 \ov (q-1)b}\Bigg[ \tht_{12}\tht_{34}\left( f^{\perp}(\tht_1, \tht_2; \tht_3, \tht_4) + f^{\parallel}(\tht_1, \tht_2; \tht_3, \tht_4)\right)\nn
&-{3 \ov (-k'_c(2))}\Bigg( \ubrace{{\cos {\tht \ov 2}\cos{\tht' \ov 2}-\cos \De \tht_+ \ov 2 \sin {\tht \ov 2}\sin{\tht' \ov 2}}}_{\chi^{-1}-{1 \ov 2}}+{\sin \tht-\tht \ov 2 \pi \sin^2{\tht \ov 2}}\left( 1-{\tht' \ov 2 \tan{\tht' \ov 2}}\right)+{\sin \tht'-\tht' \ov 2 \pi \sin^2{\tht' \ov 2}}\left( 1-{\tht \ov 2 \tan{\tht \ov 2}}\right)\nn
&+{1 \ov \pi^2}\left( -2 + {\tht \tht' \ov 2 \tan{\tht \ov. 2}\tan{\tht' \ov 2}}+{\tht^2 \ov 2 \sin^2{\tht \ov 2}}\left( 1-{\tht' \ov 2 \tan {\tht' \ov 2}}\right)+{\tht'^2 \ov 2 \sin^2{\tht' \ov 2}}\left( 1-{\tht \ov 2 \tan {\tht \ov 2}}\right)\right) \Bigg) \Bigg]
\end{align}
where the marked conformal term cancels \eqref{f_par_plus}, which is the sum of the double pole term in the OPE expansion of $f^{\perp}$ (see \eqref{f_perp1}) and $f^{\parallel}$, multiplied by $\tht_{12}\tht_{34}$. The $\theta,\theta'\to 0$ asymptotics of the full four-point function (including $\tF^{(-1)}$ given in \eqref{F-1}) is as follows:
\begin{equation}
\frac{\tF(\tht_1, \tht_2; \tht_3, \tht_4)}
{\tG_*(\tht_1, \tht_2)\tG_*(\tht_3, \tht_4)}
\approx \frac{\theta\theta'^2+\theta^2\theta'}{12\pi(q-1)b}
+\left(\frac{\beta J}{144\pi^2q^2\alpha_S}
-\frac{1}{8\pi^2(q-1)b\,(-k'_{\cc}(2))}\right)\theta^2\theta'^2.
\end{equation}
This is in agreement with equation \eqref{4p-thermo}.

In the OTO region with $2\pi > \tht_1 > \tht_3 > \tht_2 > \tht_4>0$, 
\begin{equation}
\begin{aligned}
&\left[{\tF(\tht_1, \tht_2; \tht_3, \tht_4) \ov \tG_*(\tht_1, \tht_2)\tG_*(\tht_3, \tht_4)}\right]^{(0)}\approx {1 \ov (q-1)b}\Bigg[ \tht_{12}\tht_{34}\left( f^{\perp}(\tht_1, \tht_2; \tht_3, \tht_4) + f^{\parallel}(\tht_1, \tht_2; \tht_3, \tht_4)\right)
\\
&\hspace{20pt}-{3 \ov (-k'_c(2))}{1 \ov 2\pi\sin{\tht \ov 2}\sin{\tht' \ov 2}}\left( (\pi- 2 \De \tht_+)\cos \De \tht_+ + \left( 2-{\pi-\tht \ov \tan{\tht \ov 2}}-{\pi-\tht' \ov \tan {\tht' \ov 2}}\right)\sin \De \tht_+\right)\Bigg]
\end{aligned}
\end{equation}
where we have only shown the terms that grow exponentially in real time for $\Delta\theta_+=it+O(1)$. Fitting the large $t$ asymptotics as $C^{-1}(e^{i\tkap(\pi/2-\Delta\theta_+)}-e^{i(\pi/2-\Delta\theta_+)}) \approx -C^{-1}(1-\tkap)\Delta\theta_{+}e^{-i\Delta\theta_+}$ allows for the extraction of the correction to the Lyapunov exponent, $1-\tkap\sim(\beta J)^{-1}$. In the above the contribution to such large $t$ asymptotics from the term $f^{\parallel}$ and the term $\sim \De \tht_+ \cos \De \tht_+$ cancel. Thus only the conformal part of the four-point function $f^{\perp}$ contributes to the subleading exponent. The exponent was extracted in \cite{MS16}; we do not know of an intuitive way to obtain this quantity.

\section{Dilaton gravity with conformal fields}

The goal of this section is to construct a gravity dual of the reparametrization mode in the SYK model. We will guess the 2D theory from qualitative arguments, study its general properties, and find the effective action in terms of boundary degrees of freedom.

\subsection{The choice of the model}

The authors in \cite{Jen16, MSY16, EMV16} obtained the Schwarzian action from a two-dimensional dilaton gravity with suitable boundary conditions. We will use a slightly different model, which includes the metric tensor $g$, the dilaton $\Phi$, and certain matter fields. The Euclidean action is
\begin{equation}\label{dilact}
\wideboxed{
I[g,\Phi,\ldots]
=\frac{1}{4\pi}\int_{D} \bigl(-\Phi R+U(\Phi)\bigr)\sqrt{g}\,d^2x
-\frac{1}{2\pi}\int_{\partial D}\Phi K\sqrt{g_{\ph\ph}}\,d\ph
+I_{\Matter}[g,\ldots]
}
\end{equation}
where $D$ is the unit disk, $\partial D$ its boundary, $\ph$ the angular coordinate, and $K$ the extrinsic curvature. The boundary term is needed for consistency, see Appendix~\ref{sec_pure_dilaton}. The normalization is such that $\Phi$ has the meaning of entropy (particularly, when evaluated at a black hole horizon.)

Note that adding a constant to $\Phi$ changes the action by a constant; rescaling the metric is equivalent to rescaling the dilaton potential $U$. Thus, we may assume without loss of generality that $U$ has the following expansion near $\Phi=0$:
\begin{equation}\label{U_expansion}
U(\Phi)= -2\Phi-\alpha\Phi^2+\cdots.
\end{equation}
Let us also suppose that $\Phi\gg 1$ so that the bulk can be treated classically, yet $\Phi$ is sufficiently small to allow the use of the above expansion.  When comparing with the SYK model, $\Phi$ is proportional to $N$, whereas $\alpha\sim N^{-1}$. In the most natural setting, the dilaton diverges at the boundary, but we avoid that by introducing a cutoff, $\Phi|_{\partial D}=\Phi_*$. This procedure is similar to putting a UV cutoff at $\tau\sim J^{-1}$.  Essentially, $\Phi_*$ is the dilaton value at which the potential $U(\Phi)$ becomes strongly nonlinear, that is, $\alpha\Phi_*\sim 1$. To make the problem more tractable, we will sometimes assume that $\alpha\Phi_*\ll 1$; this should not affect the general form of the result but only some coefficients.

The previously mentioned papers \cite{Jen16, MSY16, EMV16} used a linear dilaton potential, $U(\Phi)=-2\Phi$ and did not include matter fields (for the most part). This special case is known as the Jackiw-Teitelboim gravity \cite{Ja85, Te83} and has been studied in detail (in the Lorentzian signature) by Almheiri and Polchinski \cite{AlPo14}. In the Euclidean case, the classical solution is the hyperbolic plane with $R=-2$, and the dilaton satisfies the equation $\nabla^2\Phi-2\Phi=0$ as well as some other equations.

We would like to reproduce the effective action on $\partial D$ by integrating out bulk degrees of freedom. In particular, we are interested in the non-local term, which is related to the $h=2$ pole in the conformal 4-point function. In the gravity dual, this term might correspond to a massive scalar field. Using the relation between the scaling dimension on the boundary and mass in the bulk \cite{GKP98,Wi98}, $h=1/2\pm\sqrt{1/4+m^2}$, we find that $m^2=2$. In fact, the field in question can be the dilaton because the latter has a similar equation of motion. However, pure dilaton gravity with an arbitrary potential $U$ obeys Birkhoff's theorem \cite{L-MKu94}, which says that any classical solution is equivalent to a rotationally symmetric one up to a coordinate change. In effect, the bulk solutions are rigid, and all dynamics happen at the boundary. Thus, the desired non-local term is unlikely to appear unless the problem is modified, \eg by adding some matter fields.

The type of matter that will be used represents certain geometric observables. In fact, no new fields are necessary if we give up diffeomorphism covariance. Recall that the effective action in the SYK model involves the conformal time $\ph$ as well as the physical time $\tau\propto\int\sqrt{g_{\ph\ph}}\,d\ph$. The bulk analogue of $\ph$ is a complex coordinate $z$ with respect to which the metric is conformal:
\begin{equation}\label{conf_metric}
d\ell^2=e^{2\rho}\,dz\,d\bar{z},\qquad |z|\le1.
\end{equation}
It is a well-known fact that any Riemannian metric on the unit disk can be represented in this form and that such a representation is unique up to linear fractional maps $z\mapsto\frac{az+b}{cz+d}$. Now, we can define $\ph$ in terms of the boundary value of $z$, \ie $z|_{\partial D}=e^{i\ph}$. Thus, $\ph(\tau)$ is a non-local observable that depends on the metric in the whole disk.

The simplest approach is to consider the variational problem in the class of metrics \eqref{conf_metric}. It has fewer equations of motion and more solutions than the usual, generally covariant problem with the fields $g$ and $\Phi$. Specifically, the energy-momentum tensor of the dilaton field need not vanish; only its trace has to be zero. This result may seem paradoxical because any metric is conformal in suitable coordinates, and an arbitrary variation $\delta g$ is equivalent to a conformal variation up to an infinitesimal coordinate change. However, such a coordinate change also acts on the boundary. If the boundary metric is not constrained, then indeed, the conformal problem is equivalent to the generally covariant one. But if $g_{\ph\ph}|_{\partial D}$ as a function of $\ph$ is fixed, then only a subset of general variations is allowed in the conformal case, and threfore, more stationary configurations exist.

To restore the diffeomorphism covariance, we define the model not using the conformal coordinates $(z,\bar{z})$, but rather, a pair of complex conjugate scalar field $(w,\bar{w})$ such that $w=z$ (and hence $\bar{w}=\bar{z}$) on-shell. The corresponding term $I_{\Matter}$ is designed to constrain $w$, $\bar{w}$ by means of Lagrange multiplier fields, $Q_{+}$ (with spin $+1$) and $Q_{-}$ (with spin $-1$). Specifically,
\begin{equation}\label{dilact_M}
I_{\Matter}[g,w,\bar{w},Q_{+},Q_{-}]
=-\frac{1}{\pi} \int_{D} \bigl(Q_{+}\nabla_{-}w+Q_{-}\nabla_{+}\bar{w}\bigr)
\sqrt{g}\,d^2x,
\end{equation}
where $\nabla_{+}$, $\nabla_{-}$ are proportional to the partial derivatives with respect to $z$ and $\bar{z}$, see below. The equation of motion $\nabla_{-}w=0$ implies that $w$ is a holomorphic function of $z$. If we also require that $w|_{\partial D}=e^{i\ph}$, then $w=z$ in the whole disk.

Let us summarize the problem. We consider Euclidean action \eqref{dilact} with matter term \eqref{dilact_M}, where $(w,\bar{w})$ and $(Q_{+},Q_{-})$ are complex conjugate pairs. The boundary conditions are
\begin{equation}\label{boundcond}
\Phi|_{\partial D}=\Phi_*,\qquad w|_{\partial D}=e^{i\ph},
\end{equation}
whereas $g_{\ph\ph}|_{\partial D}$ is set to an arbitrary function of $\ph$. The goal is to find the effective action, \ie the stationary value of $I$ with respect to bulk degrees of freedom.

\subsection{The operators $\nabla_{+}$, $\nabla_{-}$ and other geometric objects}

Let $(v_1(x),v_2(x))$ be an orthonormal frame (``tetrad'') that smoothly depends on the point $x$, and let $z=x^1+ix^2$ and $\bar{z}=x^1-ix^2$. We also define the vectors $v_{+}$, $v_{-}$ that are related to $v_1$, $v_2$ as $(\partial_{z},\partial_{\bar{z}})$ to $(\partial_{x^1},\partial_{x^2})$:
\begin{equation}
v_{+}=\frac{1}{2}(v_1-iv_2),\qquad v_{-}=\frac{1}{2}(v_1+iv_2).
\end{equation}
The dual frame is denoted by $(\theta^{+},\theta^{-})$. Thus,
\begin{equation}
g_{\alpha\beta}v_{a}^{\alpha}v_{b}^{\beta}=\eta_{ab},\qquad
g_{\alpha\beta}=\eta_{ab}\theta^{a}_{\alpha}\theta^{b}_{\beta},\qquad
\text{where}\quad
\begin{pmatrix}\eta_{++} & \eta_{+-}\\ \eta_{-+} & \eta_{--}\end{pmatrix}
=\frac{1}{2}\begin{pmatrix} 0 & 1\\ 1 & 0\end{pmatrix}.
\end{equation}
In particular, if the metric is conformal, \ie $d\ell^2=e^{2\rho}\bigl((dx^1)^2+(dx^2)^2\bigr) =e^{2\rho}dz\,d\bar{z}$, then the frames $(v_1,v_2)$ and $(v_+,v_-)$ can be chosen as follows:
\useshortskip{\begin{equation}
\begin{pmatrix} v_1^1 & v_2^1\\ v_1^2 & v_2^2 \end{pmatrix}
=\begin{pmatrix} v_+^{z} & v_-^{z}\\ v_+^{\bar{z}} & v_-^{\bar{z}} \end{pmatrix}
=e^{-\rho} \begin{pmatrix} 1 & 0\\ 0 & 1 \end{pmatrix}.\qquad\qquad
\figbox{1.0}{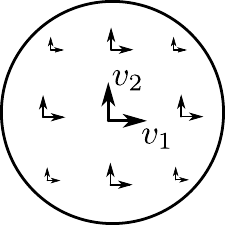}
\end{equation}}

In two Euclidean dimensions, spin corresponds to irreducible representations of $\SO(2)$ or its universal cover. Such representations are one-dimensional and characterized by a number $\nu$. By definition, the infinitesimal counterclockwise rotation $\Lambda$ acts on the spin as the multiplication by $-i\nu$. A \emph{$\nu$-spinor on $D$} is represented relative to $(v_+,v_-)$ by a complex-valued function $\psi$. A general gauge transformation, \ie the counterclockwise rotatation of the local frame by angle $\xi=\xi(x)$,
\begin{equation}\label{gauge_tr}
(v_{+},v_{-}) \to \bigl(e^{i\xi}v_{+},\,e^{-i\xi}v_{-}\bigr)
\end{equation}
takes $\psi$ to $e^{i\nu\xi}\psi$. A spin connection is described by the set of coefficients
\begin{equation}
\tensor{\omega}{_\alpha^a_b}=\omega_{\alpha}\tensor{\Lambda}{^a_b},\qquad
\text{where}\quad
\begin{pmatrix}
\tensor{\Lambda}{^+_+} & \tensor{\Lambda}{^+_-}\\
\tensor{\Lambda}{^-_+} & \tensor{\Lambda}{^-_-}
\end{pmatrix}
=\begin{pmatrix} i & 0\\ 0 & -i
\end{pmatrix}.
\end{equation}
The gauge transformation \eqref{gauge_tr} changes $\omega_{\alpha}$ to
$\omega_{\alpha}+\partial_{\alpha}\xi$. The covariant derivative of a $\nu$-spinor is given by the equation
\begin{equation}
D_{\alpha}\psi=(\partial_\alpha-i\nu\omega_\alpha)\psi.
\end{equation}
Finally, we define the operators that increase or decrease the spin by $1$:
\begin{equation}
\nabla_{\pm}:\: \text{$\nu$-spinors}\to \text{($\nu\pm1$)-spinors},\qquad\quad
\nabla_{\pm}\psi=v_{\pm}^{\alpha}D_{\alpha}\psi.
\end{equation}
Note that $\nabla_{+}$ and $\nabla_{-}$ are gauge-invariant, \ie they commute with gauge transformations. The commutator between these operators is given by the curvature:
\begin{equation}
(\nabla_{+}\nabla_{-}-\nabla_{-}\nabla_{+})\psi=-\frac{\nu}{4}R\psi,\qquad
\text{where $\psi$ is a $\nu$-spinor}.
\end{equation}

For convenience, we give some explicit formulas in conformal coordinates:
\begin{gather}
(\omega_{z},\omega_{\bar{z}})
=\bigl(-i\partial_{z}\rho,\,i\partial_{\bar{z}}\rho\bigr),
\displaybreak[0]\\[5pt]
R=-2\nabla^2\rho=-8e^{-2\rho}\partial_{z}\partial_{\bar{z}}\rho,
\displaybreak[0]\\[5pt]
\nabla_{+}\psi
=e^{(-1+\nu)\rho}\,\partial_{z}(e^{-\nu\rho}\psi),\qquad
\nabla_{-}\psi
=e^{(-1-\nu)\rho}\,\partial_{\bar{z}}(e^{\nu\rho}\psi)\qquad
(\text{$\psi$ is a $\nu$-spinor}).
\end{gather}
The metric and extrinsic curvature on the boundary of the unit disk are given by these expressions:
\begin{equation}
g_{\ph\ph}=e^{2\rho},\qquad
K=g_{\ph\ph}^{-1/2}(1+\omega_{\ph})=e^{-\rho}+n^{\gamma}\nabla_{\gamma}\rho,
\end{equation}
where $n$ is the unit normal vector.

\subsection{Variation of the action}

Recall that the Euclidean action $I=I[g,\Phi,w,\bar{w},Q_+,Q_-]$ is as follows:
\begin{equation}\label{dilact_full}
I=\frac{1}{4\pi}\int_{D}
\bigl(-\Phi R+U(\Phi)\bigr) \sqrt{g}\,d^2x
-\frac{1}{2\pi}\int_{\partial D}\Phi K\sqrt{g_{\ph\ph}}\,d\ph
-\frac{1}{\pi}\int_{D}
\bigl(Q_{+}\nabla_{-}w+Q_{-}\nabla_{+}\bar{w}\bigr) \sqrt{g}\,d^2x.
\end{equation}
In this subsection, we consider the boundary values of $g_{\ph\ph}$, $\Phi$, and $w$ as arbitrary functions of $\ph$, with the technical assumption that $w|_{\partial D}$ maps the unit circle to the boundary of some complex domain while preserving orientation. The more specific conditions \eqref{boundcond} will be imposed later.
  
Taking the variational derivative of the action with respect to $\Phi$, $(w,\bar{w})$, and $(Q_{+},Q_{-})$ gives the following equations of motion:
\begin{equation}\label{eom_R}
\wideboxed{
R=U'(\Phi)
}\vspace{-8pt}
\end{equation}
\begin{alignat}{2}\label{z_is_conformal}
\nabla_{-}w&=0,\qquad &\nabla_{+}\bar{w}&=0,
\\[3pt]
\label{eom_Q}
\nabla_{-}Q_{+}&=0,\qquad &\nabla_{+}Q_{-}&=0.
\end{alignat}
Equation \eqref{z_is_conformal} implies that $z\mapsto w(z)$ is a conformal map from disk $D$ to the previously mentioned domain. The last pair of equations is, actually, subsumed by those for the energy-momentum tensor,
\begin{equation}
(T_{\Grav}+T_{\Matter})_{\mu\nu}=0,
\end{equation}
where $T_{\Grav}$ and $T_{\Matter}$ are the gravitational part (due to the dilaton) and the matter contribution, respectively.

The energy-momentum tensor for the pure dilaton gravity with $U(\Phi)=0$ is given by \eqref{T_pure} in Appendix~\ref{sec_pure_dilaton}. In two dimensions, the Einstein tensor $R_{\alpha\beta}-\frac{1}{2}R\kern1pt g_{\alpha\beta}$ vanishes. Adding the contribution from $U(\Phi)$, we get this result:
\begin{equation}\label{TG}
(T_{\Grav})_{\mu\nu}=\frac{1}{2\pi}\biggl(
\bigl(\nabla_{\mu}\nabla_{\nu}-g_{\mu\nu}\nabla^2\bigr)\Phi
-\frac{1}{2}\kern1pt g_{\mu\nu}\kern1pt U(\Phi)
\biggr),
\end{equation}
that is,
\useshortskip{\begin{gather}
(T_{\Grav})_{+-}=(T_{\Grav})_{-+}
=\frac{1}{4}\tensor{(T_{\Grav})}{^\mu_\mu},\qquad\quad
\tensor{(T_{\Grav})}{^\mu_\mu}=-\frac{1}{2\pi}\bigl(\nabla^2\Phi+U(\Phi)\bigr),
\displaybreak[0]\\[5pt]
\begin{aligned}
(T_{\Grav})_{++}=e^{-2\rho}(T_{\Grav})_{zz}
=\frac{1}{2\pi}\,\partial_{z}(e^{-2\rho}\partial_{z}\Phi)
=\frac{1}{2\pi}\nabla_{+}\nabla_{+}\Phi,\\
\label{TG1}
(T_{\Grav})_{--}=e^{-2\rho}(T_{\Grav})_{\bar{z}\bar{z}}
=\frac{1}{2\pi}\,\partial_{\bar{z}}(e^{-2\rho}\partial_{\bar{z}}\Phi)
=\frac{1}{2\pi}\nabla_{-}\nabla_{-}\Phi.
\end{aligned}
\end{gather}}
The energy-momentum tensor of matter has two nonzero components, $(T_{\Matter})_{++}=-\frac{1}{2\pi}Q_{+}(\nabla_{+}w)$ and $(T_{\Matter})_{--}=-\frac{1}{2\pi}Q_{-}(\nabla_{-}\bar{w})$. Thus, the condition $T_{\Grav}+T_{\Matter}=0$ can be written as follows:
\begin{gather}
\label{eom_Phi}
\wideboxed{
\nabla^2\Phi+U(\Phi)=0
}
\displaybreak[0]\\[5pt]
\label{Q_TG}
Q_{+}=2\pi(\nabla_{+}w)^{-1}(T_{\Grav})_{++},\qquad
Q_{-}=2\pi(\nabla_{-}\bar{w})^{-1}(T_{\Grav})_{--}.
\end{gather}
Note that equations \eqref{TG1}, \eqref{eom_Phi} and \eqref{eom_R} (where $R=-2\nabla^2\rho$) imply that
\begin{equation}
\nabla_{-}(T_{\Grav})_{++}=e^{-3\rho}\partial_{\bar{z}}(T_{\Grav})_{zz}=0,\qquad
\nabla_{+}(T_{\Grav})_{--}=e^{-3\rho}\partial_{z}(T_{\Grav})_{\bar{z}\bar{z}}=0,
\end{equation}
hence $(T_{\Grav})_{zz}$ and $(T_{\Grav})_{\bar{z}\bar{z}}$ are holomorphic function of $z$ and $\bar{z}$, respectively. It is now easy to see that \eqref{eom_Q} follows from the other equations of motion.\smallskip

Let us also calculate the total variation of the action when that the bulk equations are satisfied but the boundary values of $g_{\ph\ph}$, $\Phi$, and $(w,\bar{w})$ change:
\begin{equation}\label{totvarI}
\wideboxed{
\delta I=\int_{\partial D}
\biggl(E(\delta\rho)-\frac{K}{2\pi}(\delta\Phi)
-T_{\perp w}(\delta w)-T_{\perp\bar{w}}(\delta\bar{w})\biggr)
\sqrt{g_{\ph\ph}}\,d\ph
}
\end{equation}
where
\begin{equation}
E=-\frac{1}{2\pi}n^{\gamma}\kern1pt\nabla_{\gamma}\Phi
\end{equation}
may be called the ``surface energy'' and
\begin{equation}
T_{\perp w}=n^{\gamma}(\nabla_{+}w)^{-1}(T_{\Grav})_{\gamma+},\qquad
T_{\perp\bar{w}}=n^{\gamma}(\nabla_{-}\bar{w})^{-1}(T_{\Grav})_{\gamma-}
\end{equation}
are components of the gravitational energy-momentum tensor just near the boundary. The latter are linear combinations of
\begin{equation}
T_{\perp\perp}=e^{2i\ph}(T_{\Grav})_{++}+e^{-2i\ph}(T_{\Grav})_{--},\qquad
T_{\perp\parallel}=ie^{2i\ph}(T_{\Grav})_{++}-ie^{-2i\ph}(T_{\Grav})_{--},
\end{equation}
which generate normal deformations and diffeomorphisms of the circle, respectively. These are explicit formulas for $T_{\perp\perp}$ and $T_{\perp\parallel}$:
\begin{equation}\label{Tpp_explicit}
T_{\perp\perp}
=KE-\frac{1}{4\pi}\,U(\Phi)-\frac{1}{2\pi}\,\partial_{\ell}^2\Phi,\qquad
T_{\perp\parallel}
=-\partial_{\ell}E-\frac{K}{2\pi}\,\partial_{\ell}\Phi,\qquad
\text{where}\quad d\ell=\sqrt{g_{\ph\ph}}\,d\ph.
\end{equation}

Note that $T_{\Grav}$ in the whole disk can be expressed in terms of $T_{\perp\parallel}$ on its boundary. For a motivating idea, we recall that any variation of the bulk metric is equivalent to a conformal variation up to a coordinate change. Let us consider an infinitesimal coordinate transformation that is conformal except at point $z$:
\begin{equation}
\delta z'=a\left(
\epsilon\,\frac{(1-\bar{z}z')^3}{z'-z}
-\bar{\epsilon}\,\frac{(z'-z)^3}{1-\bar{z}z'}
\right),\qquad
\delta\bar{z}'=a\left(
-\epsilon\,\frac{(\bar{z}'-\bar{z})^3}{1-z\bar{z}'}
+\bar{\epsilon}\,\frac{(1-z\bar{z}')^3}{\bar{z}'-\bar{z}}
\right),
\end{equation}
where $\epsilon$ and $\bar{\epsilon}$ are infinitely small parameters and $a=\frac{1}{2\pi}(1-z\bar{z})^{-3}$. These equations are obtained from the condition that $\partial_{\bar{z}'}(\delta z')$ and $\partial_{z'}(\delta\bar{z}')$ are proportional to the delta-function at point $z$ and that the vertor field $(\delta z',\delta\bar{z}')$ is tangent to the unit circle. On the one hand, the variation of the action under this transformation is equal to $\epsilon\,T_{zz}(z) +\bar{\epsilon}\,T_{\bar{z}\bar{z}}(\bar{z})$. But it can also be expressed in terms of the vector field on the boundary, which should be integrated against $T_{\perp\parallel}$. From these considerations, we find that
\begin{equation}
(T_{\Grav})_{zz}(z)=\frac{1}{2\pi i(1-z\bar{z})^3}
\int_{\partial D}\frac{(1-\bar{z}e^{i\ph})^3}{e^{i\ph}(e^{i\ph}-z)}\,
g_{\ph\ph}T_{\perp\parallel}\,d\ph,
\end{equation}
and similarly for $(T_{\Grav})_{\bar{z}\bar{z}}(\bar{z})$. This equation can be verified by using the fact that $g_{\ph\ph}T_{\perp\parallel}$ is equal to $ie^{2i\ph}(T_{\Grav})_{zz} -ie^{-2i\ph}(T_{\Grav})_{\bar{z}\bar{z}}$, where $(T_{\Grav})_{zz}$ is holomorphic and $(T_{\Grav})_{\bar{z}\bar{z}}$ is antiholomorphic.

\subsection{Static solutions and thermodynamics}

This section is largely based on the work by other authors. Static solutions (in the Lorentzian signature) were studied in \cite{L-MKu94}, and the connection to thermodynamics was made in unpublished notes by Maldacena \cite{Mal16}.

We now use the boundary condition $\Phi|_{\partial D}=\Phi_*$, whereas the fields $w$, $\bar{w}$, $Q_+$, $Q_-$ will not play any role. There is no constraint on the metric, except that the total boundary length is fixed:
\begin{equation}
\int_{\partial D}d\ell=\int_{\partial D}\sqrt{g_{\ph\ph}}\,d\ph=L.
\end{equation}
This setting corresponds to thermal equilibrium, where $L$ is the inverse temperature. The corresponding solutions will be called ``static''. Since the first variation of the action is zero whenever $\int_{\partial D}(\delta\rho)\sqrt{g_{\ph\ph}}\,d\ph=0$, the surface energy $E$ is constant. It follows that $T_{\perp\parallel}$ and $T_{\Grav}$ are zero. 

The vanishing of $T_{\Grav}$ has two important consequences. First, $\xi^{\mu} =\epsilon^{\mu\nu}\nabla_{\nu}\Phi$ is a Killing vector \cite{BaOL91}. Second, the following quantity is constant \cite{L-MKu94}:
\begin{equation}
C=-g^{\mu\nu}(\nabla_{\mu}\Phi)(\nabla_{\nu}\Phi)+W(\Phi),
\end{equation}
where
\begin{equation}
\wideboxed{
W(\Phi)=-\int_{0}^{\Phi}U(\phi)\,d\phi
}
\end{equation}
Both statements are verified by direct calculation:
\begin{gather}
\Ld_{\xi}\Phi=0,\qquad\quad
(\Ld_{\xi}g)_{\alpha\beta}
=2\pi\epsilon^{\mu\nu} \bigl((T_{\Grav})_{\alpha\nu}g_{\beta\mu}
+(T_{\Grav})_{\beta\nu}g_{\alpha\mu}\bigr),\\[5pt]
\nabla_{\mu}C
=-4\pi\bigl(
(T_{\Grav})_{\mu\alpha}-g_{\mu\alpha}\tensor{(T_{\Grav})}{^\gamma_\gamma}
\bigr) g^{\alpha\beta}(\nabla_{\beta}\Phi),
\end{gather}
where we have used \eqref{TG}.

When written in conformal coordinates $(z,\bar{z})$, the Killing vector $\xi$ represents an infinitesimal conformal map, \ie $\xi^{z}$ is a holomorphic function of $z$. Furthermore, $\xi$ is tangent to the unit circle because $\Phi|_{\partial D}$ is constant. Hence, $\xi^z=-i(c_{-1}+c_{0}z+c_{1}z^2)$, where $c_{m}^*=c_{-m}$. There are four cases: (1)~the quadratic function $\xi^z$ has one zero inside and the other outside the unit circle; (2)~both zeros lie on the unit circle; (3)~there is one double zero on the unit circle; (4)~$\xi$ is identically zero. In the last case, $\Phi=\Phi_*$ in the whole disk, which is only possible if $U(\Phi_*)=0$. Cases 2 and 3 are excluded because such $\xi$ cannot preserve a nonsingular metric on the circle. Hence, $\xi$ is nontrivial but vanishes at some point $z_0$ inside the unit disk. By applying a linear fractional map $z\mapsto\frac{az+b}{cz+d}$, we can arrange that $z_0=0$. Thus, $\xi$ is an infinitesimal rotation about the origin.

We will now work in the polar coordinates $(r,\ph)$ such that $z=re^{i\ph}$. Due to the rotational symmetry, $\Phi$ and $\rho$ depend only on $r$. Thus, \begin{gather}
\xi^{r}=0,\qquad\quad
\xi^{\ph}=-e^{-2\rho}\,\partial_{r}\Phi
=-cr,\\[5pt]
C=-e^{-2\rho}(\partial_{r}\Phi)^2+W(\Phi)=W(\Phi_0),
\end{gather}
where $c$ is some constant and $\Phi_0=\Phi|_{r=0}$. The value of $c$ can be determined from the condition that $\Phi$ is smooth at the center of the disk, or more exacly, that $\Phi-\Phi_0\sim r^2$ for small $r$. In this way, we obtain the following equations:
\begin{equation}
\partial_{r}\Phi=\frac{W(\Phi)-W(\Phi_0)}{cr},\qquad
e^{2\rho}=\frac{W(\Phi)-W(\Phi_0)}{c^2r^2},\qquad
\text{where}\quad c=\frac{-U(\Phi_0)}{2}.
\end{equation}
Note that the dilaton equation of motion $R=U'(\Phi)$ (where $R=-2e^{-2\rho}(\partial_{r}^2 +r^{-1}\partial_{r})\rho$) follows automatically.

Now, let us express some quantities that are relevant to thermodynamics, namely, the boundary length, energy, extrinsic curvature, and total action:
\begin{gather}
L=2\pi e^{\rho}|_{r=1}
=4\pi\frac{\sqrt{W(\Phi_*)-W(\Phi_0)}}{-U(\Phi_0)},
\displaybreak[0]\\[5pt]
E=-\frac{1}{2\pi}(e^{-\rho}\partial_{r}\Phi)|_{r=1}
=-\frac{1}{2\pi}\sqrt{W(\Phi_*)-W(\Phi_0)},
\displaybreak[0]\\[5pt]
K=(e^{-\rho}\bigl(1+r\partial_{r}\rho)\bigr)\big|_{r=1}
=\frac{-U(\Phi_*)}{2\sqrt{W(\Phi_*)-W(\Phi_0)}},
\displaybreak[0]\\[5pt]
I=\frac{1}{2}\int\bigl(-\Phi U'(\Phi)+U(\Phi)\bigr)e^{2\rho}r\,dr
-\frac{L}{2\pi}\Phi_*K
=\frac{2(W(\Phi_*)-W(\Phi_0))}{U(\Phi_0)}-\Phi_0.
\end{gather}
The minimum dilaton value $\Phi_0$ is interpreted as entropy, $L$ as the inverse temperature, and we may write $I=LF$, where $F$ is the free energy. As expected, these relations hold:
\begin{equation}
I=LE-\Phi_0,\qquad
\delta I = E\,\delta L-\frac{LK}{2\pi}\,\delta\Phi_*.
\end{equation}
The last equation is a special case of \eqref{totvarI}.

It is desirable to eliminate the dependence on the cutoff $\Phi_*$ as much as possible and to compare the results with the SYK model. For these purposes let us define the renormalized quantities
\begin{equation}\label{renorm_beta}
\tilde{\beta}=b_1^{-1}L,\qquad
\tilde{E}=b_1(E+b_2),\qquad \tilde{I}=I+b_2L,
\end{equation}
where
\begin{equation}
b_1=4\pi\sqrt{W(\Phi_*)},\qquad b_2=\frac{\sqrt{W(\Phi_*)}}{2\pi}.
\end{equation}
We have fixed the normalization of $b_2$ by requiring that $b_2L$ subtracts the leading divergence as $\Phi_* \to \infty$ in $I$. The normalization of $b_1$ is not significant and has been chosen arbitrarily. Then
\begin{equation}
\tilde{\beta} =\frac{1}{-U(\Phi_0)}\sqrt{1-\frac{W(\Phi_0)}{W(\Phi_*)}},\qquad
\tilde{E} =2W(\Phi_*)\left(1-\sqrt{1-\frac{W(\Phi_0)}{W(\Phi_*)}}\right),\qquad
\tilde I =\tilde{\beta}\tilde{E}-\Phi_0.
\end{equation}
The interesting limit is when $W(\Phi_*)\gg W(\Phi_0)$, which is roughly the same as $\Phi_*\gg\Phi_0$. In this case,
\begin{equation}
\wideboxed{
\tilde{\beta}\approx\frac{1}{-U(\Phi_0)},\qquad
\tilde{E}\approx W(\Phi_0),\qquad
\tilde{I}\approx\frac{W(\Phi_0)}{-U(\Phi_0)}-\Phi_0
}
\end{equation}
To summarize, the energy is a function of entropy, $\tilde{E}=W(\Phi_0)$. The other equations follow from that and thermodynamic identities.\smallskip

We are primarily interested in the case $U(\Phi)=-2\Phi-\alpha\Phi^2$,\,\, $W(\Phi)=\Phi^2+\frac{\alpha}{3}\Phi^3$, where $\alpha$ is small and $\Phi_*$ is large. In the crudest approximation, $\tilde{\beta}\approx 1/(2\Phi_0)$. These numbers will be used as small parameters:
\begin{equation}
\alpha/\tilde{\beta} \sim \alpha\Phi_0 \ll 1,\qquad\quad
(\tilde{\beta}\Phi_*)^{-1} \sim \Phi_0/\Phi_* \ll 1,\qquad\quad
\tilde{\beta}\sim 1/\Phi_0\ll 1.
\end{equation}
The first two of them quantify nonlinearity at the center of the disk due to the $-\alpha\Phi^2$ term in the dilaton potential and the proximity of the cutoff. (Note that their ratio $\alpha\Phi_*$ is arbitrary.) The last condition is needed to make sure that quantum fluctuations are small. In terms of the SYK model, the parameters $\alpha/\tilde{\beta}$ and $\tilde{\beta}$ are analogous to $(\beta J)^{-1}$ and $\beta J/N$, respectively. Under the stated assumptions, we obtain the following expression:
\begin{gather}
\label{tI_static}
\tilde{I}(\tilde{\beta}) = -\frac{1}{4\tilde{\beta}}
\biggl(
1-\frac{\alpha}{6\tilde{\beta}}
+O\Bigl((\alpha/\tilde{\beta})^2+(\tilde\beta\Phi_*)^{-2}\Bigr)
\biggr).
\end{gather}

\subsection{Effective boundary action} \label{sec: effbdact}

We now find the effective boundary action resulting from evaluating on-shell the bulk action~\eqref{dilact} with the matter term~\eqref{dilact_M}. After integrating out the Lagrange multiplier fields $Q_{\pm}$, the action can be regarded as a functional of $g$ and $\Phi$, where the metric $g$ is conformal in coordinates $(z,\bar{z})$ such that $z|_{\partial D}=e^{i\vp}$. Thus
\begin{equation}\label{dilact1}
I=\frac{1}{4\pi}\int_{D} \bigl(-\Phi R+U(\Phi)\bigr)\sqrt{g}\,d^2x
-\frac{1}{2\pi}\int_{\partial D}\Phi K\sqrt{g_{\ph\ph}}\,d\ph,
\end{equation}
where $U(\Phi)= -2\Phi-\alpha\Phi^2$ and boundary conditions are 
\begin{equation} \label{diskbc}
g_{\vp\vp}=\vep(\vp)^{-2},\qquad
\Phi|_{\partial D}=\Phi_*.
\end{equation}
Note that $\vep(\vp)$ is \emph{not} the same as in the SYK model. The comparison should be in terms of renormalized quantities, which we define by analogy with equation~\eqref{renorm_beta}:
\begin{equation}
\tvep(\vp)=4\pi\sqrt{W(\Phi_*)}\,\vep(\vp),\qquad
\tilde{I}=I+\frac{\sqrt{W(\Phi_*)}}{2\pi}\,L.
\end{equation}
Thus $\tilde{\beta}=\int\tvep^{-1}d\vp$ while $L=\int\vep^{-1}d\vp$. We will derive the following expression for the effective action, assuming $\alpha\Phi_*\ll 1$:
\begin{equation}
\label{dilact_eff}
\wideboxed{
\begin{aligned}
\tilde{I} =
&-\frac{1}{4\pi} \int\frac{d\vp}{2\pi}\,\tvep^{-1}
\biggl(\frac{\tvep^2}{2}-\frac{\tvep'^2}{2}+\tvep\tvep''\biggr)
\\[3pt]
&-\frac{\alpha}{8\pi^2}\left[
\int\frac{d\vp_1}{2\pi}\frac{d\vp_2}{2\pi}\,
\frac{\tvep(\vp_1)\tvep(\vp_2)}{\vp_{12}^4}
\left(\ln\biggl(\frac{\vp_{12}^2}{\alpha^2\tvep(\vp_1)\tvep(\vp_2)}\biggr)
+\tilde{c}\right)\right]_{\text{fin.}}
+O\bigl(\Phi_*^{-2}\tvep^3\bigr)
\end{aligned}
}
\end{equation}
where $\vp_{12}=2\sin\bigl((\ph_1-\ph_2)/2\bigr)$, the subscript fin. was defined in \eqref{Inlfin}, and
\begin{equation}
\label{dilact_eff_c}
\tilde{c}=\frac{6}{\alpha\Phi_*}+2\ln(4\pi\alpha\Phi_*)-\frac{2}{3}.
\end{equation}
The first integral in the expression for the action and the first term in $\tilde{c}$ are found by solving the $\alpha=0$ problem. (Note that the $\alpha$ contained in the second term of $\tilde{c}$ cancels with that in the logarithm.) The non-local action and the other terms in $\tilde{c}$ follow from the $\alpha \Phi^2$ term in the dilaton potential $U$. 

When the value of $\alpha$ is nonzero but small, we use perturbation theory. It would be more natural to assume that $\alpha\Phi_*\sim 1$ or even take the limit $\Phi_*\to\infty$. In this regime, strong nonlinearity at $\Phi\gtrsim\alpha^{-1}$ is expected to provide an effective cutoff, whereas $\Phi_*$ would become irrelevant. Thus, the error bound should be $O(\alpha^{2}\tvep^3)$ and $\tilde{c}$ replaced with a constant. However, we have no means of calculating it. Up to this unknown constant, the matching of the action \eqref{dilact_eff} in this non-linear regime with the SYK action in \eqref{I_loc_conf} and \eqref{I_nl_conf} is self-evident, with $N^{-1}\tvep \sim \al_S \vep_{\rm{SYK}}$ and $N \al \sim \ga/\al_S^2$. In the language of AdS/CFT, the dilaton $\Phi$ is dual to the Schwarzian operator $\OO$ \eqref{Odef}: the on-shell solution for $\Phi$ takes the form of the source $\vep_{\rm{SYK}}$ integrated against the bulk-boundary propagator for a bulk scalar field with mass $m^2=2$ (see \eqref{Phi_bbprop}, \eqref{fapprox}), and $\vep_{\rm{SYK}}$ sources $\OO$ in the Schwarzian action \eqref{Iepsrho}. 

Before proceeding with the derivation of \eqref{dilact_eff}, let us check it in the static case where $\tvep=2\pi/\tilde{\beta}$ does not depend on $\vp$. The first term gives $-\frac{1}{4\pi}\left({\tvep \ov 2}\right)=-{1 \ov 4\tilde{\beta}}$ and the second $-\frac{\alpha}{8\pi^2}(2\tvep^2)\bigl(-\frac{1}{24}\bigr) ={\alpha \ov 24\tilde{\beta}^2}$, where the coefficient $-\frac{1}{24}$ corresponds to $\left[\int{d\vp \ov 2 \pi}\left(2 \sin{\vp \ov 2}\right)^{-4}\abs{2 \sin{\vp \ov 2}}\right]_{\text{fin.}}$, see below \eqref{Inlfinexp}. This result is in agreement with \eqref{tI_static}.

\subsubsection{The $\alpha=0$ case}

\begin{figure}
\centerline{\includegraphics[scale=1.0]{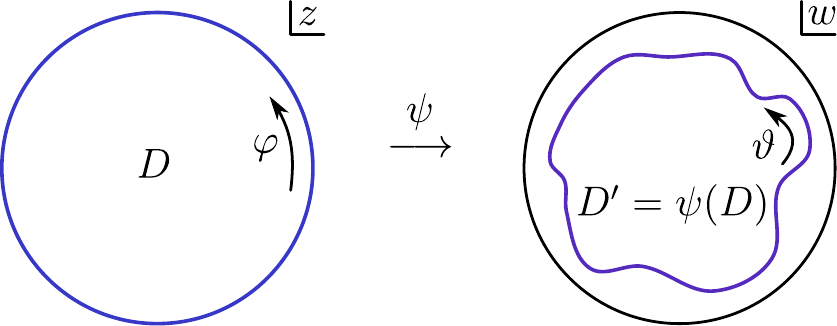}}
\caption{Map of the physical space (disk $D$) to the region $D'$ of the Poincare disk.}\label{fig: psimap}
\end{figure}

If $\alpha=0$, the equations of motion and the on-shell action become
\begin{equation}
R = -2,\qquad \nabla^2 \Phi= 2 \Phi,\qquad\quad
I_{\text{on-shell},\,\alpha=0}
=-\frac{\Phi_*}{2\pi}\int_{\partial D} K\,d\ell,
\end{equation}
where $d\ell=\sqrt{g_{\vp\vp}}\,d\vp$ is the boundary length element. The equation $R=-2$ implies the hyperbolic plane geometry; thus, this action describes the deformation energy of a curve (``1D membrane'') in the hyperbolic plane.\footnote{By the Gauss-Bonnet theorem, $\int Kdl$ equals the encolsed area plus $2\pi$. Maximazing the area over curves of a given length is equivalent to minimizing the length while keeping the area fixed. The latter setting may be described as a ``droplet'' with a surface tension.} The metric is given by the Poincare disk model in suitable coordinates:\footnote{The coordinates $(w,\wc)$ should not be confused with the matter fields that were denoted by the same letters but are currently equal to $(z,\bar{z})$.}
\begin{equation} \label{Pdisk}
d\ell^2=\frac{4}{(1- w \wc)^2}\,dw\, d\wc, \qquad  \abs{w}<1.
\end{equation}
The physical space (represented by the unit disk $D$) is mapped to a subregion $D'$ of the Poincare disk by some conformal map $\psi$, see Figure~\ref{fig: psimap}.
We will use modified polar coordinates $(y,\vt)$ such that
\begin{equation}\label{yvt_coord}
w=\sqrt{y}\,e^{i\vt}, \quad\: \wc=\sqrt{y}\,e^{-i\vt},\qquad
d\ell^2=(1-y)^{-2}\bigl(y^{-1}\kern1pt dy^2+4y\,d\vt^2\bigr).
\end{equation}

The boundary of $D'$ is described by the equation $y=y_*(\vt)$ with some function $y_*$. We also define analogously to $\vep(\vp)=d\vp/dl$ on $D$ (\cf \eqref{diskbc}) 
\begin{equation}
\de(\vt)=\frac{d\vt}{d\ell}
\end{equation}
so that $\int K d\ell=\int \de^{-1}K d\vt$. In this notation,
\begin{equation}
\de^{-1}=\frac{2\sqrt{y_*}}{1-y_*}\sqrt{1+\gamma^2},\qquad
\de^{-1}K=\frac{1+y_*}{1-y_*}+\frac{d}{d\vt}\arctan\gamma,\qquad
\text{where}\quad \gamma=-\frac{1}{2y_*}\frac{dy_*}{d\vt}.
\end{equation}
We assume that the boundary is close to the unit circle, and therefore, $\de\ll 1$,\, $\vt\approx\vp$. So both $y_*$ and $K$ can be expressed in powers of $\de$ and its derivatives with respect to $\vt$:
\begin{equation}\label{ystar_K}
y_*=1-2\de+2\de^2+O\left(\de^3\right),\qquad\quad
\de^{-1}K=\de^{-1}\left(1+\frac{\de^2}{2}-\frac{\de'^2}{2}+O\left(\de^4\right)\right)
+\text{full derivative}.
\end{equation}
Integrating $\de^{-1}K$ against $d\vt$ gives
\begin{equation}\label{osdilact0}
I_{\text{on-shell},\,\alpha=0}
=-\frac{\Phi_*}{2\pi}L
-\Phi_{*}\int \frac{d\vt}{2\pi}\,\de^{-1}
\left(\frac{\de^2}{2}-\frac{\de'^2}{2}+\de\de''\right)
+O(\Phi_*\de^3).
\end{equation}
The expression in parentheses is equal to $\Sch(e^{i\vt},\ell)$, which is in agreement with the result in \cite{Jen16, MSY16, EMV16}.

However, the action needs to be represented in terms of $\vp$ rather than $\vt$. These variables are related by a conformal map $\psi$, which is uniquely defined if we require that $\psi(0)=0$ and $\psi'(0)$ is real. Under these conditions,\vspace{-5pt}
\begin{alignat}{2}
\psi(z)&=ze^{f(z)},\qquad &
f(z)&=\sum_{k=0}^{\infty}c_{k}z^{k},\quad c_0\in\RR,\\[2pt]
y_*&=|\psi(e^{i\vp})|^2=e^{2u(\vp)},\qquad &
u(\vp)&=\Re f(e^{i\vp}),\\[5pt]
\vt&=\Im(\ln\psi(e^{i\vp}))=\vp+v(\vp),\qquad &
v(\vp)&=\Im f(e^{i\vp}).
\end{alignat}
For our purposes, it is sufficient to know that
\begin{equation}
\de\approx\vep\approx-u(\vp),\qquad \vt-\vp=v(\vp),
\end{equation}
where the approximate equality holds in the leading order in $\vep$ and the functions $u$, $v$ are related by the Hilbert transform:
\begin{equation}
v(\ph)=\text{p.v.}\int\frac{d\ph_1}{2\pi}\,
\cot\biggl(\frac{\ph-\ph_1}{2}\biggr)\,u(\ph_1).
\end{equation}

Now, we use the transformation law of the Schwarzian
\begin{equation}
\Sch(p,s)=\Sch(p,r)\biggl(\frac{dr}{ds}\biggr)^2+\Sch(r,s)
\end{equation}
to find the difference between $\Sch(e^{i\vt},\ell)=\frac{\de^2}{2}-\frac{\de'^2}{2}+\de\de''$ and the analogous function $\Sch(e^{i\vp},\ell)$ in terms of $\vep$:
\begin{equation}
\begin{aligned}
&\Sch(e^{i\vt},\ell)-\Sch(e^{i\vp},\ell)
=\frac{1}{2}\biggl(\frac{d\vt}{d\ell}\biggr)^2
-\frac{1}{2}\biggl(\frac{d\vp}{d\ell}\biggr)^2
+\Sch(\vt,\vp)\biggl(\frac{d\vp}{d\ell}\biggr)^2\\[2pt]
&\qquad\quad
\approx\bigl(v'(\vp) + v'''(\vp)\bigr)\biggl(\frac{d\vp}{d\ell}\biggr)^2
\approx 12\left[\int\frac{d\vp_1}{2\pi}
\biggl(2\sin\biggl(\frac{\vp-\vp_1}{2}\biggr)\biggr)^{-4}
\vep(\vp_1)\right]_{\text{fin.}}
\biggl(\frac{d\vp}{d\ell}\biggr)^2.
\end{aligned}
\end{equation}
Thus,
\begin{equation}\label{osdilact0f}
\begin{aligned}
I_{\text{on-shell},\,\alpha=0}
=&-\frac{\Phi_*}{2\pi}L
-\Phi_*\int \frac{d\vp}{2\pi}\,\vep^{-1}
\left(\frac{\vep^2}{2}-\frac{\vep'^2}{2}+\vep\vep''\right)\\[3pt]
&-12\Phi_*\left[\int\frac{d\vp_1}{2\pi}\frac{d\vp_2}{2\pi}\,
\frac{\vep(\vp_1)\vep(\vp_2)}{\vp_{12}^4}\right]_{\text{fin.}}
+O(\Phi_*\vep^3).
\end{aligned}
\end{equation}
Applying the renormalization $\tvep=4\pi\Phi_*\vep$,\, $\tilde{I}=I+\Phi_*L/(2\pi)$, we obtain the first term in equation~\eqref{dilact_eff} and the first term in~\eqref{dilact_eff_c}.

\subsubsection{Terms proportional to $\alpha$}

The variation of the on-shell action with respect to $\alpha$ with fixed boundary data satisfies the equation $\delta I_{\text{on-shell}}=\left({\p I \over \p \alpha}\right)\de \alpha$. Hence, in the linear order in $\alpha$, we have
\begin{equation} \label{osdilact}
I_{\text{on-shell}}-I_{\text{on-shell},\,\alpha=0}
\approx -{\al \ov 4 \pi}\biggl(\int_{D'} d^2x \sqrt{g}\,
\Phi^2\biggr)\bigg|_{\alpha=0}.
\end{equation}
To calculate the integral, we will use the Poincare metric in the $(w,\wc)$ or $(y,\vt)$ coordinates and solve the equation of motion for the dilaton, $\nabla^2\Phi=2\Phi$. The general solution can be written in terms of Fourier modes,
\begin{equation}\label{Phi_fm}
\Phi(y, \vt)=\Phi_* \sum_{n} f_{n}\,b_{n}(y)e^{in\vt},\qquad
b_n(y)=y^{|n|/2}(1-y)^{-1}\bigl(1+y+|n|(1-y)\bigr),
\end{equation}
where the coefficients $f_n$ are arbitrary, or using the boundary-to-bulk propagator:
\begin{equation}\label{Phi_bbprop}
\Phi(w, \wc)=\Phi_*\int\frac{d\vt_1}{2\pi}\,
\frac{(1-w\wc)^2}{(1-we^{-i\vt_1})^2(1-\wc e^{i\vt_1})^2}\,f(\vt_1),\qquad
\text{where}\quad
f(\vt)=\sum_{n} f_{n}e^{in\vt}.
\end{equation}
The main technical difficulty is that the integral of $\Phi^2$ diverges near the boundary of the unit disk, and thus has to be restricted to the region $D'$. To isolate the divergent terms and to locate the boundary of $D'$ (\ie the line where $\Phi(y,\vt)=\Phi_*$), we expand $\Phi$ in powers of $1-y$. This is the expansion of the $n$-th Fourier mode:
\begin{equation}\label{bulk_modes}
b_n(y)=\frac{2}{1-y}\left(
1-\frac{1}{2}(1-y)-\frac{n^2}{8}(1-y)^2
+\frac{|n|(|n|-2)(2|n|+1)}{48}(1-y)^3+\cdots
\right).
\end{equation}
Summing over $n$ yields this expression:
\begin{equation}\label{Phi_expansion}
\Phi(y,\vt)=\frac{2\Phi_*}{1-y} \left(
f-\frac{f}{2}(1-y)+\frac{f''}{8}(1-y)^{2}
+\biggl(\frac{f''}{16}+\frac{h}{2}\biggr)(1-y)^{3}
+\cdots\right),
\end{equation}
where
\begin{equation}
h(\vt)=\left[\int\frac{d\vt_1}{2\pi}
\biggl(2\sin\biggl(\frac{\vt-\vt_1}{2}\biggr)\biggr)^{-4} f(\vt_1)
\right]_{\text{fin.}}
=\sum_{n}\frac{|n|(n^2-1)}{12}\,f_{n}e^{in\vt}.
\end{equation}

Let us outline the subsequent strategy. In the first approximation,
\begin{equation} \label{fapprox}
f(\vth)\approx \frac{1-y_*(\vt)}{2}\approx \de(\vt)\approx \eps(\vp),\qquad 
\vt\approx\vp;
\end{equation}
the necessary corrections will be determined later. We focus on two types of terms in the effective action: all leading terms defined with $O(\vep^2)$ precision and the non-local terms proportional to $\vep^2$. The former arise from the divergent part of the integral of $\Phi^2$ over the unit disk that is truncated at $y=y_*(\vt)$. Such terms are expressed as boundary integrals of some local quantities. They will cancel upon the renormalization, leaving only the local term from the $\al=0$ action, but we want to get them correctly as a consistency check. Nonlocal contributions to the integral generally come from the central part of the disk. There is also a mixed contribution, wherein the non-local correction to the relelation between $f(\vth)$ and $y_*(\vt)$ due to the function $h$ in \eqref{Phi_expansion} slightly changes the magnitude of the most significant local term. All calculations are done up to local order $O(\vep^2)$ terms. However, the non-local contribution coming from the central part of the disk will turn out to have the same form as the non-local action we found in \eqref{Inlfin}, and there can be no local terms of order $\vep^2$ coming from the UV completion of such an action, as we argued in footnote~\ref{nolocarg} on page~\pageref{nolocarg}. Thus the final expression we obtain will be $O(\vep^3)$ accurate.

\paragraph{Local terms:} These are obtained by writing $\Delta I =I_{\text{on-shell}}-I_{\text{on-shell},\,\alpha=0}$ as an integral in the $(y,\vt)$ coordinates and expanding the integrand in powers of $1-y_*$. We keep all negative powers up to and including $(1-y_*)^{-2}$, which can be obtained from the first three terms in \eqref{Phi_expansion}. Thus,
\begin{equation}\label{locact1}
\begin{aligned}
\Delta I&=-\alpha\int\frac{d\vt}{2\pi}\int_{0}^{y_*(\vt)}
dy\,\frac{\Phi^2}{(1-y)^2}\\[3pt]
&\approx -\alpha\Phi_*^2 \int\frac{d\vt}{2\pi}\left(
\frac{f^2}{6}\biggl(\frac{1-y_*}{2}\biggr)^{-3}
-\frac{f^2}{2}\biggl(\frac{1-y_*}{2}\biggr)^{-2}
+\frac{f^2+ff''}{2}\biggl(\frac{1-y_*}{2}\biggr)^{-1}\right),
\end{aligned}
\end{equation}
where $y_*$ satisfies the equation $\Phi(y_*(\vt),\vt)=\Phi_*$. On the other hand, $y_*$ can be expressed in terms of $\de=d\vt/d\ell$ using the Poincare metric. (Such an expression was previously obtained in \eqref{ystar_K} but here we need more accuracy.) We proceed with the calculation:
\begin{gather}
\label{ystar}
\frac{1-y_*}{2} 
=\de\left(1-\de+\frac{1}{2}\bigl(\de^2+\de'^2\bigr)+O(\de^3)\right),\\[3pt]
f=\de\left(1-\frac{1}{2}\bigl(\de^2-\de'^2+\de\de''\bigr)+O\bigl(\de^3\bigr)\right),
\displaybreak[0]\\[5pt]
\label{dilact_local}
(\Delta I)_{\loc} = -\frac{\alpha\Phi_*^2}{6} \int\frac{d\vt}{2\pi}\,\de^{-1}\left(
1+\frac{\de^2}{2}-\frac{\de'^2}{2}+2\de\de''+O(\de^3)\right)
=\frac{\alpha\Phi_*}{6}\,I_{\text{on-shell},\,\alpha=0}.
\end{gather}
(The coefficient $2$ in front of $\de\de''$ is not important because this expression is multiplied by $\de^{-1}$ and becomes a full derivative.)

\begin{figure}[t]
\centerline{\includegraphics[scale=1.0]{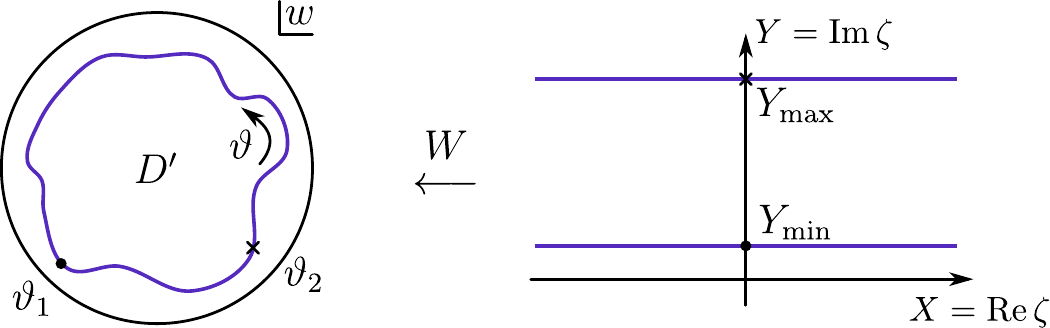}}
\caption{To evaluate the non-local part of the action, we identify the domain of the Poincare disk with the upper half-plane and approximate $D'$ as shown on the right-hand-side.}
\label{fig: maptoplane}
\end{figure}

\paragraph{Nonlocal contribution to the integral:} Let us express $\Phi(w,\wc)$ using the boundary-to-bulk propagator from \eqref{Phi_bbprop} and perform the integral over the region $D'$:
\begin{equation}
\Delta I
=-\frac{\alpha}{4\pi}\int_{D'} d^2x \sqrt{g}\,\Phi^2
=-\alpha\Phi_*^2 \int{d\vt_1 \ov 2 \pi}\int{d\vt_2 \ov 2 \pi}\,
\textit{ker}(\vt_1,\vt_2)\,f(\vt_1)f(\vt_2),
\end{equation}
where
\begin{equation}\label{nlkernel}
\textit{ker}(\vt_1,\vt_2)=\frac{1}{\pi}\int_{D'} (d\Im w) (d\Re w) \left(
\frac{1-w\wc}{(1-w e^{-i\vt_1}) (1-\wc e^{i\vt_1})
(1-w e^{-i\vt_2}) (1-\wc e^{i\vt_2})}
\right)^2.
\end{equation}
We only consider the contributions from pairs of point $\vt_1$, $\vt_2$ that are sufficiently far apart. In this case, the integral diverges logarithmically as $w$ approaches $e^{i\vt_1}$ or $e^{i\vt_2}$, with the cutoff determined by $y_*(\vt_1)$ or $y_*(\vt_2)$, respectively. The integral can be evaluated using a conformal map $W$ from the upper half-plane $\{\zeta:\, \Im \zeta\geq 0\}$ to the unit disk $\{w:\, \abs{w} \leq 1\}$ such that $W(0)=e^{i\vt_1}$ and $W(\infty)=e^{i\vt_2}$, see Figure~\ref{fig: maptoplane}. Specifically,
\begin{equation}
w=W(\zeta)=\frac{e^{i\vt_2/2}\zeta-e^{i\vt_1/2}}
{e^{-i\vt_2/2}\zeta-e^{-i\vt_1/2}}\,,
\end{equation}
 This leads to
\begin{equation} \label{Keval}
\textit{ker}(\vt_1,\vt_2)= {1 \ov \pi}\,\vt_{12}^{-4}
\int_{W^{-1}(D')} (d\Re\zeta)(d\Im\zeta)\,
\left(\frac{i(\zeta-\bar{\zeta})}{\zeta\bar{\zeta}}\right)^2
\approx 2\vt_{12}^{-4}
\int_{Y_{\text{min}}}^{Y_{\text{max}}} Y^{-1}dY,
\end{equation}
where
\begin{equation}
\vt_{12}=2\sin\frac{\vt_1-\vt_2}{2},\qquad
Y=\Im\zeta,\qquad
Y_{\text{min}}=\frac{1-y_*(\vt_1)}{2|\vt_{12}|},\quad
Y_{\text{max}}=\frac{2|\vt_{12}|}{1-y_*(\vt_2)}.
\end{equation}
Thus, we find that the on-shell action contains the non-local term
\begin{equation}\label{dilact_nl}
(\Delta I)_{\nloc}
=-2\alpha \Phi_*^2\left[\int{d\vt_1 \ov 2 \pi}{d\vt_2 \ov 2 \pi}\,
\frac{\de(\vt_1)\de(\vt_2)}{\vt_{12}^4}\,
\ln\frac{\vt_{12}^2}{\de(\vp_1)\de(\vp_2)}\right]_{\text{fin.}}.
\end{equation}

\paragraph{Mixed contribution:} Let us revisit the $1-y$ expansion of $\Phi$ near the boundary, see \eqref{Phi_expansion}. This time we are interested in the effect of the non-local term containing $h$, which was previously ignored. If the boundary $y=y_*(\vt)$ is fixed, then the addition of the $h$ term changes the value of $f$ by
\begin{equation}
f_h=-\frac{h}{2}(1-y_*)^3\approx -4h\de^3.
\end{equation}
Replacing $f$ by $f+f_h$ in \eqref{locact1} modifies the result by this amount:
\begin{equation}\label{dilact_mixed}
(\Delta I)_{\text{mixed}}
=-\frac{\alpha\Phi_*^2}{6}\int\frac{d\vt}{2\pi}\,2f\kern1pt f_h\,\de^{-3}
=\frac{4}{3}\alpha\Phi_*^2\left[\int{d\vt_1 \ov 2 \pi}{d\vt_2 \ov 2 \pi}\,
\frac{\de(\vt_1)\de(\vt_2)}{\vt_{12}^4}\right]_{\text{fin.}}.
\end{equation}

\paragraph{Putting everything together:} First, we add the local term~\eqref{dilact_local} to the expression~\eqref{osdilact0} or~\eqref{osdilact0f} for the $\alpha=0$ effective action. Then we take into account the non-local and mixed contributions. Because they are higher order, it is safe to replace in them $\de$ with $\vep$ and $\vt$ with $\vp$. The result is as follows:
\begin{align}
I_{\text{on-shell}}
=&-\biggl(1+\frac{\alpha\Phi_*}{6}\biggr)\Phi_* \left(
\int \frac{d\vp}{2\pi}\,\vep^{-1}
\left(1+\frac{\vep^2}{2}-\frac{\vep'^2}{2}+\vep\vep''\right)
+12\left[\int\frac{d\vp_1}{2\pi}\frac{d\vp_2}{2\pi}\,
\frac{\vep(\vp_1)\vep(\vp_2)}{\vp_{12}^4}\right]_{\text{fin.}}\right)
\nonumber\\[3pt]
&-2\alpha \Phi_*^2\left[\int\frac{d\vp_1}{2\pi}\frac{d\vp_2}{2\pi}\,
\frac{\vep(\vp_1)\vep(\vp_2)}{\vp_{12}^4}
\left(\ln\biggl(\frac{\vp_{12}^2}{\vep(\vt_1)\vep(\vt_2)}\biggr)-\frac{2}{3}
\right)\right]_{\text{fin.}}+O(\Phi_*\vep^3).
\end{align}
Here we assume that $\alpha\Phi_*\ll 1$ so that any term proportional to $\alpha\Phi_*^2\vep^3$ is within the error bound. As already mentioned, local terms of $O(\alpha\Phi_*^2\eps^2)$, which would be part of the UV completion of the non-local integral above, are forbidden. The renormalization $\tvep=4\pi\sqrt{W(\Phi_*)}\,\vep$, $\tilde{I}=I+\sqrt{W(\Phi_*)}\,L/(2\pi)$ where $\sqrt{W(\Phi_*)}\approx\bigl(1+\frac{\alpha\Phi_*}{6}\bigr)\Phi_*$ yields the expression \eqref{dilact_eff}. 

\section{Open questions}

\begin{enumerate}
\item We have divided the degrees of freedom of the SYK model, described by the function $\tG$, into the soft mode $\vp$ and $h\not=2$ discrete and continuous series representations of $\PSL(2,\RR)$ (see Section~\ref{sec_gensoft}). While the soft mode is responsible for the leading four-point function, the next order terms are mixed. Is there a better way to separate the variables? The more specific problem is that the definition of $\vp$ or the related physical observable $\calO(\theta)=\Sch(e^{i\vp},\theta)$ is non-local. Perhaps one could define $\calO(\theta)$ as the coefficient in front of the $(\theta_1-\theta_2)^2$ term in the expansion of $\tG(\theta_1,\theta_2)$, but we found it difficult to implement this idea.
\item Is it possible to understand the correction to the Lyapunov exponent as some sort of friction coefficient in an effective model? What does it correspond to in the bulk picture?
\item As Witten noticed \cite{Wi16}, ``the average of a quantum system over quenched disorder is not really a quantum system''. However, the replica-diagonal effective action works pretty well. When does it start producing nonsensical results such as violation of a unitarity bound? (We do not mean the non-unitarity of the $S$-matrix discussed in Section~\ref{sec_OTOC}, which is a failure of a much cruder model.) How bad are the resulting problems? If they are mild, should we consider violation of unitarity in the real world?
\item Assuming that the replica-diagonal action is consistent under given circumstances, how do we construct an effective Hilbert space not using quenched disorder?
\end{enumerate}

\section*{Acknowledgements}
We thank Juan Maldacena, Douglas Stanford, and Yingfei Gu for useful discussions. We gratefully acknowledge the support by the Simons Foundation through the ``It from Qubit'' program. A.K.\ is supported by the Simons Foundation under grant~376205 and by the Institute of Quantum Information and Matter, a NSF Frontier center funded in part by the Gordon and Betty Moore Foundation. The work of J.S.\ was supported in part by the Natural Sciences and Engineering Research Council of Canada and by the Simons Foundation through grant 376206. It was performed in part at the Aspen Center for Physics, which is supported by National Science Foundation grant PHY-1607611. 

\appendix

\section{Derivation of the effective action $I[\Sigma,G]$}
\label{sec_replicas}

Let us consider the $q=4$ case of the SYK Hamiltonian:
\begin{equation}
\hat{H} =-\frac{1}{4!}
\sum_{j,k,l,m}J_{jklm}\hat{\chi}_j\hat{\chi}_k\hat{\chi}_l\hat{\chi}_m,
\qquad\quad
\overline{J_{jklm}^2}=\frac{3!J^2}{N^3}.
\end{equation}
The exact form of probability distribution is not very important when $N$ is large, but we will assume that it is Gaussian. The averaging of an arbitrary function $f$ over $\V{J}=(J_{jklm})$ can be performed as follows:
\useshortskip{\begin{gather}
\overline{f(\V{J})}=\int\calD\V{B}\,
f\biggl(\sqrt{\tfrac{3!J^2}{N^3}}\,\V{B}\biggr),\\[3pt]
\text{where}\quad
\calD\V{B}
=\exp\left(-\frac{1}{2}\sum_{j<k<l<m}B_{jklm}^2\right)
\prod_{j<k<l<m}\!\frac{dB_{jklm}}{\sqrt{2\pi}}.
\end{gather}}

The average value of the free energy is given by the formula
\begin{equation}
\beta\overline{F}=-\overline{\ln Z}
=-\lim_{M\to 0}\frac{\ln\overline{Z^{M}}}{M}.
\end{equation}
The standard prescription is to find $\overline{Z^{M}}$ for integer $M$ in an analytic form and then take $M$ to $0$. For each realization $\V{J}$ of the disorder, $Z(\V{J})^M$ is equal to the partition function of $M$ replicas of the model. Thus, we consider an extended set of variables:
\begin{equation}
\V{\chi}
=\bigl(\chi_j^\alpha:\: j=1,\dots,N,\:\, \alpha=1,\dots,M\bigr).
\end{equation}
In the functional integral formalism, one should actually use the Grassmann variables $\chi_j^\alpha(\tau)$ parametrized by $\tau\in[0,\beta]$ with the antiperiodic boundary conditions, $\chi_j^\alpha(\beta)=-\chi_j^\alpha(0)$. We proceed with the calculation.
\begin{align}
\nonumber
\overline{Z^M}&=\int\calD\V{B}\,\int\calD\V{\chi}\,
\exp\left(\sum_{\alpha}\int_0^\beta\! d\tau\biggr(
-\frac{1}{2}\sum_{j}\chi_j^\alpha\,\partial_\tau\chi_j^\alpha
+\sum_{j<k<l<m}\sqrt{\tfrac{3!J^2}{N^3}}B_{jklm}\,
\chi_j^\alpha\chi_k^\alpha\chi_l^\alpha\chi_m^\alpha\biggl)\right)
\\[3pt]
\nonumber
&=\int\calD\V{\chi}\, \exp\left(
-\frac{1}{2}\sum_{\alpha,j}\int\!d\tau\,
\chi_j^\alpha\,\partial_\tau\chi_j^\alpha
+\frac{3!J^2}{2N^3}\sum_{j<k<l<m}
\biggl(\sum_{\alpha}\int\!d\tau\,
\chi_j^\alpha(\tau)\chi_k^\alpha(\tau)\chi_l^\alpha(\tau)\chi_m^\alpha(\tau)
{\biggl)\!}^2\, \right)
\\[3pt]
\label{action4}
&=\int\calD\V{\chi}\, \exp\Biggr(
-\frac{1}{2}\sum_{\alpha,j}\int\!d\tau\,
\chi_j^\alpha\,\partial_\tau\chi_j^\alpha
+\frac{NJ^2}{8}\sum_{\alpha,\beta}\iint\!d\tau\,d\tau'\,
{\,\underbrace{\!
\biggl(\frac{1}{N}\sum_{j} \chi_j^\alpha(\tau)\chi_j^\beta(\tau')\biggl)\!}
_{\Xi_{\alpha\beta}(\tau,\tau')}}^4\, \Biggr).
\end{align}

Here, we are presented with the problem of decoupling the nonlinear term $\Xi^4$, where $\Xi=\Xi_{\alpha\beta}(\tau,\tau')=\frac{1}{N}\sum_{j} \chi_j^\alpha(\tau)\chi_j^\beta(\tau')$. It can be solved using this identity:
\begin{equation}\label{fdec}
f(\Xi)=\int_{-\infty}^{+\infty} dx\,f(x)\,\delta(x-\Xi)
=\frac{N}{2\pi}\int_{-\infty}^{+\infty} dx \int_{-\infty}^{+\infty}dy\:
f(x)\,e^{iNy(x-\Xi)}.
\end{equation}
The last expression may be interpreted as an integral over a particular real plane in the two-dimensional complex space. The integration surface can be rotated to ensure fast decay at infinity. For example, one may use the substitution $x=u+iv$,\, $y=i(u-iv)$, where $(u,v)$ runs over $\RR^2$. This improves the convergence as the leading term in the exponent, $iNyx=-N(u^2+v^2)$ is strongly negative. Applying this method to functions like $f(\Xi)=e^{\Xi^4}$ is still risky, but we will proceed anyway. The integral \eqref{fdec} should be incorporated into the larger expression \eqref{action4} and evaluated in the large $N$ limit using the saddle point approximation.

There is, actually, a separate instance of $\Xi$ for each degree of freedom, \ie a combination of $\alpha$, $\beta$, and $\tau>\tau'$. (The case $\tau<\tau'$ is redundant because $\Xi_{\alpha\beta}(\tau,\tau') =-\Xi_{\beta\alpha}(\tau',\tau)$.) The corresponding instances of $x$ and $y$ will be denoted by $-G_{\alpha\beta}(\tau,\tau')$ and $-i\Sigma_{\alpha\beta}(\tau,\tau')$, respectively. We now apply \eqref{fdec} to $f(\Xi)=\exp\bigl(\frac{NJ^2}{4}\Xi^4\bigr)$:
\begin{align}\label{expXi4}
&\exp\Biggl(\frac{NJ^2}{4}
\sum_{\alpha,\beta}\,\int\limits_{\tau>\tau'}\! d\tau\,d\tau'\,
\Xi_{\alpha\beta}(\tau,\tau')^4\Biggr)\\[2pt]
&\quad=\int\calD\Sigma\: \calD G\,
\exp\Biggl(N\sum_{\alpha,\beta}\,\int\limits_{\tau>\tau'}\! d\tau\,d\tau'
\biggl(\frac{J^2}{4}G_{\alpha\beta}(\tau,\tau')^4
-\Sigma_{\alpha\beta}(\tau,\tau')
\Bigl(G_{\alpha\beta}(\tau,\tau')+\Xi_{\alpha\beta}(\tau,\tau')\Bigr)
\!\biggr)\!\Biggr),\nonumber
\end{align}
where the integration measure $\calD\Sigma\,\calD G$ includes the factor $iN/(2\pi)$ for each degree of freedom, so that
\begin{equation}
\wideboxed{
\int\calD\Sigma\: \calD G\,\exp\biggl(-\frac{N}{2}\sum_{\alpha,\beta}\,
\int d\tau\,d\tau'\,
\Sigma_{\alpha\beta}(\tau,\tau')\,G_{\alpha\beta}(\tau,\tau')\biggr)
=1
}
\end{equation}

It remains to combine \eqref{action4} and \eqref{expXi4}, and substitute $\frac{1}{N}\sum_{j} \chi_j^\alpha(\tau)\chi_j^\beta(\tau')$ for $\Xi_{\alpha\beta}(\tau,\tau')$. In the resulting expression, the integral over $\V{\chi}$ factors into $N$ identical integrals, each corresponding to a particular value of $j$. Thus, the result can be written in terms of a smaller set of Grassmann variables, $\chi=(\chi^\alpha:\, \alpha=1,\dots,M)$:
\begin{align*}
\overline{Z^M}=\int\calD \Sigma\: \calD G \Biggl(\int\calD\chi
\exp\Biggl(\!
&-\frac{1}{2}\sum_{\alpha}\int d\tau\,\chi^\alpha\,\partial_\tau\chi^\alpha
-\frac{1}{2}\sum_{\alpha,\beta}\int d\tau\,d\tau'\,
\Sigma_{\alpha\beta}(\tau,\tau')\,\chi^\alpha(\tau)\chi^\beta(\tau')\\[2pt]
&+\frac{1}{2}\sum_{\alpha,\beta}\int d\tau\,d\tau'\biggl(
\frac{J^2}{4}G_{\alpha\beta}(\tau,\tau')^4
-\Sigma_{\alpha\beta}(\tau,\tau')\,G_{\alpha\beta}(\tau,\tau')\biggr)
\!\Biggr)\!{\Biggr)\!}^N.
\end{align*}
The integral over the Grassmann variables $\chi^\alpha(\tau)$ is equal to the Pfaffian of the operator $-\partial_\tau-\hat{\Sigma}$. It is, strictly speaking, UV-divergent, but we may use this regularization:
\begin{equation}
\Pf\bigl(-\partial_\tau-\hat{\Sigma}\bigr)
=2^{M/2}\Pf(-\partial_\tau-\hat{\Sigma})/\Pf(-\partial_\tau),
\end{equation}
where the ratio of the two Pfaffians is defined unambigously. In the large $N$ limit, the outer integrals $\calD\Sigma\,\calD G$ can be performed by finding a saddle point. The result is as follows:
\begin{equation}\label{minmax_replicas}
-\ln\overline{Z^M}=-\ln\biggl(\int\calD \Sigma\: \calD G\:
\exp\bigl(-I^{(M)}[\Sigma,G]\bigr)\!\biggr)
\approx\min_{\Sigma}\,\max_{G}I^{(M)}[\Sigma,G],
\end{equation}
where
\begin{equation}\label{effact_replicas}
\wideboxed{
I^{(M)}[\Sigma,G]
=N\Biggl(\!-\ln\Pf(-\partial_\tau-\hat{\Sigma})
+\frac{1}{2}\sum_{\alpha,\beta}\int\! d\tau\,d\tau'
\biggl(\Sigma_{\alpha\beta}(\tau,\tau')\,G_{\alpha\beta}(\tau,\tau')
-\frac{J^2}{4}G_{\alpha\beta}(\tau,\tau')^4\biggr)\!\Biggr)
}
\end{equation}
The maximum is attained at the $G$ that satisfies the equation $\Sigma(\tau,\tau')=J^2 G(\tau,\tau')^3$, and the minimization over $\Sigma$ yields the equation $(-\partial_\tau-\hat\Sigma)^{-1}=\hat{G}$. Thus, the saddle point values of $G$ and $\Sigma$ are exactly the Green function and the self-energy in the mean field approximation.

Note that the functional integral in \eqref{minmax_replicas} gives all perturbative $1/N$ corrections to the saddle point. However, it might not capture nonperturbative effects because its derivation involved manipulation of formally divergent integrals. Perhaps one can get the correct result by carefully choosing the integration surface; this question requires additional study.\smallskip

The most natural solution for the minimum over $\Sigma$ is diagonal in replicas:
\begin{equation}
\Sigma_{\alpha\beta}(\tau,\tau')=\Sigma(\tau,\tau')\,\delta_{\alpha\beta}.
\end{equation}
With this ansatz, taking the $M\to 0$ limit is trivial, and \eqref{minmax_replicas}, \eqref{effact_replicas} are simplified as follows:
\begin{gather}
\label{minmax_rd}
-\overline{\ln Z}\approx -\ln\overline{Z}
= -\ln\biggl(\int\calD \Sigma\: \calD G\:
\exp\bigl(-I[\Sigma,G]\bigr)\!\biggr)
\approx\min_{\Sigma}\,\max_{G}I[\Sigma,G],\\[5pt]
\label{effact_rd}
I[\Sigma,G]
=N\Biggl(-\ln\Pf(-\partial_\tau-\hat{\Sigma})
+\frac{1}{2}\int d\tau\,d\tau'
\biggl(\Sigma(\tau,\tau')\,G(\tau,\tau')
-\frac{J^2}{4}G(\tau,\tau')^4\biggr)\!\Biggr).
\end{gather}
In the mean field approximation, the action \eqref{effact_rd} is equivalent to the full action $I^{(M)}$ (with $M\to 0$ replicas), provided $T>T_{\text{glass}}$. 

Let us briefly discuss the use of the replica-diagonal approximation beyond mean field. The functional integral in \eqref{minmax_rd} gives the disorder-averaged partition function $\overline{Z}$. The high temperature expansion of $-\ln\overline{Z}$ includes all connected diagrams, whereas the expansion of $\beta F=-\overline{\ln Z}$ consists of those diagrams that are connected along fermionic lines. The leading diagram that belongs to the first but not the second class is this one:
\begin{equation}
\figbox{1.0}{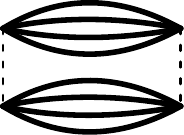}\quad\propto N^{-2}.
\end{equation}
It is, actually, an infinite sum of diagrams because each thick solid line represents $G_*$ (denoted by $G$ in \eqref{G_hte}). Thus, the replica-diagonal approximation gives the free energy with $O(N^{-2})$ accuracy.

It is interesting to derive the high temperature expansion of $-\ln\overline{Z}$ from action \eqref{effact_rd} rather than the original Hamiltonian. This way, one gets the same set of diagrams, but with a different interpretation. Instead of carrying site indices $j$, $k$, etc., the fermionic lines are grouped (essentially, by that index) and organized into $2n$-gons. We begin with breaking the exponent $-I[\Sigma,G]$ in the functional integral into the main part,
\begin{equation}
-I_0[\Sigma,G]
=-\frac{N}{2}\int d\tau\,d\tau'\,\Sigma(\tau,\tau')\,G(\tau,\tau')
\end{equation}
and two perturbation terms, containing the Pfaffian and $G^4$. The expectation value of any polynomial in $\Sigma$ and $G$ with weight $\exp(-I_0[\Sigma,G])$ is easily calculated. For example,
\begin{equation}\label{SG0avrege}
\bcorr{\Sigma(\tau_1,\tau_2)\,G(\tau_3,\tau_4)}_{0}
=N^{-1}\Bigl(\delta(\tau_1-\tau_3)\,\delta(\tau_2-\tau_4)
-\delta(\tau_1-\tau_4)\,\delta(\tau_2-\tau_3)\Bigr).
\end{equation}

Now, let us Taylor-expand the perturbation terms, in particular,
\begin{equation}
N\ln\frac{\Pf(-\partial_\tau-\hat{\Sigma})}{\Pf(-\partial_\tau)}
=-N\biggl(
\begin{array}[t]{@{}l@{}}\displaystyle
\frac{1}{2}\,\underbrace{\Trr(\hat{G}_{\bb}\hat{\Sigma})}
+\frac{1}{4}\,\underbrace{\Trr(\hat{G}_{\bb}\hat{\Sigma})^2}
+\frac{1}{6}\,\underbrace{\Trr(\hat{G}_{\bb}\hat{\Sigma})^3}
+\cdots\\
\hspace{25pt}\figbox{1.0}{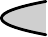}
\hspace{48pt}\figbox{1.0}{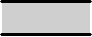}
\hspace{50pt}\figbox{1.0}{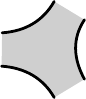}
\end{array}
\biggr),\vspace{-3pt}
\end{equation}
In these diagrams, the solid lines represent $\hat{G}_{\bb}=(-\partial_\tau)^{-1}$, and a copy of $\hat{\Sigma}$ is attached to each open end (\ie a side without a solid line). As is usual, the number of symmetries ($2,4,6,\ldots$) appears in the denominator. The elementary diagrams in the above expression (without the $\Sigma$'s or the symmetry coefficients) will be called ``sheets''. The factor of $-N$ is still included. Such sheets can be connected at the open ends by four-fold ``seams'', depicted as dotted lines. The seams arise from the perturbation term $N(J^2/4)G^4$. For the use in the diagrammatic calculus, the coefficient in that expression should be multiplied by the combinatorial factor $4!$ and also by $N^{-4}$; the last factor comes from \eqref{SG0avrege}. Thus, each seam carries the weight $3!J^2/N^{3}$, of which we associate $J^2$ with the dotted line itself and keep $3!/N^{3}$ separate.

A closed, connected diagram represents a contribution to $\ln\overline{Z}$. A diagram with $n$ open ends corresponds to the correlation function
\begin{equation}\label{corfunc}
\V{G}(\tau_{1},\tau_{1}',\dots,\tau_{n},\tau_{n}')
=(-1)^n\sum_{j_1,\dots,j_n}
\bcorr{\TT\chi_{j_1}(\tau_{1})\chi_{j_1}(\tau_{1}')
\dots \chi_{j_n}(\tau_{n})\chi_{j_n}(\tau_{n}')}.
\end{equation}
The overall sign of a diagram is obtained using the following recipe:
\begin{enumerate}
\item Orient all sides of each sheet clockwise or counterclockwise; orient each seam.
\item A solid line from $l$ to $k$ represents $G_{\bb}(\tau_k,\tau_l)$.
\item Each open end should be oriented from $\tau_{s}'$ to $\tau_{s}$, or else a minus sign is introduced. Each orientation conflict between a sheet and an adjacent seam gives a minus sign.
\end{enumerate}
To illustrate these rules, we apply them to one diagram that contributes to $\V{G}(\tau,\tau')=NG(\tau,\tau')$:
\begin{equation}
\frac{1}{3!}\,\figbox{1.0}{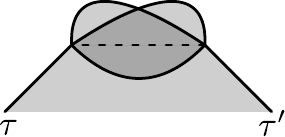}\:
=\:\frac{1}{3!}\,(-1)^4 \figbox{1.0}{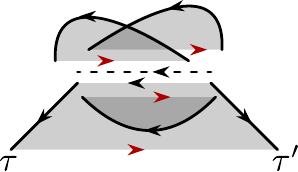} (-N)^4\frac{3!}{N^3}\:
=\:N\,\figbox{1.0}{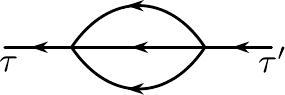}\:.
\end{equation}

\begin{figure}[t]
\centering
\includegraphics[scale=1]{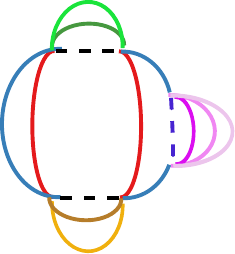}\qquad \qquad \includegraphics[scale=0.63]{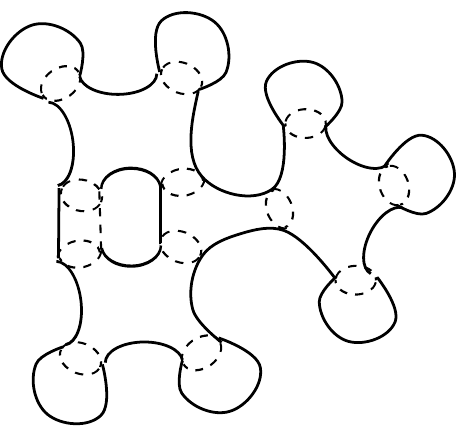}
\centering
\caption{Left: A Feynman diagram in the free energy of the disorder-averaged $q=4$ SYK model, given by $-\beta^{-1} \overline{\ln Z}=-\beta^{-1}\lim_{M \to 0}M^{-1}Z^M$ with $Z^M$ as in \eqref{action4}. Site indices for fermions have been distinguished with color. Right: Corresponding three-dimensional handlebody.}\label{fig: handlebody}
\end{figure}

Finally, let us describe the mapping of Feynman diagrams to three-dimensional handlebodies whose genus counts the powers of $N$ in the diagram. We can work with diagrams with respect to the full disorder-averaged partition function, \ie which are connected along fermionic lines. For diagrams with external fermions we count the powers of $N$ as in \eqref{corfunc}. Then the power of $N$ in a diagram is given by $(1-q)(\text{number of disorder-averaged interactions})+(\text{number of fermion indices})$. An Euler-like interpretation of this counting is
\begin{align}
&\underbrace{(\# \text{ of interactions})}_{V/2} - \underbrace{(\#\text{ of times a fermion threads an interaction})}_{=q (\# \text{ of interactions})=E/2} + \underbrace{(\# \text{ of fermion indices})}_{F/2}
\nonumber\\[2pt]
&=\underbrace{(\# \text{ of factors of $N$})}_{1-g}.
\end{align}
That $V$ and $E$ are not independent implies we do not have the most general tiling of surfaces. In fact, each diagram naturally maps to a three-dimensional handlebody with corresponding genus $g$ as follows: We map each disorder-averaged interaction and fermion site index to a $3$-ball. If a fermion line threads an interaction, we connect corresponding $3$-balls with a $1$-handle; see Figure~\ref{fig: handlebody}. For the two-dimensional diagrams constructed above with sheets and seams, a handlebody corresponding to a diagram can be visualized as the tubular neighborhood of an embedding of it in $\mathbb{R}^3$. 

\section{Energy-momentum tensor for pure dilaton gravity}\label{sec_pure_dilaton}

The pure dilaton gravity on a manifold $X$ of arbitrary dimensionality is described by this Euclidean action:
\begin{equation}\label{dilaton_pure}
I=-\frac{1}{4\pi}\int_{X} \Phi R\,\sqrt{g}\,dx
-\frac{1}{2\pi}\int_{\partial X}\Phi K\sqrt{\bar{g}}\,d\bar{x}.
\end{equation}
The extrinsic curvature is defined as $K=\nabla_{\bar{\nu}}n^{\bar{\nu}}$, where $n$ is the unit normal vector and the bar indicates boundary variables. The second term in \eqref{dilaton_pure} is necessary to set up the variational problem with the Dirichlet boundary conditions. For this problem to be well-defined, it is necessary that the variation of the action with fixed $\Phi|_{\partial X}$ and $g_{\alpha\beta}|_{\partial X}$ be an integral over the bulk without any boundary term.

We will use the tetrad formalism. Let $v(x)$ be an orthonormal local frame at point $x$ and $\theta(x)$ the dual frame, so that $g_{\alpha\beta}=\eta_{ab}\,\theta^a_\alpha \theta^b_\beta$, where $\eta$ is the unit matrix. The covariant derivative acts on Greek (tangent space) indices but not Latin (fixed frame) indices. That is, one should distinguish between $\nabla_{\mu}u^{c}=\partial_{\mu}u^{c}$ and $(\nabla_{\mu}u)^{c} =\partial_{\mu}u^{c} +\tensor{\omega}{_\mu^c_b}u^{b}$, but $\nabla_{\mu}u^{\nu} =(\nabla_{\mu}u)^{\nu} =v_c^{\nu}(\nabla_{\mu}u)^{c}$.

Let us consider a variation of $v_a^\alpha(x)$ that vanishes at the boundary. The spin connection coeffitients $\tensor{\omega}{_\nu^a_b}$ are initially regarded as independent variables, and then expressed in terms of the metric:
\begin{equation}
\tensor{\omega}{_\gamma_\alpha_\beta} 
=\frac{1}{2}\bigl(
\tensor{C}{_\gamma_\alpha_\beta}
-\tensor{C}{_\alpha_\beta_\gamma} 
+\tensor{C}{_\beta_\gamma_\alpha}\bigr),\qquad
\text{where}\quad
\tensor{C}{_\alpha_\beta^l}
=\partial_\alpha\theta^l_\beta-\partial_\beta\theta^l_\alpha.
\end{equation}
In general, this procedure gives the following result:
\begin{align}
\label{varact1}
\delta I&=\int_{X} \tensor{\tilde{T}}{^\mu_\nu}\theta^c_\mu(\delta e^\nu_c)
\sqrt{g}\,dx
-\frac{1}{2}\int \tensor{J}{^\nu^a^b}(\delta\tensor{\omega}{_\nu_a_b})
\sqrt{g}\,dx\\[2pt]
\label{varact2}
&=\int_{X} \tensor{T}{^\mu_\nu}\theta^c_\mu(\delta e^\nu_c)
\sqrt{g}\,dx,
\end{align}
where $\tilde{T}$ is the canonical energy-momentum tensor, $J$ the spin current, and $T$ the full (Belinfante-Rosenfeld) energy-momentum tensor:
\begin{equation}
\tensor{T}{^\alpha^\beta}
=\tensor{\tilde{T}}{^\alpha^\beta} -\frac{1}{2}\nabla_{\gamma}\bigl(
J^{\alpha\beta\gamma} -J^{\beta\gamma\alpha} +J^{\gamma\alpha\beta}
\bigr).
\end{equation}

We now carry out the calculations for the concrete action \eqref{dilaton_pure}. These are the expressions for the variation of local quantities:
\begin{gather}
\delta\tensor{R}{_\mu_\nu^a_b}
=\tensor{(\nabla_\mu(\delta\omega_\nu))}{^a_b}
-\tensor{(\nabla_\nu(\delta\omega_\mu))}{^a_b},\qquad\quad
\delta R=2R^b_\nu(\delta v_{b}^{\nu})
+2\nabla_\mu(v_{a}^{\mu}v_{b}^{\nu}(\delta\tensor{\omega}{_\nu^a^b})),
\displaybreak[0]\\[5pt]
\delta\sqrt{g}=-\theta^{b}_{\nu}(\delta v_{b}^{\nu}),
\displaybreak[0]\\[5pt]
\delta K =v_{a}^{\nu} (\delta\tensor{\omega}{_\nu^a_b})n^b\qquad
(\text{if } \delta v_{a}^{\nu}|_{\partial X}=0).
\end{gather}
When calculating the variation of action, the boundary term that comes from the integration of $\Phi \nabla_\mu(v_{a}^{\mu}v_{b}^{\nu} (\delta\tensor{\omega}{_\nu^a^b}))$ by parts cancels the one from the extrinsic curvature. The result has the form \eqref{varact1}, where
\begin{equation}
\tilde{T}_{\alpha\beta}
=-\frac{1}{2\pi}\Phi\Bigl(R_{\alpha\beta}
-\frac{1}{2}R\kern1pt g_{\alpha\beta}\Bigr),\qquad
J_{\alpha\beta\gamma}=\frac{1}{2\pi}\Bigl(
(\nabla_{\gamma}\Phi)g_{\alpha\beta} -(\nabla_{\beta}\Phi)g_{\alpha\gamma}
\Bigr).
\end{equation}
Thus,
\begin{equation}\label{T_pure}
\wideboxed{
T_{\alpha\beta}=\frac{1}{2\pi}\biggl(
-\Phi\Bigl(R_{\alpha\beta}-\frac{1}{2}R\kern1pt g_{\alpha\beta}\Bigr)
+\bigl(\nabla_{\alpha}\nabla_{\beta}-g_{\alpha\beta}\nabla^2\bigr)\Phi
\biggr)
}
\end{equation}

\bibliography{soft_mode_refs}

\providecommand{\href}[2]{#2}\begingroup\raggedright\begin{thebibliography}{10}

\bibitem{SaYe93}
S.~Sachdev and J.~Ye, \emph{{Gapless spin fluid ground state in a random,
  quantum Heisenberg magnet}},
  \href{https://doi.org/10.1103/PhysRevLett.70.3339}{\emph{Phys. Rev. Lett.}
  {\bfseries 70} (1993) 3339},
  [\href{https://arxiv.org/abs/cond-mat/9212030}{{\ttfamily
  cond-mat/9212030}}].

\bibitem{Kit.KITP.1}
A.~Kitaev, ``Hidden correlations in the {Hawking} radiation and thermal
  noise.'' Talk at KITP
  \url{http://online.kitp.ucsb.edu/online/joint98/kitaev/}, February, 2015.

\bibitem{Kit.KITP}
A.~Kitaev, ``A simple model of quantum holography.'' Talks at KITP
  \url{http://online.kitp.ucsb.edu/online/entangled15/kitaev/} and
  \url{http://online.kitp.ucsb.edu/online/entangled15/kitaev2/}, April and May,
  2015.

\bibitem{PaGe99}
O.~{Parcollet} and A.~{Georges}, \emph{{Non-Fermi-liquid regime of a doped Mott
  insulator}}, \href{https://doi.org/10.1103/PhysRevB.59.5341}{\emph{Phys. Rev.
  B} {\bfseries 59} (Feb., 1999) 5341--5360},
  [\href{https://arxiv.org/abs/cond-mat/9806119}{{\ttfamily
  cond-mat/9806119}}].

\bibitem{GePaSa00}
A.~{Georges}, O.~{Parcollet} and S.~{Sachdev}, \emph{{Quantum fluctuations of a
  nearly critical Heisenberg spin glass}},
  \href{https://doi.org/10.1103/PhysRevB.63.134406}{\emph{Phys. Rev. B}
  {\bfseries 63} (Apr., 2001) 134406},
  [\href{https://arxiv.org/abs/cond-mat/0009388}{{\ttfamily
  cond-mat/0009388}}].

\bibitem{BaAlKa16}
D.~Bagrets, A.~Altland and A.~Kamenev, \emph{{Sachdev--Ye--Kitaev model as
  Liouville quantum mechanics}},
  \href{https://doi.org/10.1016/j.nuclphysb.2016.08.002}{\emph{Nucl. Phys.}
  {\bfseries B911} (2016) 191--205},
  [\href{https://arxiv.org/abs/1607.00694}{{\ttfamily 1607.00694}}].

\bibitem{Sach10}
S.~Sachdev, \emph{{Strange metals and the AdS/CFT correspondence}},
  \href{https://doi.org/10.1088/1742-5468/2010/11/P11022}{\emph{J. Stat. Mech.}
  {\bfseries 1011} (2010) P11022},
  [\href{https://arxiv.org/abs/1010.0682}{{\ttfamily 1010.0682}}].

\bibitem{DtH85}
T.~Dray and G.~'t~Hooft, \emph{{The gravitational shock wave of a massless
  particle}}, \href{https://doi.org/10.1016/0550-3213(85)90525-5}{\emph{Nucl.
  Phys.} {\bfseries B253} (1985) 173--188}.

\bibitem{tH87}
G.~'t~Hooft, \emph{{Strings From Gravity}},
  \href{https://doi.org/10.1088/0031-8949/1987/T15/019}{\emph{Phys. Scripta}
  {\bfseries T15} (1987) 143}.

\bibitem{tH90}
G.~'t~Hooft, \emph{{The black hole interpretation of string theory}},
  \href{https://doi.org/10.1016/0550-3213(90)90174-C}{\emph{Nucl. Phys.}
  {\bfseries B335} (1990) 138--154}.

\bibitem{tH96}
G.~'t~Hooft, \emph{{The Scattering matrix approach for the quantum black hole:
  An Overview}}, \href{https://doi.org/10.1142/S0217751X96002145}{\emph{Int. J.
  Mod. Phys.} {\bfseries A11} (1996) 4623--4688},
  [\href{https://arxiv.org/abs/gr-qc/9607022}{{\ttfamily gr-qc/9607022}}].

\bibitem{LaOv69}
A.~I. Larkin and Y.~N. Ovchinnikov, \emph{Quasiclassical method in the theory
  of superconductivity}, {\emph{Soviet Physics, JETP} {\bfseries 28} (1969)
  1200--1205}.

\bibitem{ShSt13}
S.~H. Shenker and D.~Stanford, \emph{{Black holes and the butterfly effect}},
  \href{https://doi.org/10.1007/JHEP03(2014)067}{\emph{JHEP} {\bfseries 03}
  (2014) 067}, [\href{https://arxiv.org/abs/1306.0622}{{\ttfamily 1306.0622}}].

\bibitem{ShSt14}
S.~H. Shenker and D.~Stanford, \emph{{Stringy effects in scrambling}},
  \href{https://doi.org/10.1007/JHEP05(2015)132}{\emph{JHEP} {\bfseries 05}
  (2015) 132}, [\href{https://arxiv.org/abs/1412.6087}{{\ttfamily 1412.6087}}].

\bibitem{Kit.BPS}
A.~Kitaev, ``Hidden correlations in the {Hawking} radiation and thermal
  noise.'' Talk at Breakthrough Prize Symposium
  \url{https://www.youtube.com/watch?v=OQ9qN8j7EZI}, December, 2014.

\bibitem{SeSu08}
Y.~Sekino and L.~Susskind, \emph{{Fast scramblers}},
  \href{https://doi.org/10.1088/1126-6708/2008/10/065}{\emph{JHEP} {\bfseries
  10} (2008) 065}, [\href{https://arxiv.org/abs/0808.2096}{{\ttfamily
  0808.2096}}].

\bibitem{MSS15}
J.~Maldacena, S.~H. Shenker and D.~Stanford, \emph{{A bound on chaos}},
  \href{https://doi.org/10.1007/JHEP08(2016)106}{\emph{JHEP} {\bfseries 08}
  (2016) 106}, [\href{https://arxiv.org/abs/1503.01409}{{\ttfamily
  1503.01409}}].

\bibitem{BrMo80}
A.~J. Bray and M.~A. Moore, \emph{Replica theory of quantum spin glasses},
  \href{https://doi.org/10.1088/0022-3719/13/24/005}{\emph{Journal of Physics
  C: Solid State Physics} {\bfseries 13} (1980) L655}.

\bibitem{PoRo16}
J.~Polchinski and V.~Rosenhaus, \emph{{The spectrum in the Sachdev-Ye-Kitaev
  model}}, \href{https://doi.org/10.1007/JHEP04(2016)001}{\emph{JHEP}
  {\bfseries 04} (2016) 001},
  [\href{https://arxiv.org/abs/1601.06768}{{\ttfamily 1601.06768}}].

\bibitem{MS16}
J.~Maldacena and D.~Stanford, \emph{{Remarks on the Sachdev-Ye-Kitaev model}},
  \href{https://doi.org/10.1103/PhysRevD.94.106002}{\emph{Phys. Rev.}
  {\bfseries D94} (2016) 106002},
  [\href{https://arxiv.org/abs/1604.07818}{{\ttfamily 1604.07818}}].

\bibitem{Jen16}
K.~Jensen, \emph{{Chaos in AdS$_2$ Holography}},
  \href{https://doi.org/10.1103/PhysRevLett.117.111601}{\emph{Phys. Rev. Lett.}
  {\bfseries 117} (2016) 111601},
  [\href{https://arxiv.org/abs/1605.06098}{{\ttfamily 1605.06098}}].

\bibitem{MSY16}
J.~Maldacena, D.~Stanford and Z.~Yang, \emph{{Conformal symmetry and its
  breaking in two dimensional nearly Anti-de-Sitter space}},
  \href{https://doi.org/10.1093/ptep/ptw124}{\emph{PTEP} {\bfseries 2016}
  (2016) 12C104}, [\href{https://arxiv.org/abs/1606.01857}{{\ttfamily
  1606.01857}}].

\bibitem{EMV16}
J.~Engels{\"o}y, T.~G. Mertens and H.~Verlinde, \emph{{An investigation of
  AdS$_{2}$ backreaction and holography}},
  \href{https://doi.org/10.1007/JHEP07(2016)139}{\emph{JHEP} {\bfseries 07}
  (2016) 139}, [\href{https://arxiv.org/abs/1606.03438}{{\ttfamily
  1606.03438}}].

\bibitem{JeSuYo16}
A.~Jevicki, K.~Suzuki and J.~Yoon, \emph{{Bi-local holography in the SYK
  model}}, \href{https://doi.org/10.1007/JHEP07(2016)007}{\emph{JHEP}
  {\bfseries 07} (2016) 007},
  [\href{https://arxiv.org/abs/1603.06246}{{\ttfamily 1603.06246}}].

\bibitem{randmat}
J.~S. Cotler, G.~Gur-Ari, M.~Hanada, J.~Polchinski, P.~Saad, S.~H. Shenker
  et~al., \emph{{Black holes and random matrices}},
  \href{https://doi.org/10.1007/JHEP05(2017)118}{\emph{JHEP} {\bfseries 05}
  (2017) 118}, [\href{https://arxiv.org/abs/1611.04650}{{\ttfamily
  1611.04650}}].

\bibitem{JeSu16}
A.~Jevicki and K.~Suzuki, \emph{{Bi-local holography in the SYK model:
  perturbations}}, \href{https://doi.org/10.1007/JHEP11(2016)046}{\emph{JHEP}
  {\bfseries 11} (2016) 046},
  [\href{https://arxiv.org/abs/1608.07567}{{\ttfamily 1608.07567}}].

\bibitem{BaAlKa17}
D.~Bagrets, A.~Altland and A.~Kamenev, \emph{{Power-law out of time order
  correlation functions in the SYK model}},
  \href{https://doi.org/10.1016/j.nuclphysb.2017.06.012}{\emph{Nucl. Phys.}
  {\bfseries B921} (2017) 727--752},
  [\href{https://arxiv.org/abs/1702.08902}{{\ttfamily 1702.08902}}].

\bibitem{StWi17}
D.~Stanford and E.~Witten, \emph{{Fermionic Localization of the Schwarzian
  Theory}}, \href{https://doi.org/10.1007/JHEP10(2017)008}{\emph{JHEP}
  {\bfseries 10} (2017) 008},
  [\href{https://arxiv.org/abs/1703.04612}{{\ttfamily 1703.04612}}].

\bibitem{Sach15}
S.~Sachdev, \emph{{Bekenstein-Hawking entropy and strange metals}},
  \href{https://doi.org/10.1103/PhysRevX.5.041025}{\emph{Phys. Rev.} {\bfseries
  X5} (2015) 041025}, [\href{https://arxiv.org/abs/1506.05111}{{\ttfamily
  1506.05111}}].

\bibitem{FuGaMaSa16}
W.~Fu, D.~Gaiotto, J.~Maldacena and S.~Sachdev, \emph{{Supersymmetric
  Sachdev-Ye-Kitaev models}}, \href{https://doi.org/10.1103/PhysRevD.95.069904,
  10.1103/PhysRevD.95.026009}{\emph{Phys. Rev.} {\bfseries D95} (2017) 026009},
  [\href{https://arxiv.org/abs/1610.08917}{{\ttfamily 1610.08917}}].

\bibitem{GrRo16}
D.~J. Gross and V.~Rosenhaus, \emph{{A generalization of Sachdev-Ye-Kitaev}},
  \href{https://doi.org/10.1007/JHEP02(2017)093}{\emph{JHEP} {\bfseries 02}
  (2017) 093}, [\href{https://arxiv.org/abs/1610.01569}{{\ttfamily
  1610.01569}}].

\bibitem{Wi16}
E.~Witten, \emph{{An SYK-like model without disorder}},
  \href{https://arxiv.org/abs/1610.09758}{{\ttfamily 1610.09758}}.

\bibitem{KlGr16}
I.~R. Klebanov and G.~Tarnopolsky, \emph{{Uncolored random tensors, melon
  diagrams, and the Sachdev-Ye-Kitaev models}},
  \href{https://doi.org/10.1103/PhysRevD.95.046004}{\emph{Phys. Rev.}
  {\bfseries D95} (2017) 046004},
  [\href{https://arxiv.org/abs/1611.08915}{{\ttfamily 1611.08915}}].

\bibitem{Gu16}
Y.~Gu, X.-L. Qi and D.~Stanford, \emph{{Local criticality, diffusion and chaos
  in generalized Sachdev-Ye-Kitaev models}},
  \href{https://doi.org/10.1007/JHEP05(2017)125}{\emph{JHEP} {\bfseries 05}
  (2017) 125}, [\href{https://arxiv.org/abs/1609.07832}{{\ttfamily
  1609.07832}}].

\bibitem{PeSpVo17}
C.~Peng, M.~Spradlin and A.~Volovich, \emph{{A Supersymmetric SYK-like Tensor
  Model}}, \href{https://doi.org/10.1007/JHEP05(2017)062}{\emph{JHEP}
  {\bfseries 05} (2017) 062},
  [\href{https://arxiv.org/abs/1612.03851}{{\ttfamily 1612.03851}}].

\bibitem{L-MKu94}
D.~Louis-Martinez and G.~Kunstatter, \emph{Birkhoff's theorem in
  two-dimensional dilaton gravity},
  \href{https://doi.org/10.1103/PhysRevD.49.5227}{\emph{Phys. Rev. D}
  {\bfseries 49} (May, 1994) 5227--5230}.

\bibitem{GrRo17}
D.~J. Gross and V.~Rosenhaus, \emph{{The bulk dual of SYK: cubic couplings}},
  \href{https://doi.org/10.1007/JHEP05(2017)092}{\emph{JHEP} {\bfseries 05}
  (2017) 092}, [\href{https://arxiv.org/abs/1702.08016}{{\ttfamily
  1702.08016}}].

\bibitem{SL2R}
A.~Kitaev, \emph{Notes on {$\widetilde{\mathrm{SL}}(2,\mathbb{R})$}
  representations},  \href{https://arxiv.org/abs/1711.08169}{{\ttfamily
  1711.08169}}.

\bibitem{Rep76}
J.~Repka, \emph{Tensor products of unitary representations of $\mathrm{SL}_2
  \left(\mathbb{R} \right)$},
  \href{https://doi.org/10.1090/S0002-9904-1976-14223-1}{\emph{Bull. Amer.
  Math. Soc.} {\bfseries 82} (11, 1976) 930--932}.

\bibitem{Rep78}
J.~Repka, \emph{Tensor products of unitary representations of
  $\mathrm{SL}_2(\mathbb{R})$},
  \href{https://doi.org/10.2307/2373909}{\emph{American Journal of Mathematics}
  {\bfseries 100} (1978) 747--774}.

\bibitem{RSS14}
D.~A. Roberts, D.~Stanford and L.~Susskind, \emph{{Localized shocks}},
  \href{https://doi.org/10.1007/JHEP03(2015)051}{\emph{JHEP} {\bfseries 03}
  (2015) 051}, [\href{https://arxiv.org/abs/1409.8180}{{\ttfamily 1409.8180}}].

\bibitem{KVV95}
Y.~Kiem, H.~L. Verlinde and E.~P. Verlinde, \emph{{Black hole horizons and
  complementarity}},
  \href{https://doi.org/10.1103/PhysRevD.52.7053}{\emph{Phys. Rev.} {\bfseries
  D52} (1995) 7053--7065},
  [\href{https://arxiv.org/abs/hep-th/9502074}{{\ttfamily hep-th/9502074}}].

\bibitem{Pol15}
J.~Polchinski, \emph{{Chaos in the black hole S-matrix}},
  \href{https://arxiv.org/abs/1505.08108}{{\ttfamily 1505.08108}}.

\bibitem{FeVe63}
R.~P. Feynman and F.~L. Vernon, Jr., \emph{{The Theory of a general quantum
  system interacting with a linear dissipative system}},
  \href{https://doi.org/10.1016/0003-4916(63)90068-X}{\emph{Annals Phys.}
  {\bfseries 24} (1963) 118--173}.

\bibitem{ShSt13.1}
S.~H. Shenker and D.~Stanford, \emph{{Multiple Shocks}},
  \href{https://doi.org/10.1007/JHEP12(2014)046}{\emph{JHEP} {\bfseries 12}
  (2014) 046}, [\href{https://arxiv.org/abs/1312.3296}{{\ttfamily 1312.3296}}].

\bibitem{Ja85}
R.~Jackiw, \emph{Lower dimensional gravity},
  \href{https://doi.org/10.1016/0550-3213(85)90448-1}{\emph{Nuclear Physics B}
  {\bfseries 252} (1985) 343 -- 356}.

\bibitem{Te83}
C.~Teitelboim, \emph{{Gravitation and Hamiltonian structure in two space-time
  dimensions}}, \href{https://doi.org/10.1016/0370-2693(83)90012-6}{\emph{Phys.
  Lett.} {\bfseries 126B} (1983) 41--45}.

\bibitem{AlPo14}
A.~Almheiri and J.~Polchinski, \emph{{Models of AdS$_{2}$ backreaction and
  holography}}, \href{https://doi.org/10.1007/JHEP11(2015)014}{\emph{JHEP}
  {\bfseries 11} (2015) 014},
  [\href{https://arxiv.org/abs/1402.6334}{{\ttfamily 1402.6334}}].

\bibitem{GKP98}
S.~S. Gubser, I.~R. Klebanov and A.~M. Polyakov, \emph{{Gauge theory
  correlators from noncritical string theory}},
  \href{https://doi.org/10.1016/S0370-2693(98)00377-3}{\emph{Phys. Lett.}
  {\bfseries B428} (1998) 105--114},
  [\href{https://arxiv.org/abs/hep-th/9802109}{{\ttfamily hep-th/9802109}}].

\bibitem{Wi98}
E.~Witten, \emph{{Anti-de Sitter space and holography}}, {\emph{Adv. Theor.
  Math. Phys.} {\bfseries 2} (1998) 253--291},
  [\href{https://arxiv.org/abs/hep-th/9802150}{{\ttfamily hep-th/9802150}}].

\bibitem{Mal16}
J.~Maldacena. Private communication.

\bibitem{BaOL91}
T.~Banks and M.~O'Loughlin, \emph{{Two-dimensional quantum gravity in Minkowski
  space}}, \href{https://doi.org/10.1016/0550-3213(91)90547-B}{\emph{Nucl.
  Phys.} {\bfseries B362} (1991) 649--664}.

\end{thebibliography}\endgroup
\bibliographystyle{JHEP}
 
\end{document}